\documentclass[aps,prd,nofootinbib,onecolumn,superscriptaddress,preprintnumbers,balancelastpage,longbibliography,nobibnotes]{revtex4-1}

\usepackage[dvipsnames]{xcolor}
\usepackage{graphicx,color,amsmath,amssymb,flushend,bm,mathrsfs,comment}
\definecolor{linkcolor}{rgb}{0.0, 0.28, 0.67}

\usepackage[
   colorlinks=true,
    urlcolor=linkcolor,
   anchorcolor=linkcolor,
    citecolor=linkcolor,
    filecolor=linkcolor,
    linkcolor=linkcolor,
    menucolor=linkcolor,
    linktocpage=true,
    pdfproducer=medialab,
    pdfa=true
]{hyperref}

\usepackage[capitalise]{cleveref}
\usepackage{tikz}
\usepackage{ulem}
\usepackage{float}
\usepackage{amsmath}
\usepackage{physics}
\usepackage{simpler-wick}
\usepackage{amsfonts}
\usepackage{amssymb}
\usepackage[mathscr]{euscript}
\usepackage{setspace}% http://ctan.org/pkg/setspace
\usepackage{lipsum}% http://ctan.org/pkg/lipsum
\usepackage{slashed}
\usepackage{cancel}
\usepackage{multirow}
\usepackage[utf8]{inputenc}
\usepackage{mathtools}
\usepackage{braket}
\usepackage{bbold}
\usepackage{amsmath}
\usepackage{soul}
\usepackage{makecell}
\usepackage{chngcntr} % to change figure and table labels in appendices

\usepackage{placeins}

\usepackage{fontawesome5}
\usepackage{orcidlink}

\usetikzlibrary{calc}
\usepackage{siunitx}
\DeclareSIUnit{\year}{yr}
\DeclareSIUnit{\parsec}{pc}
\DeclareSIUnit{\eV}{e\kern-.05em V}
\DeclareSIUnit{\Jansky}{Jy}
\DeclareSIUnit{\sr}{sr}

\newcommand{\nocontentsline}[3]{}
\newcommand{\tocless}[2]{\bgroup\let\addcontentsline=\nocontentsline#1{#2}\egroup}

\def\dbar{{\mathchar'26\mkern-12mu \dd}}

\newcommand{\bea}{\begin{eqnarray}\begin{aligned}}
\newcommand{\eea}{\end{aligned}\end{eqnarray}}

\newcommand{\const}{\text{const}}

\newcommand{\Mpl}{M_\text{pl}}

\newcommand{\alphaem}{\alpha_\text{em}}

\newcommand{\eff}{\text{eff}}

\newcommand{\atm}{\text{atm}}

\newcommand{\eV}{\text{eV}}

\newcommand{\mev}{\text{MeV}}
\newcommand{\gev}{\text{GeV}}

\newcommand{\Eq}[1]{Eq.~(\ref{eq:#1})}

\newcommand{\formfactorreduce}{\widetilde{\mathcal{F}}}

\newcommand{\tot}{\text{tot}}

\newcommand{\psiinc}{\psi_\text{inc}}
\newcommand{\psisc}{\psi_\text{sc}}

\newcommand{\psiout}{\psi_\text{out}}

\newcommand{\psiint}{\psi_\text{int}}

\newcommand{\km}{\text{km}}

\newcommand{\bg}{\text{bg}}

\newcommand{\Sec}[1]{Sec.~\ref{sec:#1}}
\newcommand{\Subsec}[1]{Sec.~\ref{subsec:#1}}
\newcommand{\Appx}[1]{Appendix~\ref{appx:#1}}

\newcommand{\Fig}[1]{Fig.~\ref{fig:#1}}

\newcommand{\sigmak}{\sigma_k}

\newcommand{\mM}{m_\text{M}}
\newcommand{\mMtest}{m_{\text{M},\mathcal{T}}}
\newcommand{\mMpt}{m_{\text{M},\text{Pt}}}
\newcommand{\mMti}{m_{\text{M},\text{Ti}}}
\newcommand{\mMsat}{m_{\text{M},\text{sat}}}
\newcommand{\mMatm}{m_{\text{M},\text{atm}}}
\newcommand{\mMsource}{m_{\text{M},\mathcal{S}}}
\newcommand{\mMearth}{m_{{\rm M},\oplus}}

\newcommand{\vecr}{\mathbf{r}}
\newcommand{\vecrhat}{\hat{\mathbf{r}}}
\newcommand{\veckhat}{\hat{\mathbf{k}}}
\newcommand{\vecx}{\mathbf{x}}

\newcommand{\vecz}{\mathbf{z}}
\newcommand{\veca}{\mathbf{a}}
\newcommand{\veck}{\mathbf{k}}

\newcommand{\vecv}{\mathbf{v}}

\newcommand{\vecF}{\mathbf{F}}

\newcommand{\vecnabla}{\pmb{\nabla}}

\newcommand{\vectheta}{\boldsymbol{\theta}}

\newcommand{\cyg}{\text{cyg}}

\newcommand{\testmass}{\mathcal{T}}
\newcommand{\sourcemass}{\mathcal{S}}

\newcommand{\esc}{\text{esc}}

\newcommand{\vecX}{\hat{\mathbf{X}}}
\newcommand{\vecY}{\hat{\mathbf{Y}}}
\newcommand{\vecZ}{\hat{\mathbf{Z}}}

\newcommand{\EP}{\text{EP}}

\newcommand{\kg}{\text{kg}}

\newcommand{\obs}{\text{obs}}

\newcommand{\formfactorV}{\mathcal{F}}

\newcommand{\sph}{\text{sph}}

\newcommand{\mhat}{{\hat{m}}}

\newcommand{\ti}{\text{Ti}}
\newcommand{\pt}{\text{Pt}}

\newcommand{\bmtx}{\begin{pmatrix}}
\newcommand{\emtx}{\end{pmatrix}}

\newcommand{\calA}{\mathcal{A}}

\newcommand{\calE}{\mathcal{E}}

\newcommand{\calV}{\mathcal{V}}
\newcommand{\calR}{\mathcal{R}}

\newcommand{\rmass}{{\text{rm}}}
\newcommand{\bind}{{\text{bind}}}

\newcommand{\sigmapi}{\sigma_{\pi N}}

\newcommand{\Nsens}{N_\text{sens}}

\newcommand{\orb}{\text{orb}}
\newcommand{\spin}{\text{spin}}

\newcommand{\thetarhat}{\hat{\boldsymbol{\theta}}_\vecr}

\newcommand{\vecL}{\mathbf{L}}

\newcommand{\LSR}{\text{LSR}}

\newcommand{\ECL}{\mathrm{ECL}}
\newcommand{\EQU}{\mathrm{EQU}}

\newcommand{\inst}{\text{inst}}

\newcommand{\sat}{\text{sat}}

\newcommand{\yr}{\mathrm{yr}}

\def\beq{\begin{equation}}
\def\eeq{\end{equation}}

\usepackage{amsmath}

\begin{document}

\title{Background-Induced Forces from Quadratically Coupled Ultralight Dark Matter}

\preprint{DESY-26-085}

\author{Thomas Bouley\orcidlink{0009-0008-9757-054X}}
\email{tbouley@uoregon.edu}
\affiliation{Humanities \& Sciences Division, Cincinnati State Technical and Community College, Cincinnati OH 45223 USA}

\author{Xucheng Gan\orcidlink{0000-0003-2834-7498}}
\email{xucheng.gan@desy.de}
\affiliation{Deutsches Elektronen-Synchrotron DESY, Notkestr. 85, 22607 Hamburg, Germany}

\author{Hailin Xu\orcidlink{0000-0002-9528-2068}}
\email{hailin.xu@sjtu.edu.cn}
\affiliation{State Key Laboratory of Dark Matter Physics, Tsung-Dao Lee Institute \& School of Physics and Astronomy, Shanghai Jiao Tong University, Shanghai 200240, China}
\affiliation{Key Laboratory for Particle Astrophysics and Cosmology (MOE) \& Shanghai Key Laboratory for Particle Physics and Cosmology, Shanghai Jiao Tong University, Shanghai 200240, China}

\author{Tien-Tien Yu\orcidlink{0000-0003-4708-809X}}
\email{tientien@uoregon.edu}
\affiliation{Department of Physics and Institute for Fundamental Science, University of Oregon, Eugene OR 97403 USA}

\graphicspath{{figs/}}

\begin{abstract}
Quadratically coupled ultralight scalar dark matter behaves as a coherent classical field whose interactions with matter can induce a composition-dependent force through the dark matter background. We present a complete calculation of this background-induced force beyond the spherically symmetric approximation. Using a partial-wave treatment of dark-matter scattering, we determine its angular dependence and derive an analytic description valid even when the dark-matter wavelength is much smaller than the Earth's radius. We show for the first time that Earth screening generates a characteristic frequency-band structure, splitting the signal into multiple sidebands that provide a distinctive experimental signature. We further show that the relative amplitudes of these sidebands vary annually as the satellite's Sun-synchronous orbital plane precesses in step with the Earth's revolution around the Sun, allowing us to construct a complete signal template. As an application of these results, we re-evaluate constraints from the MICROSCOPE mission, which currently provides the strongest laboratory limits on equivalence-principle violations from ultralight dark matter. We further show that proposed space-based equivalence-principle experiments, such as Galileo Galilei and STE-QUEST, can significantly enhance their sensitivity to ultralight scalar dark matter by incorporating the full frequency-band information.
\end{abstract}

\maketitle

\tableofcontents
\section{Introduction}
The particle nature of dark matter (DM) is one of the leading puzzles of particle physics. There is a vast zoo of potential candidates, but one particularly compelling category of candidates is ultralight dark matter (ULDM)~\cite{Antypas:2022asj}. 
For the masses of interest, the Galactic halo DM density corresponds to occupation numbers far above unity.
As a result, the DM must be bosonic and behaves as a coherent classical wave on astrophysical scales. 
This wave-like nature gives rise to a variety of distinctive signatures in precision measurements. 
Through its interactions with Standard Model (SM) fields, ULDM can induce variations in fundamental constants~\cite{Damour:2010rm,Damour:2010rp,Stadnik:2015kia,Uzan:2024ded}. Such variations can be probed using precision measurements with, for example, atomic clocks~\cite{Arvanitaki:2014faa,VanTilburg:2015oza,Hees:2016gop,Wcislo:2018ojh,kennedy2020precision,Filzinger:2023zrs,Filzinger:2023qqh}, searches for anomalous accelerations and fifth forces~\cite{Olive:2007aj,Graham:2015ifn,Hees:2018fpg,VanTilburg:2024xib,Burrage:2026loe}, atom interferometers~\cite{AEDGE:2019nxb,Badurina:2019hst,MAGIS-100:2021etm,Zhao:2021din}, pulsar timing arrays~\cite{Porayko:2018sfa,Kaplan:2022lmz,PPTA:2022eul,Gan:2025icr}, space-based laser interferometers~\cite{Morisaki:2018htj,Grote:2019uvn,Gan:2026vph}, black-hole scalar sirens~\cite{Gavilan-Martin:2026zzw}, S-star spectroscopy~\cite{Bai:2025yxm}, dark-matter echoes~\cite{Arza:2019nta,Gong:2023ilg}, and stellar mass-radius relation~\cite{Bartnick:2025lbg,Bartnick:2025lbv}.
These diverse effects have motivated a broad experimental program, making laboratory, cosmological, and astrophysical observations increasingly sensitive probes of ULDM over a wide range of masses and couplings. A review of such efforts can be found in Ref.~\cite{Banerjee:2022sqg}.

Of particular interest are scalar ULDM models that give rise to composition-dependent forces and can therefore be probed through equivalence principle (EP) tests.  
Models with linear couplings are already strongly constrained by weak-EP and fifth-force searches~\cite{Damour:2010rm,Wagner:2012ui,Berge:2017ovy,Touboul:2022yrw}. 
If a symmetry suppresses the linear interaction, however, quadratic couplings become the leading observable effect and give rise to qualitatively different phenomenology. 
In contrast to the linear case, quadratic couplings cause the ULDM field to be sensitive to the surrounding matter environment, and the DM background mediates an additional force that is affected by the finite density and size of massive bodies such as the Earth.
This is known as the background-induced force.
Since the strongest constraints for much of the scalar ULDM parameter space currently come from the MICROSCOPE EP test~\cite{Hees:2018fpg,Touboul:2022yrw}, an accurate treatment of the background-induced force is essential for reliably interpreting existing bounds and assessing the sensitivity of future experiments.

Depending on whether the interaction increases or decreases the DM mass in finite-density media, matter effects give rise to either repulsive or attractive potentials, respectively. Previous studies of the repulsive case have largely relied on spherically symmetric approximations~\cite{Hees:2018fpg,Berezhiani:2018oxf,Banerjee:2022sqg} or analyses restricted to selected scattering directions~\cite{Gan:2025nlu}, while treatments including the full angular dependence have been limited to the perturbative regime~\cite{VanTilburg:2024xib}. 
The complementary attractive case has also been explored and exhibits distinct phenomenology~\cite{Banerjee:2025dlo,delCastillo:2025rbr}. 
The motion of the Earth through the Galactic DM halo introduces a preferred direction, rendering the force intrinsically anisotropic. 
In particular, the force is no longer spherically symmetric once the ULDM de Broglie wavelength becomes comparable to the Earth's radius, corresponding roughly to DM mass $m_\phi \gtrsim 10^{-11}$ eV.
Moreover, the resulting DM wake can occur for arbitrary orientations relative to the Earth's motion through the halo.
This is particularly relevant for experiments such as MICROSCOPE, whose orbit samples a range of DM-Earth interaction geometries over time. Accurately modeling the resulting signal therefore requires accounting for the full angular dependence of the interaction, together with an appropriate averaging over the ULDM velocity distribution.

In this work we develop a scattering-theory framework that incorporates phase-space averaging, applies to both scenarios, and remains valid beyond the spherically symmetric approximation. For concreteness, we focus on the repulsive potential case. Using this framework, we compute the scalar profile and derive its optical-limit description. We further show that the resulting anisotropic force produces a characteristic frequency-band structure in EP experiments, revealing a previously unexplored signature of quadratically coupled ULDM. These results enable a consistent treatment of terrestrial and space-based probes across the relevant ULDM parameter space.

This paper is organized as follows. In \Sec{scalar_models}, we review several representative models of quadratically coupled scalar ULDM, including a low-energy effective theory, the Higgs portal, a universally coupled scalar, and the QCD axion. 
In \Sec{bg_force}, we develop a quantum mechanical scattering-theory description of the background-induced force beyond the spherical approximation. 
We introduce a background-induced-potential form factor to quantify the deviation from spherical symmetry. Our established formalism can be used to compute the background-induced force across the full parameter space.
In \Sec{microscope}, we apply this formalism to EP experiments, using the setup of the MICROSCOPE mission as an example. 
We show that the angular dependence of the force produces a distinctive frequency-band structure in the experimental signal, providing a new phenomenological signature of quadratically coupled ULDM. 
We then derive updated MICROSCOPE constraints and demonstrate that future space-based EP tests, such as Galileo Galilei and STE-QUEST, can substantially improve their sensitivity to ULDM by leveraging the full frequency-band information.
Additional technical details are presented in the appendices. Appendix~\ref{appx:dilaton_charge_appx} derives the dilaton charges relevant for ULDM couplings, Appendix~\ref{appx:sph_ansatz_appx} reviews the spherically symmetric approximation employed in previous work, Appendix~\ref{appx:partial_wave_appx} provides details of the partial-wave analysis, and Appendix~\ref{appx:geometric_optics} derives the optical-limit description and its connection to the full calculation in the high-mass, strongly coupled regime.

\section{Models}\label{sec:scalar_models}

We begin with a discussion of models of quadratically-coupled scalars, where we add a real scalar field $\phi$ to the SM. 
To eliminate the linear interaction term $\phi\,\mathcal{O}_{\rm SM}$ and ensure that the quadratic interaction is the leading contribution, we impose a $\mathbb{Z}_2$ symmetry on the
scalar sector under the transformation $\phi \to -\phi$.
The leading interactions with the SM then take the form
\begin{equation}
\label{eq:L_OSM}
\mathcal{L} \supset \frac{\phi^2}{2\Lambda^2}\,\mathcal{O}_{\rm SM}\,,
\end{equation}
where $\mathcal{O}_{\rm SM}$ is a CP-even operator in the SM Lagrangian, and $\Lambda$ denotes the suppression scale of the effective operator.

This section introduces the benchmark models used to explore the detectability of equivalence-principle (EP) violation in the upcoming sections. We proceed in three steps. In \Subsec{effective_coupling}, we present the effective scalar--SM couplings, which provide the widely adopted full-set parameterization of the varying fundamental constants. In \Subsec{higgs_portal}, we discuss the Higgs portal as a concrete UV completion that induces a specific linear combination of those couplings. Finally, we briefly comment on other scenarios, including the universal coupling and the light QCD axion. 
Throughout the remainder of this work, we focus primarily on the effective scalar--SM couplings of \Subsec{effective_coupling} and the Higgs portal of \Subsec{higgs_portal}.

\subsection{Effective Scalar--SM Couplings}\label{subsec:effective_coupling}

Following the conventions adopted in Refs.~\cite{Damour:2010rp,Damour:2010rm,Hees:2018fpg,Banerjee:2022sqg,Bouley:2022eer}, we parameterize the interactions between the scalar field and the SM by the Lagrangian
\bea
\label{eq:Ldamour}
{\cal L}_\text{int}  \supset 2\pi \frac{\phi^2}{M_{\rm pl}^2}  \bigg[\frac{d_e^{(2)}}{4 e^2}F_{\mu\nu}F^{\mu\nu}-\frac{d_g^{(2)} \beta_s}{2 g_s}G^A_{\mu\nu}G^{A\mu\nu}\vphantom{\sum_{i=u,d}} -d_{m_e}^{(2)}m_e \bar \psi_e \psi_e -\sum_{q=u,d}\big(d_{m_q}^{(2)}+\gamma_{m_q}d_g^{(2)}\big) \, m_q \, \bar\psi_q\psi_q \bigg]\,,
\eea
where $M_{\rm pl}=1.22\times 10^{19}$ GeV is the Planck mass, $A$ is the color index, $g_s$ is the QCD gauge coupling, $\beta_s = \partial g_s/\partial \log \mu$ is the QCD beta function, and $\gamma_{m_q} = - \partial \log m_q/ \partial \,\log\mu$ are the anomalous dimensions of the $u$ and $d$ quarks. The superscript ``$(2)$'' specifies that these are the {\it quadratic} (as opposed to linear) couplings of the scalar. The bare mass term of the scalar is taken to be
${\cal L}_\phi \supset - \frac{1}{2} m_\phi^2 \phi^2 \,$,
where $m_\phi$ denotes the bare scalar mass in the vacuum. 
The mass term is not protected from quadratic corrections, but there have been model-building efforts to address this by e.g. imposing a $\mathbb{Z}_N$ discrete symmetry~\cite{Hook:2018jle,Hook:2019mrd,Brzeminski:2020uhm,DiLuzio:2021pxd,DiLuzio:2021gos,Gan:2023wnp} or mirror symmetry~\cite{Delaunay:2025pho}, exploiting field-space boundaries~\cite{Becker:2025pgb}, or introducing a relaxion mechanism~\cite{Banerjee:2022sqg}.
In \Eq{Ldamour}, the effective scale $\Lambda_i \simeq \Mpl \, (1/4 \pi d_i^{(2)})^{1/2}$, where $i=e,g,m_e,m_u,m_d$. 
Note that, in the constraint plots presented in this work, we turn on only one coupling at a time while setting all other couplings to zero.

The interactions defined in Eq.~\eqref{eq:Ldamour} induce a $\phi$-dependence in the fundamental constants of the SM, parameterized by the couplings $d_i^{(2)}$. Following the renormalization-scale-invariant parametrization of Ref.~\cite{Damour:2010rp}, generalized here to the case of quadratic couplings, a background value of $\phi$ leads to the following variations of the constants:~\footnote{The subscript ``$e$'' in $d_e^{(2)}$ denotes the electric charge, since the $\phi^2 F^2$ interaction induces a variation in the fine-structure constant $\alpha_{\rm em}$. We clarify this notation to distinguish it from
$d_{m_e}^{(2)}$, the coupling multiplying $m_e \bar{\psi}_{e} \psi_e$, which governs the variation of the electron mass.}
\begin{align}
\label{eq:vary_const}
\frac{\Delta\alphaem}{\alphaem} = d_e^{(2)} \times 2\pi \frac{\phi^2}{\Mpl^2}\,, \quad \,\, \frac{\Delta\Lambda_{\rm QCD}}{\Lambda_{\rm QCD}}= d_g^{(2)} \times 2\pi\frac{\phi^2}{\Mpl^2}\,, \quad \,\, 
\frac{\Delta m_e}{m_e} =  d_{m_e}^{(2)} \times 2\pi \frac{\phi^2}{\Mpl^2}\,, \quad \,\, 
\frac{\Delta m_{q}}{m_{q}} =  d_{m_{q}}^{(2)} \times 2\pi \frac{\phi^2}{\Mpl^2}\,.
\end{align}
Here, $\alphaem \simeq 1/137$ is the fine-structure constant, $\Lambda_{\rm QCD}$ denotes the QCD confinement scale, $m_e$ denotes the electron mass, and $m_q\,\,(q=u,d)$ denote the up and down quark masses. A key feature of this parameterization is that the quantities in Eq.~\eqref{eq:vary_const} are renormalization-group~(RG) invariant, and therefore constitute genuine low-energy observables. This invariance is not accidental: it is built into the structure of the interaction Lagrangian Eq.~\eqref{eq:Ldamour} through the inclusion of the anomalous dimension $\gamma_{m_q}$ in the quark-mass coupling. Specifically, the operator coefficients $d_g^{(2)} \beta_s/2 g_s$ and 
$(d_{m_q}^{(2)} + \gamma_{m_q}\,d_g^{(2)})$ in 
Eq.~\eqref{eq:Ldamour} are chosen so that the physical variations $\Delta\Lambda_{\rm QCD}/\Lambda_{\rm QCD}$ and $\Delta m_q/m_q$ in Eq.~\eqref{eq:vary_const} are 
RG-scale independent~\cite{Damour:2010rp}.

Although the couplings defined in Eqs.~\eqref{eq:vary_const} may be used directly, it is more convenient to reorganize the up and down quark mass variation when matching onto atomic observables, since the relevant quantities are more naturally expressed in terms of the sum and difference of the up- and down-quark masses. We therefore define
\bea
\label{eq:hatm_deltam}
& \hat{m} \equiv \frac{m_u+m_d}{2}~~~~~~~\text{(symmetric),}\\
& \delta m \equiv m_d - m_u~~~~~~\text{(anti-symmetric)},
\eea
which correspond to the symmetric and antisymmetric combinations of the up- and down-quark masses, respectively. Here, $\hat{m}$ controls the pion mass through the Gell-Mann--Oakes--Renner relation given by $m_\pi^2 \propto \Lambda_{\rm QCD} \, \hat{m}$~\cite{Gell-Mann:1968hlm,Scherer:2002tk} and thereby dominates the quark-mass dependence of nuclear binding through pion exchange, while $\delta m$ characterizes isospin-symmetry breaking which thereby determines the neutron-proton mass difference. Using the definitions of $\hat{m}$ and $\delta m$ together with Eq.~\eqref{eq:vary_const}, we define the corresponding coupling strengths varying the quark masses as
\begin{align}
d_{\hat m}^{(2)}&\equiv \frac{m_d\,d_{m_d}^{(2)}+m_u\,d_{m_u}^{(2)}}{m_d+m_u}~~~~~~~\text{(symmetric),}\\
d_{\delta m}^{(2)}&\equiv \frac{m_d\,d_{m_d}^{(2)}-m_u\,d_{m_u}^{(2)}}{m_d-m_u}~~~~~~~\text{(anti-symmetric).}
\end{align}

Although the Lagrangian in \Eq{Ldamour} is widely used in the literature~(See Refs.~\cite{Damour:2010rm,Damour:2010rp,Hees:2018fpg,Bouley:2022eer} for example) to parametrize the full set of variations of physical constants, it is important to emphasize that it is formulated at the quark level, above the QCD confinement scale. To relate this framework to observables in low-energy experiments, one must understand how the variations of fundamental constants in \Eq{vary_const} affect both the rest masses of elementary particles and nuclear binding energies. This mapping is encoded in the dilaton charges, which were systematically studied in Refs.~\cite{Damour:2010rm,Damour:2010rp}. In \Appx{dilaton_charge_appx}, we provide a pedagogical introduction to this formalism and update the dilaton charges using modern experimental inputs. 

\subsection{Higgs Portal}\label{subsec:higgs_portal}

The Higgs-portal model provides an ultraviolet completion of the effective theory discussed in the previous section, and was studied in Refs.~\cite{Piazza:2010ye,Graham:2015ifn}.
The interaction Lagrangian takes the renormalizable form
\bea
\label{eq:L_Higgs_Portal}
\mathcal{L}_\text{int} \supset 4 \pi \frac{\phi^2}{\Mpl^2} \, d^{(2)}_H \, m_h^2 H^\dagger H \,,
\eea
where $H = (1/\sqrt{2})\,(0,\, v + h)^T$ in unitary gauge. This modifies the Higgs mass as
$m_h^2(\phi) = m_h^2\,\,\!\big(1 + 4\pi\,d_{H}^{(2)}\,\phi^2/\Mpl^2\big)$,
where $m_h \simeq 125\,\gev$ is the Higgs boson mass~\cite{ParticleDataGroup:2024cfk}. Assuming the Higgs self-coupling is unaffected, the Higgs vacuum expectation value is correspondingly shifted to
$v(\phi) = v\,\!\big(1 + 2\pi\,d_{H}^{(2)}\,\phi^2/\Mpl^2\big)$, where $v \simeq 246\,\gev$~\cite{ParticleDataGroup:2024cfk}.

At energies well below the Higgs mass, the $\phi$-dependent shift in the Higgs VEV induces shifts in the electron, up-quark, and down-quark masses. In addition, $\Lambda_\text{QCD}$ is indirectly modified, because the heavy quark
masses (top, bottom, charm) governing the renormalization-group flow are also shifted
by the scalar background~\cite{Piazza:2010ye,Graham:2015ifn}.
Matching to the effective theory in \Eq{Ldamour} gives
\bea
\label{eq:low_energy_dh}
d_e^{(2)} = \frac{\alphaem}{\pi} \, d_{H}^{(2)}\,, \quad \quad d_g^{(2)} =  \frac{2}{9} d_{H}^{(2)}\,, \quad \quad d^{(2)}_{m_e} = d^{(2)}_{\hat{m}} = d^{(2)}_{\delta m} = d_H^{(2)}\,.
\eea
The coupling $d_e^{(2)}$, which governs the variation of the fine-structure constant, arises from the combination of the $\phi^2 h$ vertex and the $h\gamma\gamma$ vertex~\cite{Piazza:2010ye}. Since the $h\gamma\gamma$ vertex is generated at one loop of SM particles, $d_e^{(2)}$ is suppressed by a loop factor $\alphaem/\pi$ relative to the other scalar--SM couplings, and is therefore neglected throughout this work.

Finally, we note that the scalar bare mass receives a tree-level contribution from the Higgs vacuum expectation value upon electroweak symmetry breaking. Nevertheless, the scalar can remain ultralight, provided that this contribution is
either cancelled by a mass counterterm or suppressed by the imposition of an additional symmetry~\cite{Hook:2018jle,Brzeminski:2020uhm,Banerjee:2022sqg,Gan:2023wnp}.

\subsection{Other Couplings}\label{subsec:other_coupling}

Before closing this section, we also briefly discuss two other theories that can induce the quadratic coupling considered here.
The first example is the universal coupling from scalar-tensor theory~\cite{Brans:1961sx,Damour:1992we}, as implemented for quadratically-coupled ULDM in Ref.~\cite{Sibiryakov:2020eir}, where the BBN constraints on this scenario were first derived. Below we briefly summarize the setup of Ref.~\cite{Sibiryakov:2020eir}. The model assumes that the scalar field $\phi$ couples to the Standard Model universally through a $\phi$-dependent and conformally rescaled metric, $g_{\mu\nu}(\phi) = \Omega^2(\phi)\,g_{\mu\nu}$, where $g_{\mu\nu}(\phi)$ denotes the Jordan-frame metric, while $g_{\mu\nu}$ denotes the Einstein-frame metric. The conformal factor $\Omega(\phi)$ is parameterized as
\bea
\Omega(\phi)=1 + d^{(2)}_\text{univ}\, \times 2\pi \,\frac{\phi^2}{\Mpl^2}\,,
\eea
where $d_\text{univ}^{(2)}$ is the single parameter of the model. Starting from the point-particle action in the Jordan frame and transforming to the Einstein frame as $S_\text{SM} = -m_\text{SM}\int\sqrt{g_{\mu\nu}(\phi)\,\dd x^\mu \dd x^\nu} = -\Omega(\phi)\,m_\text{SM}\int\sqrt{g_{\mu\nu}\,\dd x^\mu \dd x^\nu}$, one finds that every elementary or composite SM particle mass rescales universally as $m_\text{SM}(\phi) = \Omega(\phi)\,m_\text{SM}$. For hadrons, this universality is inherited from rescaling the QCD confinement scale, $\Lambda_\text{QCD}(\phi) = 
\Omega(\phi)\,\Lambda_\text{QCD}$, which in turn follows from requiring the UV cutoff $M_\text{UV}$ to transform in the same way as all other mass scales, namely $M_\text{UV}(\phi) =
\Omega(\phi)\,M_\text{UV}$~\cite{Sibiryakov:2020eir}. Since all masses rescale by the same universal factor $\Omega(\phi)$, the weak EP is manifestly preserved. The model is therefore described by a single dimensionless parameter $d^{(2)}_\text{univ}$. Matching to \Eq{Ldamour} and \Eq{vary_const} gives
\bea
\label{eq:low_energy_universal}
d^{(2)}_e = 0\,, \quad \quad  d^{(2)}_g = d^{(2)}_{m_e} = d^{(2)}_{\hat{m}} = d^{(2)}_{\delta m} = d^{(2)}_\text{univ} \,.
\eea
The fact that $d^{(2)}_e=0$ in this matching is physically significant: any nonzero $d^{(2)}_e$, however small, would cause the fine-structure constant $\alphaem$ to vary with $\phi$. This dependence would shift the nuclear Coulomb energy in a composition-dependent manner and thereby violate the equivalence principle~(See \Eq{binding_energy_five_term_appx} and \Eq{dilaton_charge_scalar_SM_appx} in \Appx{dilaton_charge_appx}). 
This result is protected and can be seen in the Jordan-frame in which the scalar field decouples from the SM entirely~\cite{Sibiryakov:2020eir}. 
As established in
Refs.~\cite{Fujii:1996td,Hui:2010dn,Armendariz-Picon:2011ydk,Nitti:2012ev,Sibiryakov:2020eir}, this decoupling ensures that SM quantum corrections cannot generate particle-selective, EP-violating couplings at any loop order in the SM sector, and $d^{(2)}_e$ therefore strictly vanishes to all orders in SM perturbation theory. 
EP violation can still arise from graviton loops, but is highly suppressed by inverse powers of $M_{\rm pl}$~\cite{Armendariz-Picon:2011ydk}. 
Given the matching in \Eq{low_energy_universal}, this model is therefore effectively immune to EP tests, but can still be constrained by fifth-force searches, which probe the overall coupling strength rather than composition-dependent differences.~\footnote{Strictly speaking, the above EP-conserving argument for the universal coupling applies only when the two test masses are located at the same position. In practice, a finite separation $\Delta L \sim \mathcal{O}(1)\,\text{cm}$ can induce a tidal differential acceleration due to the spatial variation of the ULDM field sourced by Earth's matter through the universal coupling to the SM sector. However, this effect is suppressed as $\Delta a_\bg/a_\bg \lesssim \Delta L/R_\oplus \ll 1$, where $a_\bg$ denotes the acceleration induced by the background-induced force and the Earth's radius $R_\oplus$ sets the characteristic scale over which the local DM field profile varies.} For this reason, we do not consider this scenario further in the remainder of this work.

The second example is the light QCD axion~\cite{Hook:2017psm,DiLuzio:2021pxd,
DiLuzio:2021gos}, which also induces EP violation, with two distinct microscopic origins: the proton--neutron mass difference arising from isospin violation~\cite{Ubaldi:2008nf,GrillidiCortona:2015jxo,Kumamoto:2024wjd}, and composition-dependent differences in nuclear binding energies~\cite{Gue:2024onx,Bauer:2024hfv}. EP searches therefore provide a complementary probe of this scenario, independent of astrophysical or cosmological assumptions. Nevertheless, we do not pursue this case further for two reasons. First, as the axion enters the strongly-coupled regime, Earth undergoes a phase transition and scalar self-interactions become non-negligible~\cite{Hook:2017psm}. Second, as shown in 
Refs.~\cite{Gan:2025icr,Gue:2025nxq}, the parameter space within reach of current EP constraints is largely covered by existing astrophysical and terrestrial searches~\cite{Abel:2017rtm,Hook:2017psm,Balkin:2022qer,Witte:2025ilt,Gomez-Banon:2024oux,Bai:2025yxm,Huang:2026tgv}. For completeness, we nevertheless present in Appendix~\ref{subsec:dilaton_charge} the analytical expression for the QCD-axion charge, \Eq{axion_charge_appx}, to illustrate the microscopic origin of the EP violation.

\section{Background-Induced Force}\label{sec:bg_force}

The interactions discussed in the previous section can  affect the behavior of both SM objects and the scalar field.
The interactions in Eq.~\eqref{eq:vary_const} lead to modifications of the mass of an object proportional to $\langle \phi^2 \rangle$, where $\langle \cdots \rangle$ denotes the ensemble average over the DM distribution. Because the object mass depends on $\langle \phi^2 \rangle$, moving an object from a region of low $\langle \phi^2 \rangle$ to one of higher $\langle \phi^2 \rangle$ requires extra work. Equivalently, $\langle \phi^2 \rangle$ acts as an effective potential for the object, giving rise to a force and thus an induced acceleration. On the flip-side, when the scalar field $\phi$ propagates through ordinary matter, it can interact coherently with many particles if its de Broglie wavelength exceeds the interatomic spacing. This coherence modifies the propagation and distribution of $\phi$, known as the \emph{matter effect}. The matter-effect phenomenon appears in diverse contexts, including the Mikheyev-Smirnov-Wolfenstein~(MSW)
effect~\cite{Wolfenstein:1977ue,Mikheyev:1985zog,Mikheyev:1986wj,Botella:1986wy,Notzold:1987ik,Mirizzi:2009td,Huang:2023nqf}, the Meissner effect in superconductors~\cite{Meissner:1933ela}, and electromagnetic reflection in
 metals~\cite{Jackson:1998nia}.
The matter effect for quadratically-coupled ULDM was discussed in detail in Ref.~\cite{Gan:2025nlu}.

In the literature, this has been called the {\it wake force}~\cite{VanTilburg:2024xib}, in which the disturbance in the background field, a wake, is produced by the source and sensed by the
neighboring test object.
Because such a force is induced by the DM background, in contrast to the background-independent vacuum force~\cite{Moody:1984ba,Feinberg:1989ps,Ferrer:2000hm,Damour:2002mi,Olive:2007aj,Damour:2010rp,Damour:2010rm,Berge:2017ovy,Fichet:2017bng,Bauer:2023czj,Grossman:2025jub},
it is also referred to as the {\it background-induced force}~\cite{Ghosh:2022nzo,Blas:2022ovz,Barbosa:2024pkl,Ghosh:2024qai,Grossman:2025cov,Cheng:2025fak,Gan:2025nlu}. From a phenomenological standpoint of long-range interactions, the same underlying physical effect has also been investigated as a medium-dependent fifth force or long-range force~\cite{Ferrer:2000hm,Hees:2018fpg,Berezhiani:2018oxf,Banerjee:2022sqg}. Throughout this work, we use the terminology {\it background-induced force}. 

The interplay between these two above-mentioned effects -- the dependence of the $\phi$ mass on the SM background and the converse dependence of SM parameters on $\phi$ -- gives rise to rich phenomenology, in particular a {\it background-induced potential~(a.k.a. wake potential)} imposed on the test object, which is given by
\beq
\label{eq:background_potential}
V_\bg(\vecr) = -\frac{\rho_\phi}{m^2_\phi}\frac{(\mMtest^2 \, \calV_\testmass)(\mMsource^2 \, \calV_\sourcemass)}{4\pi r}\times \formfactorV(\vecr) \,,
\eeq
which leads to a background-induced, or wake force\footnote{Here $\vecF_\bg|_{\testmass \leftarrow \sourcemass} \equiv - \vecnabla_\testmass \,(V_\bg|_{\testmass \leftarrow \sourcemass})$ denotes the background-induced force sourced by $\sourcemass$ and experienced by $\testmass$. The corresponding reverse force, $\vecF_\bg|_{\sourcemass \leftarrow \testmass} \equiv - \vecnabla_\sourcemass \,(V_\bg|_{\sourcemass \leftarrow \testmass})$, does not satisfy reciprocity, i.e.,
$\vecF_\bg|_{\testmass \leftarrow \sourcemass} + \vecF_\bg|_{\sourcemass \leftarrow \testmass} \neq \mathbf{0}$. This reflects momentum exchange between the two-body system and the scalar background. Such non-reciprocity is present in the ULDM results of Refs.~\cite{VanTilburg:2024xib,Gan:2025nlu}, as reflected by the fact that, in general, $\formfactorV(\theta_\vecr) \neq \formfactorV(\theta_\vecr+\pi)$. Similar discussions of the non-reciprocity of the force also arise in the context of optical binding~\cite{Rieser:2022nye,Rudolph:2024cey}. Since $M_\sourcemass \gg M_\testmass$, the force on $\sourcemass$ is negligible, and we use $V_\bg$ and $\vecF_\bg$ as shorthand for $V_\bg|_{\testmass \leftarrow \sourcemass}$ and $\vecF_\bg|_{\testmass \leftarrow \sourcemass}$ throughout this work.}
\bea
\label{eq:background_force_gradient}
\vecF_\bg = - \vecnabla \, V_\bg \,,
\eea
where ``bg'' stands for ``background.'' 
Note that $V_\bg \propto \rho_\phi$, the density of $\phi$, and thus the potential and corresponding force vanishes in the absence of the scalar field background. The subscript ``$\sourcemass$'' denotes the source object that modifies the scalar background, while ``$\testmass$'' denotes the test object that experiences the scalar-induced force.
In this work, $\sourcemass$ is the Earth (denoted by $\oplus$) while $\testmass$ is the test mass of the experiment (see e.g.~\ref{tab:dilaton_charges_table}).
$\calV_\sourcemass$ and $\calV_\testmass$ denote the volumes of the source and test objects, respectively. $\mMsource$ and $\mMtest$ correspond to the induced scalar mass due to the interactions with $\sourcemass$ and $\testmass$, respectively.
$\formfactorV(\vecr)$ is the spatially-dependent potential form factor that describes the deviation of the background-induced potential from a Newtonian-like $1/r$ potential. For the remainder of this work, we will set $\sourcemass=\oplus.$
%%%
In what follows, we will describe the computation of the background-induced potential. 
\begin{table}[t]
\centering
\renewcommand{\arraystretch}{1.7}
\setlength{\tabcolsep}{9.0pt}
\begin{tabular}{|c|c|cccccc|}
\hline
 & Object & $(Q_\calA)_g$ & $(Q_\calA)_{e}$ & $(Q_\calA)_{m_e}$ & $(Q_\calA)_{\hat{m}}$ & $(Q_\calA)_{\delta m}$ & $(Q_\calA)_H$ \\
 & $\calA$ &  & $[\times 10^{-3}]$ & $[\times 10^{-4}]$ & $[\times 10^{-3}]$ & $[\times 10^{-4}]$ &  $[\times 10^{-3}]$   \\
\hline
Source Mass \, $\sourcemass$   & $\oplus$   & $1$ & $2.0~(1.9)$ & $2.7$ & $97.1\,(81.0)$ & $0.3\,(0.4)$ & $297.9\,(285.4)$ \\
\hline
\multirow{3}{*}{Test Mass\, $\testmass$}
          & $\ti$              & $1$ & $2.3$ & $2.5$ & $98.8\,(82.7)$ & $0.9\,(1.3)$ & $299.4\,(286.9)$ \\
          & $\pt$              & $1$ & $4.2$ & $2.2$ & $101.5\,(85.4)$ & $2.2\,(3.2)$ & $301.6\,(289.1)$ \\
          & $\ti-\pt$   & $0$ & $-1.9$ & $0.3$ & $-2.7$ & $-1.3\,(-1.9)$ & $-2.2$ \\
\hline
\end{tabular}
\caption{Numerical values of the dilaton charges for the source mass (the Earth, $\oplus$) and the test masses (Ti and Pt) in the MICROSCOPE experiment. ``Ti'' denotes the Titanium alloy and ``Pt'' denotes the Platinum alloy. The row ``$\ti-\pt$'' represents the difference in dilaton charge between the two test masses. Analytical expressions and the numerical evaluation of these charges are given in \Appx{dilaton_charge_appx}. For our fiducial values, we use the updated hadronic inputs listed in Table~\ref{tab:asymmetry_parameters}. The values in parentheses are obtained using the same analytical dilaton-charge formulas, but with the hadronic inputs adopted by Damour and Donoghue~(DD 2010)~\cite{Damour:2010rp,Damour:2010rm}. Entries without parentheses are unchanged at the displayed precision. For the constraints shown in Fig.~\ref{fig:microscope_constraint}, we adopt the updated hadronic and nuclear inputs as our fiducial benchmark values, keeping them fixed in the numerical analysis. Although we have verified that uncertainties in the hadronic and nuclear inputs do not lead to visible changes in Fig.~\ref{fig:microscope_constraint}, we still list them for conceptual clarity.
}
\label{tab:dilaton_charges_table}
\end{table}

\subsection{Modified Mass of Ordinary Matter}\label{subsec:modified_mass_ordinary}

We begin by expanding the discussion of how a scalar background modifies the masses of macroscopic objects. As shown in \Eq{vary_const}, when the scalar field acquires a non-zero value, it shifts the fundamental constants. 
As a result, the atomic mass of $\calA$ changes because the scalar field modifies both the rest masses of its constituent particles and the nuclear binding energy, as shown in \Appx{dilaton_charge_appx}.
Consequently, the mass of the ordinary matter varies as~\footnote{To avoid notational confusion, throughout the text we use capital $M$ to denote the masses of macroscopic objects, such as the Earth and test masses. By contrast, lowercase $m$ always denotes a scalar mass, either the bare scalar mass $m_\phi$ or the matter-induced scalar mass $\mM$.}
\bea
\label{eq:macroscopic_mass_variation}
\frac{\Delta M_\calA}{M_\calA} = \alpha^{(2)}_\calA \times \frac{2\pi \phi^2}{\Mpl^2} \quad \quad \quad \quad \quad (\,\calA = \oplus, \testmass\,)\, .
\eea
Here, $\alpha_\calA^{(2)}$ is the effective coupling between the scalar and the object $\calA$. For the full sets of the effective scalar--SM coupling given by \Subsec{effective_coupling}, we have
\bea
\label{eq:coupling_scalar_SM_main}
\alpha^{(2)}_\calA = (Q_\calA)_g \, d^{(2)}_g + (Q_\calA)_e \, d^{(2)}_e + (Q_\calA)_{m_e} \, (d^{(2)}_{m_e}-d^{(2)}_g) + (Q_\calA)_{\hat{m}} \, (d^{(2)}_{\hat{m}}-d^{(2)}_g)+ (Q_\calA)_{\delta m} \, (d^{(2)}_{\delta m}-d^{(2)}_g)\, ,
\eea
where the $(Q_{\calA})_i$ are the ``dilaton charges"~\cite{Damour:2010rp,Damour:2010rm} that quantify the material-dependent response of $M_\calA$ to the dimensionless scalar--SM coupling $d_i^{(2)}$. For example, in the renormalizable Higgs portal (\Subsec{higgs_portal}), we have
\bea
\label{eq:coupling_scalar_Higgs_main}
\alpha^{(2)}_\calA 
&=\left\{(2/9) + (\alphaem/\pi) (Q_{\calA})_e + (7/9)
\left[ (Q_{\calA})_{m_e} + (Q_{\calA})_{\hat{m}} + (Q_{\calA})_{\delta m} \right]\right\} \, d^{(2)}_H\\
&\equiv(Q_\calA)_{H} \, d^{(2)}_H\, .
\eea

In \Appx{dilaton_charge_appx}, we present the derivation of $(Q_\calA)_{H}$ and more generally the dilaton charges, together with the updated numerical inputs from atomic and nuclear physics for the scalar models introduced in \Sec{scalar_models}. 
As a reminder, in this work the object $\calA$ refers to either the Earth or a test mass $\testmass$, since the scalar field modifies the masses of both. In Table \ref{tab:dilaton_charges_table}, we present the numerical values of the dilaton charges for the source, $\sourcemass=\oplus$, and the test masses, $\testmass=\{\ti,\,\pt\}$, in the MICROSCOPE experiment. ``Ti'' denotes the titanium alloy $\mathrm{Ti}/\mathrm{Al}/\mathrm{V}\,[90:6:4]$, while ``Pt'' denotes the platinum alloy $\mathrm{Pt}/\mathrm{Rh}\,[90:10]$, where the numbers in square brackets denote the mass percentages of the constituent elements. To compute the Earth dilaton charge, we approximate the Earth composition by $\text{Fe}/\text{Mg}\text{Si}\text{O}_3\,[32:68]$, where $\text{Fe}$ represents the iron-core component and $\text{Mg}\text{Si}\text{O}_3$ represents the
mantle component~\cite{Seager:2007ix}. This two-component toy model captures the dominant Earth elemental composition, $\text{Fe}\,(32\%)$, $\text{O}\,(30\%)$, $\text{Si}\,(16\%)$, and $\text{Mg}\,(15\%)$, up to subdominant elements at the few-percent level~\cite{mcdonough2003compositional}.\footnote{Compared with the $\text{Fe}/\text{Si}\text{O}_2\,[32:68]$ approximation used in earlier EP-test literature~\cite{Damour:2010rp,Hees:2018fpg,Banerjee:2022sqg}, this choice does not qualitatively change the resulting dilaton charges, but provides a more realistic representation of the Earth's magnesium~(Mg)
component.}

The modification of the Earth and test masses has two physical consequences. First, it changes the free energy of the system, which is dominated by the Earth in our setup since $M_{\oplus}\gg M_{\testmass}$. This modifies the scalar distribution inside the Earth through an induced scalar mass term $\mMearth$, as discussed next in \Subsec{modify_scalar_distribution}. Second, it affects the motion of the test object $\testmass$ through the scalar gradient. After ensemble averaging over the DM phase space, this effect is described by the background-induced potential $V_\bg$, which is determined by $\langle \phi^2 \rangle$. The resulting force on the test mass is obtained by taking the gradient of this potential, as shown in \Eq{background_force_gradient}. We discuss this second effect in \Subsec{Vbg_form_factor}.
%and \Subsec{phase_benchmark}.
\subsection{Modified Scalar Distribution}\label{subsec:modify_scalar_distribution}

The interactions between the SM and scalar in a finite-density environment also modify the scalar-field dynamics. Previous studies examined the scalar’s early-Universe cosmology, where interactions with the relativistic plasma generate matter, or thermal, effects. These effects significantly alter the scalar’s cosmological evolution and can shift the resulting constraints by orders of magnitude~\cite{Belokon:2018hrn,Sibiryakov:2020eir,Bouley:2022eer,Baryakhtar:2024rky,Baryakhtar:2025uxs,Ghosh:2025pbn} compared to the treatment with only the scalar bare mass~\cite{Stadnik:2015kia}.
Here, we study how the nonrelativistic (NR) matter density affects the scalar’s spatial distribution, rather than its cosmological time evolution.

We saw in the previous section that the mass of a macroscopic object is modified by the scalar field (\Eq{macroscopic_mass_variation}). For the object $\calA$, this object acquires an additional scalar-induced density, given by $\Delta \rho_\calA/\rho_\calA = \alpha_\calA^{(2)} (2\pi \phi^2/\Mpl^2)$, where $\Delta \rho_\calA$ can be treated as the increase of the scalar-induced free energy of the system. Since $\phi$ couples {\it quadratically} to matter, $\phi$ gets an effective mass in the presence of matter. The matter-induced scalar mass inside the object $\calA$ is thus given by
\bea
\label{eq:mM_def}
m_{\text{M},\calA}^2(\vecr)=\alpha^{(2)}_\calA \,\, \frac{4\pi \rho_\calA(\vecr)}{\Mpl^2}\,. \quad \quad \quad \quad \quad (\,\calA = \oplus, \testmass\,)
\eea
Here $\rho_\calA(\vecr)$ is the matter density of the object $\calA$, and $\alpha^{(2)}_\calA \sim d^{(2)} Q_\calA$ denotes the effective scalar coupling to $\calA$. In the notation $m_{\mathrm{M},\calA}$, ``$\mathrm{M}$'' labels the matter-induced contribution to the scalar mass, while $\calA$ specifies the object in which this effective scalar mass is generated. 
Since the Earth dominates the scalar distribution, in this section we fix $\calA=\oplus$ and solve for the scalar profile using only the Earth density distribution, neglecting the back-reaction from the test mass $\testmass$. 

In general the induced mass depends on the position so $\phi$ obeys the Klein-Gordon equation
\bea\label{eq:klein}
\Box\,\phi+m^2_{{\rm eff},\oplus}\,(\vecr) \, \phi=0\,,
\eea
where the spatially dependent scalar effective mass given by
\bea
m^2_{{\rm eff},\oplus}\,(\vecr)=m_\phi^2+\mMearth^2\,(\vecr)\,.
\eea
To solve for the scalar configuration in the presence of a spatially-dependent matter contribution, we use the following ansatz in the NR approximation:
\bea
\label{eq:phi_NR_approx}
\phi(\vecr,t;\veck) = \Re[ e^{-i \omega\,(\veck)\, t} \psi(\vecr;\veck) ]\,,
\eea
where $\veck$ labels the corresponding mode, $\omega\,(\veck) = (m_\phi^2 + |\veck|^2)^{1/2}$, $\phi$ denotes the real scalar-field, and $\psi$ denotes its NR component.~\footnote{Another commonly used convention for the NR limit is $\phi(\vecr,t;\veck)=\Re\,[e^{-i m_\phi
t}\,\widetilde{\psi}(\vecr,t;\veck)]$~\cite{Bjorken:1965sts,Greiner:1997xwk}. This is equivalent to the parameterization in \Eq{phi_NR_approx} under the transformation $\widetilde{\psi}(\vecr,t;\veck)=e^{-i E_{\eff}(\veck) t}\,\psi(\vecr;\veck)$ in the NR limit. In this convention, the Schr\"odinger-type equation takes the form $- \frac{1}{2m_\phi} \vecnabla^2 \widetilde{\psi} + V_{\eff,\oplus}(\vecr)\,\widetilde{\psi} = i \, \partial_t \widetilde{\psi}$ in the NR limit $|m_\phi^2 \, \partial_t^2 \widetilde{\psi}| \ll |m_\phi \, \partial_t \widetilde{\psi}|$.} Because $|m_\phi \, \partial_t \psi|$ and $|\partial_t^2 \psi|$ are much smaller than $|m^2_\phi \,\psi|$ in the NR limit, we know that $\psi$ is only a function of the spatial coordinate $\vecr$, rather than the time coordinate $t$. Therefore, we have the Schr\"odinger-type equation
\bea
\label{eq:Schrodinger_Eq}
- \frac{1}{2 m_\phi} \vecnabla^2 \psi + V_{\eff,\oplus}\,(\vecr) \, \psi = E_{\eff}(\veck) \, \psi\,,
\eea
where the effective potential and the effective kinetic energy is given by
\bea
V_{\eff, \oplus}\,(\vecr) = \frac{\mMearth^2\,(\vecr)}{2 \,m_\phi}\,, \quad \quad \,\,\, E_{\eff}(\veck) = \frac{\veck^2}{2 \,m_\phi}\,,
\eea
respectively. In the limit where the Earth is approximated as a uniform sphere, we have $\mMearth^2(\vecr)=\mMearth^2\,\theta(R_\oplus -\abs{\vecr})$, indicating that the matter-induced mass is non-zero only inside the Earth, i.e., for $\abs{\vecr}\leq R_\oplus$. Here, $R_\oplus \simeq 6378\,\km$, as given in Refs.~\cite{moritz2000geodetic,luzum2011iau,IAUInter-DivisionA-GWorkingGrouponNominalUnitsforStellarPlanetaryAstronomy:2015fjh}. Since the Earth-induced potential $V_{\eff,\oplus}$ modifies the propagation of the scalar, the problem reduces to a standard scattering problem. 

\subsubsection{Scattering Theory Formalism}\label{subsec:scattering}

%%%% figure classification
\begin{figure}[t]
\centering
\includegraphics[width=0.78\linewidth]{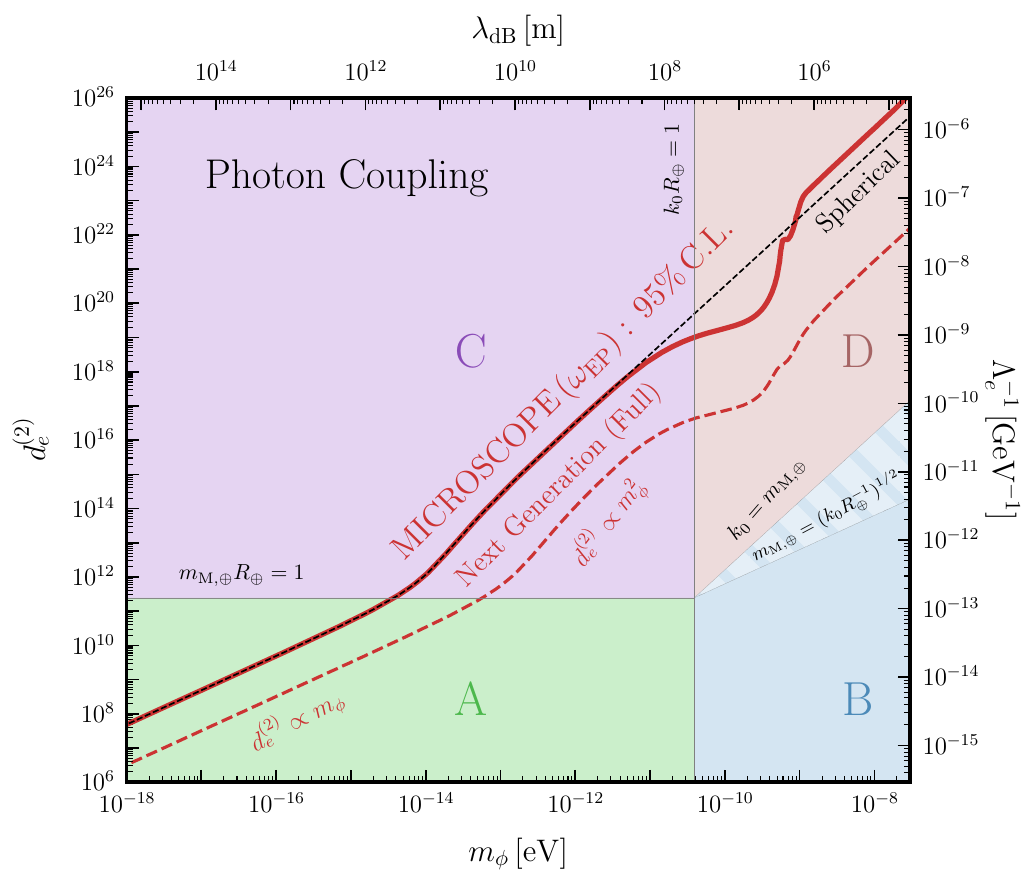}
\caption{Classification of the ULDM parameter space from \Subsec{scattering} with the scalar--photon interaction as an example, showing the $95\%\,\mathrm{C.L.}$ constraint from MICROSCOPE as the {\bf red solid} line (Sec.~\Ref{subsec:microscope_revisit}) and the next-generation projected $95\%\,\mathrm{C.L.}$ sensitivity with the benchmark parameters inspired by the Galileo Galilei proposal~\cite{Nobili:2000bzv,Nobili:2012uj,Nobili:2017cxu,Nobili:2018eym} as the {\bf red dashed} line (\Sec{future}). The {\bf black dashed} line is taking the MICROSCOPE constraint in the spherically symmetric limit (Sec.~\Ref{subsec:sph_symmetric_main}).
$k_0$ denotes the mean momentum of the ULDM and characterizes its wave-like behavior. $\mMearth$ is the scalar mass induced by the Earth, which characterizes the interaction strength between $\phi$ and the SM sector, while $R_\oplus$ is the Earth radius. In this work, the test mass $\testmass$ is always treated as point-like. EP tests that treat the Earth as the source mass, such as MICROSCOPE~\cite{Touboul:2017grn,MICROSCOPE:2019jix,MICROSCOPE:2022doy,Touboul:2022yrw} and Galileo Galilei~\cite{Nobili:2000bzv,Nobili:2012uj,Nobili:2017cxu,Nobili:2018eym}, probe regions (A), (C), and (D). To highlight the classification of matter effects induced by the Earth, we focus here on the parameter space where additional screening effects due to the satellite and the atmosphere are negligible. These effects are discussed in more detail in \Subsec{microscope_revisit} and \Appx{atm_matter_effect}.}
\label{fig:classification}
\end{figure}
%%%%%%%%%%
We can apply QM scattering theory to solve for the scalar distribution~\cite{Landau:1991wop,Sakurai:2011zz}. Both the physical picture and the computational strategy for describing the scattering process depend on the scalar-field parameter space. 
In \Fig{classification}, we present the simplified classification of the calculation in terms of the mean scalar momentum $k_0 = m_\phi \, v_0$, where $v_0$ is the average DM speed, and the Earth-induced mass $\mMearth$. 
A more detailed classification can be found in Ref.~\cite{Gan:2025nlu}.
The four regions are:
\begin{itemize}
\item {\bf(A): perturbative, low-momentum regime ($\mMearth R_\oplus\lesssim 1 ~\text{and~} k_0 R_\oplus \lesssim 1)$.} The Born approximation provides a valid description of the scattering, since the scattering is perturbative ($|\psisc| \lesssim |\psi_0|$). 
In this regime, the source can be treated as point-like since $1/k_0 \gtrsim R_\oplus$, and the geometric structure of the source (the Earth) can be neglected. 
The behavior of $\mathcal{F}$ depends on the distance $r$ to the Earth center, or more specifically $k_0 r$.
On the other hand, for the near-field region $k_0 r \lesssim 1$, the contributions from the bulk of the Earth and from different momentum modes add coherently to the potential, giving $\formfactorV=1$. For the far-field region $k_0 r \gtrsim 1$, the wave function becomes highly oscillatory. The phase cancellation induced by averaging over the finite DM phase space suppresses the potential as $\formfactorV \propto 1/(k_0 r)^2$. This suppression is known as the {\it incoherence effect}.~\footnote{The term ``incoherence effect'' here refers to the suppression caused by destructive cancellation, i.e. the loss of coherence, when integrating over the rapidly oscillating background-induced potential. This term can be viewed as the counterpart of coherent interactions between ULDM and a macroscopic object. We adopt this terminology from Refs.~\cite{Bednyakov:2018mjd,Blas:2022ovz}. The same effect is also referred to as the ``decoherence effect'' in Refs.~\cite{Grossman:2025cov,Gan:2025nlu}.}
\item {\bf(B): semi-perturbative, high-momentum regime ($\mMearth \lesssim k_0 ~\text{and~}k_0 R_\oplus\gtrsim 1$).}
The Born approximation remains applicable, and the form factor $\formfactorV$ still exhibits perturbative behavior. However, since the full perturbative condition is more restrictive ($k_0 R_\oplus\gtrsim1~\text{and~} \mMearth \lesssim(k_0/R_\oplus)^{1/2}$)~\cite{Gan:2025nlu}, we refer to this regime as ``semi-perturbative.''\,\footnote{The detailed classification subdivides Region B according to whether non-perturbative features appear in the scattering cross section $\sigma$. For the background-induced force, however, this entire region remains effectively perturbative once the ULDM momentum is large enough to overcome the potential barrier, so the simplified classification suffices for our purposes.} 
In addition, since $k_0 R_\oplus \gtrsim1$, the Earth can no longer be treated as a point-object and finite-size effects come into play.
Relative to region~(A), the finite-size effects introduce an enhancement of $\mathcal{O}(1)$--$\mathcal{O}(10)$ in $\formfactorV$.
\item {\bf(C):  non-perturbative, low-momentum regime ($\mMearth R_\oplus \gtrsim 1$ and $k_0 R_\oplus\lesssim1$).} 
The Born approximation no longer holds in this regime, and we must use non-perturbative methods to calculate $\formfactorV$. In the near-field region, $k_0 r \lesssim 1$, one may use the spherically symmetric ansatz of Refs.~\cite{Hees:2018fpg,Berezhiani:2018oxf,Banerjee:2022sqg}. 
In this approximation, the potential is suppressed by $\formfactorV \simeq 3/(\mMearth R_\oplus)^2$ compared to region (A). 
One factor of $1/(\mMearth R_\oplus)$ arises from the suppression of the scalar field at the surface of the Earth, while the additional factor $3/(\mMearth R_\oplus)$ reflects the fact that only a thin shell of thickness $\mMearth^{-1}$ contributes to the external potential, since $4\pi R_\oplus^2 \mMearth^{-1}/\mathcal{V}_\oplus = 3/(\mMearth R_\oplus)$. 
Both of these suppression factors are a result of the strong-coupling, which leads to a matter-induced {\it screening effect} of the scalar field.
For the far-field region, $k_0 r \gtrsim 1$, a complete treatment requires a full partial-wave analysis combined with phase-space averaging, as performed in this work.
\item {\bf(D): non-perturbative, high-momentum regime ($\mMearth \gtrsim k_0$ and $k_0 R_\oplus\gtrsim1$).}
As in region~(C), non-perturbative methods are required. Unlike in region~(C), however, the spherically symmetric ansatz is no longer applicable, and one must instead use a partial-wave analysis combined with phase-space averaging for both the near- and far-field regions. For fixed $\mMearth$, increasing $k_0$ alleviates the screening effect in two stages. First, once $k_0 R_\oplus \gtrsim 1$, the scalar kinetic energy is large enough to overcome the surface-gradient cost, so that the scalar wave can be pushed toward the vicinity of the potential barrier rather than being governed by the long-wavelength, spherically symmetric profile of region~(C). Second, as $k_0$ becomes comparable to $\mMearth$, the scalar wave can further penetrate the potential barrier of the Earth. Together, these two stages progressively alleviate the {\it screening effect} discussed in region~(C). We refer to this overall behavior as the {\it descreening effect}. In the high-momentum regime, the form factor is still suppressed by the {\it incoherence effect}. The descreening effect therefore appears as a partial compensation of the incoherence suppression found in regions~(A) and~(B). This saturation behavior can equivalently be understood as the approach to the optical limit discussed in \Appx{geometric_optics}.
\end{itemize}

In the partial-wave analysis that follows, we will focus in particular on the non-perturbative regions (C) and (D), although the techniques introduced are also applicable to regions (A) and (B). 

We start with the scattering between the monochromatic scalar field with momentum $\veck$ and the Earth. 
The wave function outside the Earth is represented as
\bea
\psi(\vecr;\veck) = \psiinc(\vecr;\veck) + \psisc(\vecr;\veck) \quad \quad (r\geq R_\oplus)\,,
\eea
where $\psiinc$ is the incident wave and $\psisc$ is the scattered wave. Here, the incident wave is the plane wave $\psiinc = \abs{\psi_0} \, e^{i \veck \cdot \vecr}$, where $|\psi_0| = \abs{\phi_0}$ is the amplitude of the incident scalar wave. Using a partial-wave expansion (see \Appx{partial_wave_appx}), we have
\bea
\label{eq:psi_out_monochromatic_main}
\psi(\vecr;\veck) = \abs{\psi_0} \sum_{l=0}^{\infty} (2l+1) \, i^l \, \calR_l(kr) \, P_l(\cos\theta)\,,
\eea
where $\calR_l(kr)$ denotes the radial component of each partial wave and $\cos\theta=\hat\veck\cdot\hat\vecr$. 
We focus on the wave function outside the Earth, which corresponds to the region where an Earth-orbiting satellite is located. 
Outside the Earth, the radial component is given by
\bea
\label{eq:R_l_out_main}
\calR_l(kr) = \underbrace{j_l(kr)}_{\text{inc}} + \underbrace{A_l \, h_l(kr)}_{\text{sc}} \quad \quad (r\geq R_\oplus)\, ,
\eea
where the $j_l$ term represents the incident plane wave and $A_l h_l$ represents the outgoing scattered wave. 
The scattered-wave coefficient $A_l$ is obtained by matching the boundary conditions at $r=R_\oplus$ and is derived explicitly in \Appx{partial_wave_appx}. 
Specifically, we have 
\bea
A_l = - \frac{k \, j_l(k_\oplus R_\oplus) \, j_{l+1}(k R_\oplus) - k_\oplus \, j_l(k R_\oplus) \, j_{l+1}(k_\oplus R_\oplus) }{ k \, j_l(k_\oplus R_\oplus) \, h_{l+1}(k R_\oplus) - k_\oplus \, h_l(k R_\oplus) \, j_{l+1}(k_\oplus R_\oplus) } \,,
\eea
where $k_\oplus=\sqrt{k^2 - \mMearth^2}$ is the scalar momentum inside the Earth. 
Importantly, the incident and scattered components converge at different partial-wave scales, which must be accounted for in the numerical computation.
The incident plane wave requires partial waves up to $l_\text{max} \sim kr$, whereas the scattered wave is controlled by the size of the Earth and converges at $l_\text{max} \sim k R_\oplus$. 
Therefore, numerical computations must sum the plane-wave contribution to sufficiently high $l$ even after the scattered-wave contribution has converged; truncating the incident-wave expansion prematurely will produce incorrect results.

\begin{figure}[h]
\centering
\includegraphics[width=0.5\linewidth]{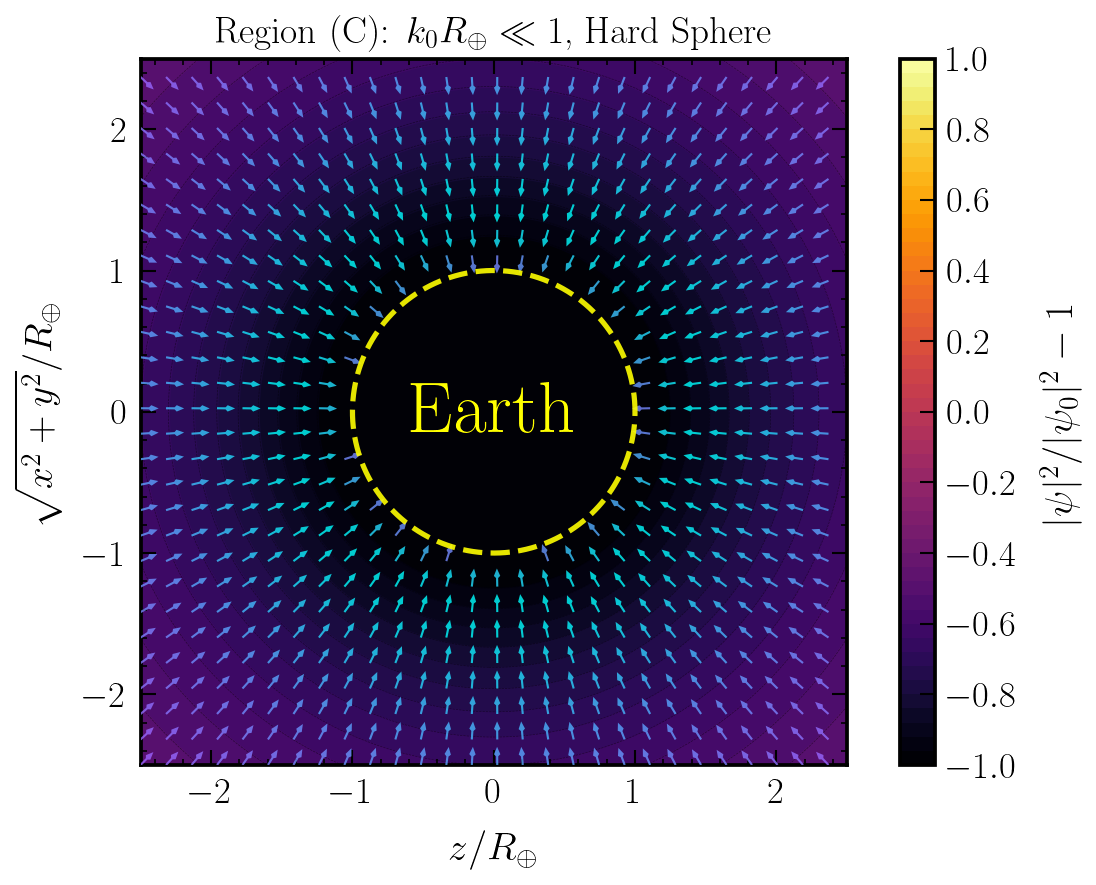}
\caption{Configuration of $|\psi|^2/|\psi_0|^2-1$ in the hard-sphere limit $\mMearth R_\oplus \gg 1$ with $k_0 R_\oplus \ll 1$, as previously discussed in Refs.~\cite{Hees:2018fpg,Berezhiani:2018oxf,Banerjee:2022sqg}. This configuration corresponds to region~(C) in \Fig{classification}. Matching the boundary conditions suppresses the scalar field at the surface of the sphere, giving $\abs{\psi(r=R_\oplus)}/\abs{\psi_0}=1/(\mMearth R_\oplus)$, which reveals the screening effect. The arrows indicate the direction of the background-induced force, while their lightness of their color represents its magnitude. In this limit, the background-induced force is an attractive central force that vanishes both at the Earth's surface and asymptotically far from the Earth.}
\label{fig:nonpeturb_sph_plt}
\end{figure}

\subsubsection{Spherically Symmetric Limit}\label{subsec:sph_symmetric_main}

Before proceeding, we discuss the $k_0 R_\oplus \ll 1$ limit where the solution to \Eq{Schrodinger_Eq} has a simple analytical form and captures the main features of the screening effect. In this limit, it is sufficient to keep only the $s$-wave ($l=0$) component and the system is spherically symmetric in the near-field region~\cite{Gan:2025nlu}. The corresponding scattered-wave coefficient can be written as
\bea
\label{eq:A0_main}
A_0 = -i \frac{\mMearth R_\oplus - \tanh(\mMearth R_\oplus)}{\mMearth R_\oplus} \, k R_\oplus + \cdots\,,
\eea
where ``$\cdots$'' denotes higher-order terms in the $kR_\oplus$ expansion. 
Noting that $h_0(k r) = - ie^{ikr}/kr$, the wavefunction outside the Earth is given by $\psi(\vecr;\veck) \simeq \abs{\psi_0} \left[e^{i \veck \cdot \vecr} - i A_0\,e^{ikr}/k r \right]$. 
For a satellite in near-Earth orbit, we have $r/R_\oplus \sim \mathcal{O}(1)$.  Therefore, in the $s$-wave limit of $k_0 R_\oplus \ll 1$, we also have $k_0 r \ll 1$, and the exterior scalar profile reduces to
\bea
\label{eq:psi_sph_config}
\psi_\sph(r) \simeq \abs{\psi_0} \bigg[ 1 - \frac{\mMearth^2 \mathcal{V}_\oplus}{4 \pi r} J_+(\mMearth R_\oplus) \bigg] \quad \quad (r\geq R_\oplus)\, ,
\eea
where the constant term is the incident plane wave and the $1/r$ term is the scattered $s$-wave contribution. 
$J_+=3\,[x-\tanh(x)]/x^3$ characterizes the screening behavior due to the induced mass and
\begin{align}
\label{eq:screening}
J_+(\mMearth R_\oplus)\simeq \begin{cases}
        1\,, & \mMearth \, R_\oplus \lesssim 1~~\text{Region (A)}\\
        3/(\mMearth R_\oplus)^2\,, & \mMearth \, R_\oplus \gtrsim 1~~\text{Region (C)}
    \end{cases}
\end{align}
and $\mathcal{V}_\oplus=4\pi R_\oplus^3/3$ is the volume of the Earth.
As an independent cross-check, the same configuration also follows from the spherically symmetric ansatz of Refs.~\cite{Hees:2018fpg,Berezhiani:2018oxf,Banerjee:2022sqg}, as reviewed in detail in \Appx{sph_ansatz_appx}. 
In the hard sphere regime (C), where $\mMearth \, R_\oplus \gg 1, k_0 R_\oplus\ll 1$, we have $\psi_\sph \simeq \abs{\psi_0}\,h/(R_\oplus+h)$, indicating that in this limit, at low altitudes the scalar field roughly linearly decreases with the height $h$. In \Fig{nonpeturb_sph_plt}, we show the field profile $\psi_\sph^2/\abs{\psi_0}^2-1$, which is spherically symmetric and exhibits the screening effect near the Earth's surface.

Connecting \Eq{psi_sph_config} back to \Eq{background_potential}, we define the spherically symmetric potential form factor
\bea
\label{eq:formfactorV_sph}
\formfactorV_\sph(r) = J_+(\mMearth R_\oplus) \, \times \left[1 - \frac{1}{2} \frac{\mMearth^2 \mathcal{V}_\oplus}{4 \pi r} J_+(\mMearth R_\oplus)\right]\,,
\eea
which encodes the screening ($\formfactorV_\sph=1$ in the unscreened limit.)
We see $J_+$ governs the behavior of $\formfactorV_\sph(r)$, and therefore \Eq{formfactorV_sph} follows the same behavior as \Eq{screening} in which the potential is suppressed compared to the unscreened limit. 
From \Eq{psi_sph_config}, we can compute the background-induced force for the spherically symmetric limit as
\bea
\label{eq:F_bg_sph}
\vecF_\bg = - \frac{\rho_\phi}{m^2_\phi}\frac{(\mMtest^2 \, \calV_\testmass)(\mMearth^2 \, \calV_\oplus)}{4\pi r^2}\times r^2 \frac{\dd}{\dd r}\left(\frac{\formfactorV_\sph(r)}{r}\right) \, \hat{\vecr}\,,
\eea
where the last term is
\bea
\label{eq:formfactorF_sph}
r^2 \frac{\dd}{\dd r}\left(\frac{\formfactorV_\sph(r)}{r}\right) = J_+(\mMearth R_\oplus) \, \times \left[1 - \frac{\mMearth^2 \mathcal{V}_\oplus}{4 \pi r} J_+(\mMearth R_\oplus)\right]\,,
\eea
which quantifies the screening effect and deviation from a Newtonian-type $1/r^2$ force.
From \Eq{F_bg_sph} and \Fig{nonpeturb_sph_plt}, we see that within the spherically symmetric ansatz, the background-induced force is always a central force. Moreover, \Eq{formfactorF_sph} is always positive, so this central background-induced force is always attractive in the spherically symmetric ansatz~\cite{Hees:2018fpg,Berezhiani:2018oxf,Banerjee:2022sqg}. 
For a satellite at a height $h$ above the Earth's surface, we have
\bea
\label{eq:formfactorF_sph_approx}
\text{screening factor:~~~}r^2 \frac{\dd}{\dd r}\left(\frac{\formfactorV_\sph(r)}{r}\right) \simeq
\left\{
\begin{aligned}
& 1 & \quad \,\, (\mMearth R_\oplus \lesssim 1)\\
& \frac{3}{(\mMearth R_\oplus)^2} \frac{h}{R_\oplus + h} & \quad \,\, (\mMearth R_\oplus \gtrsim 1)
\end{aligned}
\right. \,,
\eea
where the first line is the unscreened, perturbative case while the second line shows the screening effect in 
the non-perturbative regime $\mMearth R_\oplus \gtrsim 1$. 
As explained in \Subsec{scattering}, the factor $3/(\mMearth R_\oplus)^2$ has two origins: one factor of $1/(\mMearth R_\oplus)$ comes from the suppression of the scalar field at the Earth surface, $r=R_\oplus$, 
The additional factor $3/(\mMearth R_\oplus)$ is a geometric correction that arises when the penetration depth is smaller than the Earth's radius, causing the scaling to shift from volume-based to surface-based. 
The factor $h/(R_\oplus+h)$ reflects the roughly linear dependence of the background-induced force on the altitude $h$ in the low-altitude regime, and shows that this force vanishes at the Earth's surface in the hard-sphere limit. 

The spherically symmetric limit provides useful insight into the behavior of EP-test constraints. In \Fig{classification}, which uses the photon coupling $d_e^{(2)}$ as a representative example, the black dashed curve follows the approximate scaling
\bea
d_i^{(2)} \propto \left\{
\begin{aligned}
& m_\phi & \quad \,\, (\mMearth R_\oplus \lesssim 1)\\
& m_\phi^2 & \quad \,\, (\mMearth R_\oplus \gtrsim 1)
\end{aligned}
\right. \,,
\eea
which can be understood as follows. In the perturbative regime, $\mMearth R_\oplus \lesssim 1$, we have $r^2 \dd(\formfactorV_\sph/r)/\dd r \simeq 1$, so the background-induced force is unscreened and receives contributions from the full volumes of both the test mass and the Earth. Consequently,
\begin{equation}
|\vecF_\bg| \propto \frac{\mMtest^2 \mMearth^2}{m_\phi^2}
\propto \frac{(d_i^{(2)})^2}{m_\phi^2}~~~~~\text{for}\,\,~\mMearth R_\oplus \lesssim 1 \,,
\end{equation}
implying the constraint scales as $d_i^{(2)} \propto m_\phi$. By contrast, for $\mMearth R_\oplus \gtrsim 1$, the factor $3/(\mMearth R_\oplus)^2$ in \Eq{formfactorF_sph_approx} limits the Earth's contribution. Substituting \Eq{formfactorF_sph_approx} into \Eq{F_bg_sph} shows that the force becomes independent of $\mMearth$, yielding
\begin{equation}
|\vecF_\bg| \propto \frac{\mMtest^2}{m_\phi^2}
\propto \frac{d_i^{(2)}}{m_\phi^2} ~~~~~\text{for}\,\,~\mMearth R_\oplus \gtrsim 1\,.
\end{equation}
As a result, the constraint scales as $d_i^{(2)} \propto m_\phi^2$ in the screened regime.

Although this spherically symmetric ansatz is analytically simple and useful for describing the screening effect and the force behavior in the low-momentum limit, both the ansatz itself and the corresponding central-force picture break down once
\bea
k_0 R_\oplus \gtrsim 1 \quad \,\, \Longleftrightarrow \quad \,\, m_\phi \gtrsim 4 \times 10^{-11}\,\eV \quad \quad \text{(No Spherical Symmetry)}.
\eea
In this high-momentum regime, the scalar profile becomes intrinsically non-spherical: while the force remains attractive in the forward direction, it reverses sign in the backward direction and becomes repulsive, as we show below. Therefore, the previous central-force description is no longer applicable. A proper treatment of this high-momentum regime, together with a correct description of the time-dependent signal template, requires a full partial-wave analysis combined with ensemble averaging over the DM phase space, which we discuss next.

\subsection{Ensemble Averages and Multipole Expansion}\label{subsec:Vbg_form_factor}

In this section, we present a general formalism that combines the partial-wave method introduced in the previous section with the integration over DM phase space to compute the background-induced potential. Although this formalism applies throughout \Fig{classification}, it is particularly important in the non-perturbative, high-momentum region~(D), where the simplifying approximations used in other regions break down. 
We present two equivalent numerical methods as cross-checks: the direct partial-wave computation used in Ref.~\cite{Gan:2025nlu}, and the ensemble-averaged multipole expansion introduced in Ref.~\cite{Brzeminski:2026rox}. 
Using these methods, we compute the DM phase-space-averaged scalar profile $\langle \phi^2 \rangle$, which is a key ingredient of the background-induced potential.

\begin{figure}[t!]
\centering
\includegraphics[width=0.49\linewidth]{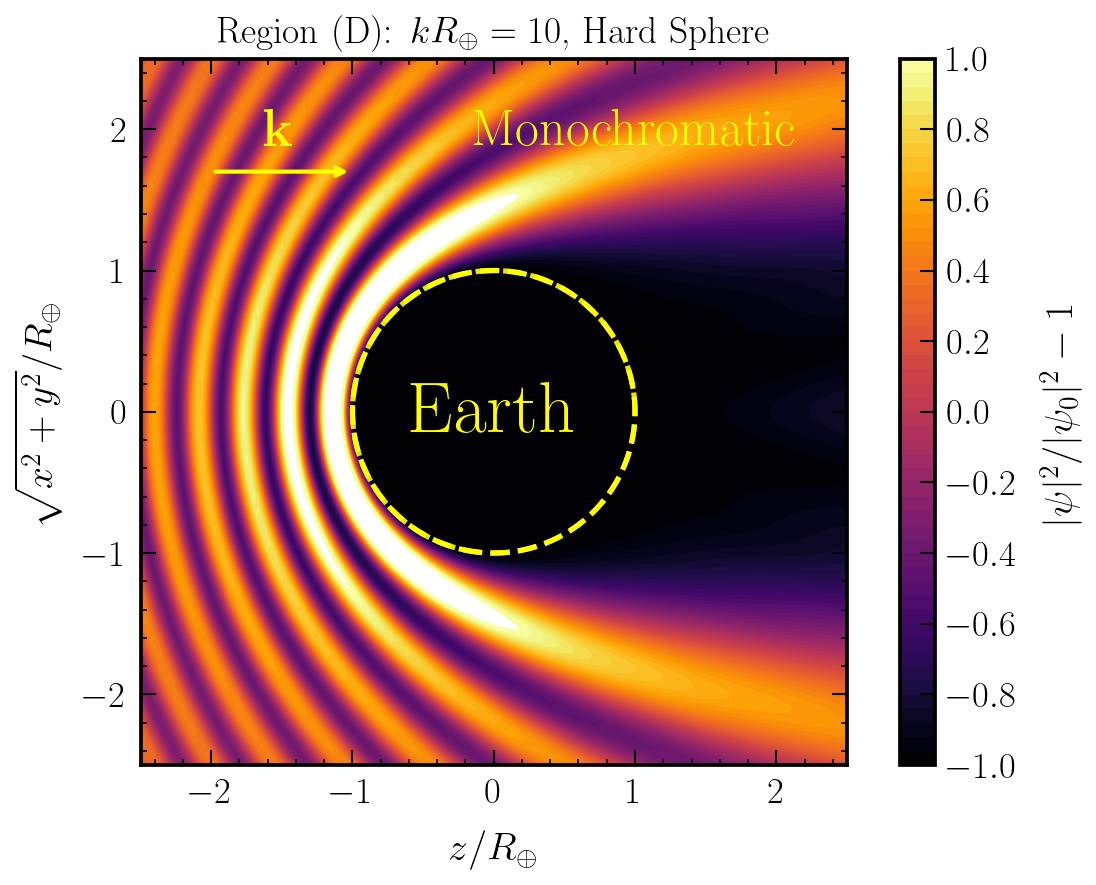}
\includegraphics[width=0.49\linewidth]{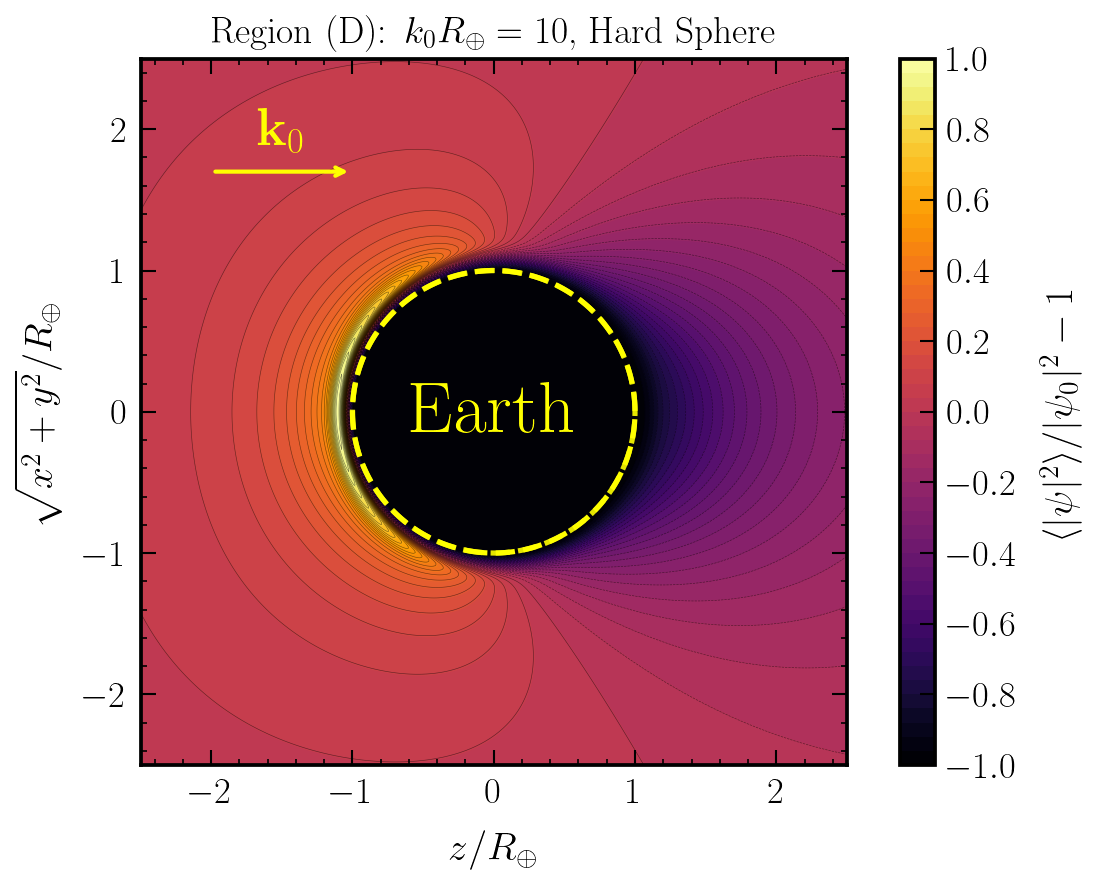}
\caption{The scalar field configurations in the hard-sphere limit at high incident momentum demonstrating the effect of ensemble averaging. {\bf Left}. The scalar field configuration $\abs{\psi}^2/\abs{\psi_0}^2-1$ with monochromatic incident momentum peaked at $\veck$, which satisfies $k R_\oplus = 10$. The profile becomes highly oscillatory and exhibits noticeable interference patterns. {\bf Right}. The phase-space averaged scalar-field configuration $\langle\abs{\psi}^2\rangle/\abs{\psi_0}^2-1$, where $\langle \cdots \rangle$ denotes the ensemble average over the DM distribution in \Eq{phase_space}. Here, the mean scalar momentum satisfies $k_0 R_\oplus = 10$. The oscillatory structures in the left panel are substantially smeared out by the phase-space integration. The leading directional dependence is dominated by the dipole term $a_1$. However, the scalar configuration deviations from a pure dipole pattern,  indicating the presence of higher-multipole contributions. 
}
\label{fig:nonpeturb_sca_nonsph_plt}
\end{figure}

\subsubsection{Phase-Space Ensemble Average}\label{sec:phase_space_ensemble_average}
%%%%%
\begin{figure}[h!]
\centering
\includegraphics[width=0.6\linewidth]{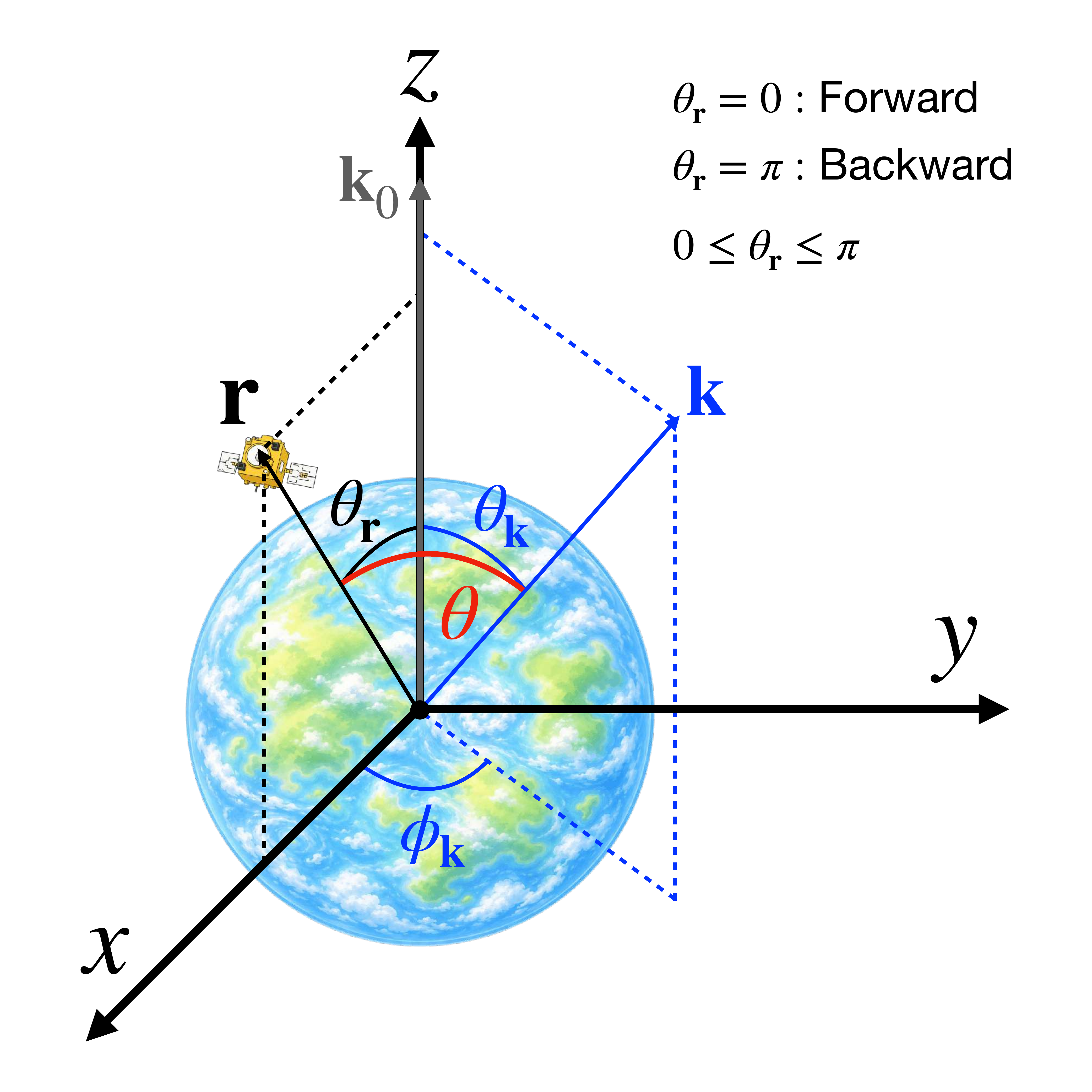}
\caption{Coordinate frame for phase space integration. The mean momentum $\veck_0$ of the DM defines the $z$-axis, and the position vector $\vecr$ lies in the $x\text{-}z$ plane at an angle $\theta_\vecr$ relative to $\veck_0$. The forward direction corresponds to $\theta_\vecr = 0$, while the backward direction corresponds to $\theta_\vecr = \pi$. A monochromatic mode in the phase space distribution is represented by $\veck$, which has a polar angle $\theta_\veck$ and an azimuthal angle $\phi_\veck$. 
$\cos\theta=\hat\veck\cdot\hat\vecr$
defines the deflection angle in QM scattering as discussed in \Subsec{scattering}.
}
\label{fig:frame}
\end{figure}
%%%%%
Up to now, the discussion has assumed a monochromatic scalar field with momentum $\veck$. We now generalize to the Maxwell--Boltzmann momentum distribution as expected for a virialized Galactic DM halo.
Following the standard stochastic description of virialized ULDM~\cite{Derevianko:2016vpm,Foster:2017hbq,Lisanti:2021vij,Hui:2021tkt,Kim:2023pkx,Cheong:2024ose}, we write the real scalar field as
\bea
\label{eq:phi_mode_expansion}
\phi(\vecx;t) = \frac{1}{2} \int \dbar^3 \veck \left[ a(\veck) \, e^{-i \omega(\veck) t} \, \psi(\vecx;\veck) + a^*(\veck)\, e^{i \omega(\veck) t}\, \psi^*(\vecx;\veck) \right]\,,
\eea
where $a(\veck)$ is a phase-space weighted complex Gaussian random coefficient, and $a^*(\veck)$ is its complex conjugate. The random coefficient satisfies
\begin{align}
\label{eq:gaussian_random}
\langle a(\veck) \, a^*(\veck') \rangle  = (2\pi)^3 \frac{f_\phi(\veck)}{n_\phi} \delta^{(3)}(\veck-\veck')\,, \quad \quad\,\, \langle a(\veck) \, a(\veck') \rangle = 0\,.
\end{align}
For a distribution sharply peaked at a single $\veck$, \Eq{phi_mode_expansion} reduces to the monochromatic field of \Eq{phi_NR_approx}, up to an overall stochastic coefficient carried by $a(\veck)$. In the fixed-amplitude approximation, this coefficient reduces to an overall random phase. In the perturbative regime (A), where the matter effect is negligible, or sufficiently far away from the Earth where the scattered contribution becomes subdominant, the spatial part of the wave function is approximately a plane wave $\psi(\vecx;\veck) \simeq \abs{\psi_0} \, e^{i \, \veck\cdot \vecx}$. In this limit, the stochastic field description reduces to the standard treatment discussed in Refs.~\cite{Derevianko:2016vpm,Foster:2017hbq,Lisanti:2021vij,Hui:2021tkt,Kim:2023pkx,Cheong:2024ose}.

From \Eq{phi_mode_expansion} and \Eq{gaussian_random}, we have the ensemble average of the squared scalar field value
\bea
\label{eq:phisq_ave_average}
2 \langle \phi^2 \rangle = \langle \abs{\psi}^2 \rangle = \frac{1}{n_\phi} \int \dbar^3 \veck \, f_\phi(\veck) \, \abs{\psi(\vecx;\veck)}^2\,.
\eea
\Eq{phisq_ave_average} shows that there is no interference between different momentum modes after ensemble averaging. This follows directly from the $\delta^{(3)}(\veck-\veck')$ term in \Eq{gaussian_random}. Equivalently, writing $a(\veck)=|a(\veck)| \, e^{i\chi(\veck)}$, the phase $\chi(\veck)$ is randomly and uniformly distributed for each momentum mode, so the cross terms between different momentum modes average to zero. The factor of ``$2$'' in \Eq{phisq_ave_average} comes from the fast time average of a real scalar field: for any specific $\veck$-mode, $\langle \phi^2\rangle = |\psi|^2 \langle \cos^2(\omega(\veck)\,t)\rangle = |\psi|^2/2$. In the parameter space of interest, the scalar oscillation frequency is much larger than the relevant experimental response frequencies, so this time average is valid for the observables considered here.

We describe the scalar phase-space distribution in the Solar rest frame using a boosted Maxwell--Boltzmann distribution truncated at the Galactic escape velocity,
\bea
\label{eq:phase_space}
f_\phi(\veck) = \frac{n_\phi }{\mathcal{N}(k_\esc)} \, \left(\frac{2\pi}{\sigma_k^2}\right)^{3/2} \exp\left[ - \frac{(\veck-\veck_0)^2}{2\sigma_k^2} \right] \Theta(k_\esc - \abs{\veck-\veck_0})\, ,
\eea
where $n_\phi$ is the scalar number density. Here, $k_\esc=m_\phi v_\esc$, with $v_\esc=544\;\text{km/s}$ the Galactic escape speed, and $\veck_0=m_\phi \vecv_0$. 
We denote the mean DM velocity in the Solar frame by $\vecv_0=-\vecv_\odot$, where $\vecv_\odot$ is the Sun's velocity relative to the Galactic rest frame, with magnitude $v_\odot \simeq 250\,$km/s~\cite{Baxter:2021pqo}. 
We neglect the Earth's orbital motion, which produces only a subleading correction. Following the SHM~\cite{Baxter:2021pqo}, we take the velocity dispersion to be $\sigma_v \simeq v_\LSR/\sqrt{2} \simeq 168$ km/s, which fixes $\sigma_k = m_\phi \sigma_v$.

The normalization condition $\int \dbar^3 \, \veck \, f_\phi(\veck) = n_\phi$ gives
\bea
\mathcal{N}(k_\esc) = \erf\left(\frac{k_\esc}{\sqrt{2} \sigmak}\right) -  \sqrt{\frac{2}{\pi}}\frac{k_\esc}{\sigmak} \exp\left(-\frac{k_\esc^2}{2 \, \sigmak^2}\right)\, .
\eea
To simplify the numerical computation, we take $k_\esc/\sigma_k \rightarrow \infty$ limit and set $\mathcal{N}(k_\esc)\simeq 1$. We have verified that this simplification has negligible effects on our final results.

In \Fig{nonpeturb_sca_nonsph_plt}, we compare the scalar profile for a monochromatic incident mode with that obtained after averaging over the phase-space distribution in \Eq{phisq_ave_average}. The left panel shows $\abs{\psi}^2/\abs{\psi_0}^2-1$ for a fixed momentum $\veck$ with $kR_\oplus=10$, where the field exhibits pronounced oscillations due to interference. The right panel shows the averaged profile $\langle\abs{\psi}^2\rangle/\abs{\psi_0}^2-1$ for $k_0R_\oplus=10$. 
Averaging over the momentum distribution in \Eq{phase_space} smooths out the oscillatory features of the monochromatic solution. In both cases, the scalar density is enhanced on the backward hemisphere ($\cos\theta_\vecr \lesssim 0$), where the ULDM wind accumulates near the Earth, and suppressed on the forward hemisphere ($\cos\theta_\vecr \gtrsim 0$), where the Earth partially shadows the incident flux. 
Here, $\cos\theta_\vecr=\hat\veck_0\cdot\hat\vecr$ defines the polar angle of $\vecr$ relative to the mean ULDM wind direction (\Fig{frame}).
As $k_0 R_\oplus$ increases further, the scalar profile approaches the optical limit and becomes only weakly dependent on $k_0 R_\oplus$. Although the optical approximation breaks down in the weak-coupling regime (B)~\cite{Huang:2024tog,Gruzinov:2024ciz,Kalia:2024xeq,VanTilburg:2024xib,Grossman:2025cov,Gan:2025nlu}, it remains valid in the strongly coupled regime (D) considered here. We also find that the shadowed region is not completely depleted after averaging over the phase-space distribution in \Eq{phase_space} and the fraction of ULDM remains at the \(\mathcal{O}(10\%)\) level. This residual density arises from the finite velocity dispersion, which allows a fraction of ULDM modes to arrive from directions substantially misaligned with the mean wind and populate the shadow region.

\begin{figure}[t]
\centering
\includegraphics[width=0.70\linewidth]{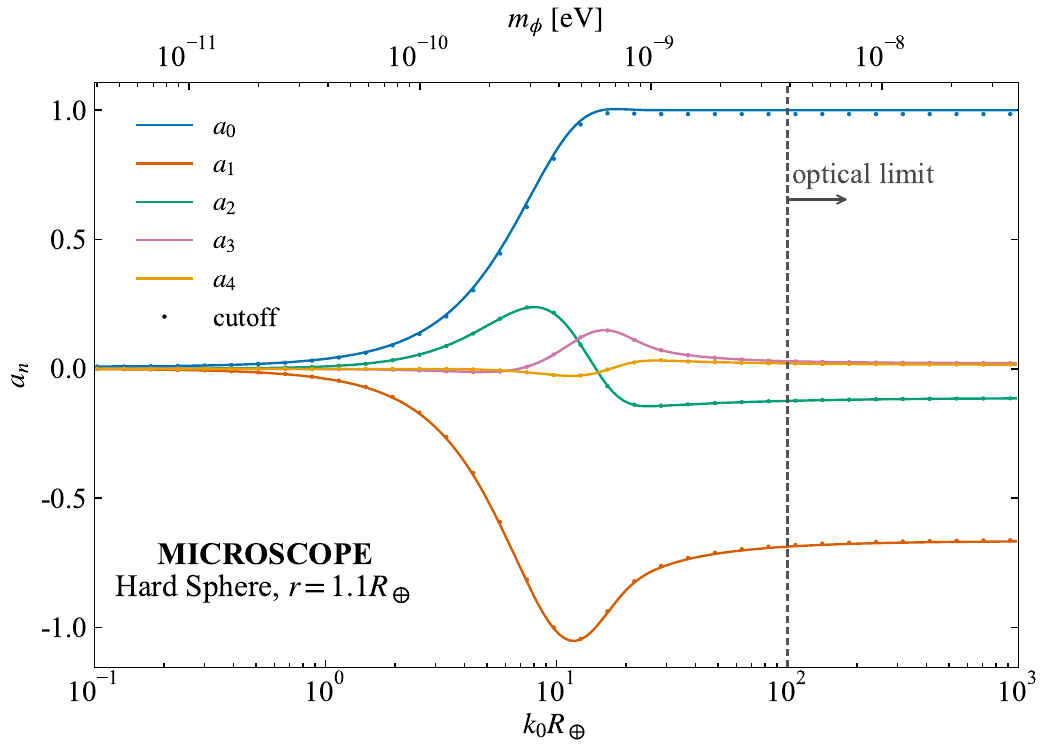}
\caption{Multipole coefficients $a_L$ of $\langle \abs{\psi}^2 \rangle$ for the first 5 moments as a function of $k_0 R_\oplus$ in the hard-sphere region (C and D) at $r=1.1\,R_\oplus$, corresponding to MICROSCOPE. {\bf Solid} lines are computed in the infinite-cutoff limit, $k_\esc/\sigma_k \rightarrow \infty$, while {\bf dots} with the same colors denote the computation including the finite cutoff $k_\esc$ in the phase-space distribution \Eq{phase_space}. 
When $k_0 R_\oplus \gtrsim 100$, the system enters the optical limit and all $a_L$ approach constant values. See \Sec{multipole} for discussion.}
\label{fig:hard_sphere_a_series}
\end{figure}

\subsubsection{Multipole Expansion}
\label{sec:multipole}
The previous section developed the formalism and provided an intuitive picture for the ensemble average $\langle \abs{\psi}^2 \rangle$, which is a key ingredient of the background-induced potential. 
The corresponding force is obtained from its spatial gradient. 
Previous calculations of $\langle \abs{\psi}^2 \rangle$ either integrated the partial-wave solution over the DM phase-space distribution~\cite{Gan:2025nlu} or relied on numerical simulations with many random realizations~\cite{VanTilburg:2024xib}.
More recently, Ref.~\cite{Brzeminski:2026rox} reformulated the result as a rapidly convergent multipole expansion, substantially simplifying the computation. 
We present this approach in this section and verify that it reproduces the direct calculation of Ref.~\cite{Gan:2025nlu} for arbitrary spatial angles in the appendices (see \Fig{direct_vs_ensemble_multipole_comparison}).

Because the system is symmetric around the mean DM momentum $\veck_0$, $\langle \abs{\psi}^2 \rangle$ is only a function of $r$ and $\cos\theta_\vecr=\hat \veck_0\cdot \hat\vecr$. 
Therefore, it can be expanded as
\bea
\label{eq:averaged_multipole_main}
\langle \abs{\psi}^2 \rangle = \abs{\psi_0}^2 \sum_{L=0}^{\infty} a_L(r;\,k_0,\mMearth) \, P_L(\cos\theta_\vecr)\, ,
\eea
 where $a_L$ is the $L$-th multipole coefficient of $\langle \abs{\psi}^2 \rangle$. 
 The sum over $L$ converges rapidly as $L_\text{max}$ increases, and we find that truncating the multipole expansion at $L_\text{max}\geq 4$ is sufficient to accurately describe $\langle \abs{\psi}^2 \rangle$.

\begin{figure}[t]
\centering
\includegraphics[width=0.70\linewidth]{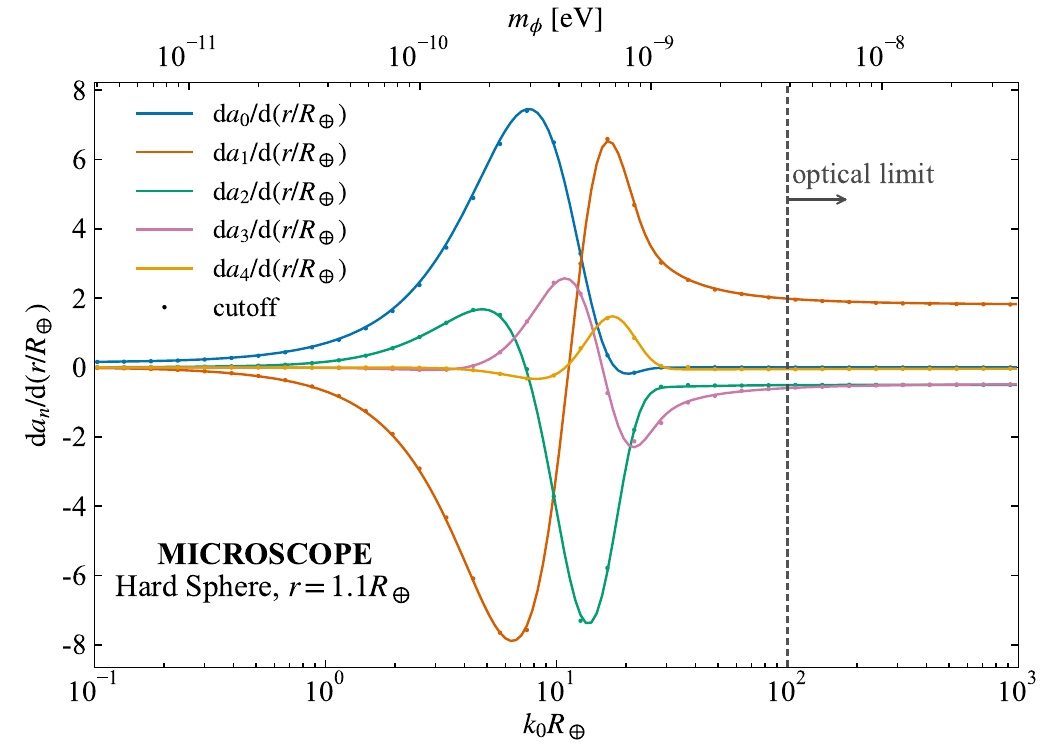}
\caption{The radial derivatives $\dd a_L/\dd(r/R_\oplus)$ for the first 5 moments as a function of $k_0 R_\oplus$ in the hard-sphere region (C and D). {\bf Solid} lines are computed in the infinite-cutoff limit, $k_\esc/\sigma_k \rightarrow \infty$, while {\bf dots} denote the computation given the finite cutoff $k_\esc$ in \Eq{phase_space}. 
When $k_0 R_\oplus \gtrsim 100$, the system enters the optical limit and $\dd a_L/\dd(r/R_\oplus)$ approach constants. See \Sec{multipole} for discussion.
}
\label{fig:hard_sphere_a_series_derivatives}
\end{figure}

To summarize this simplified method, we introduce two major reduction steps following Ref.~\cite{Brzeminski:2026rox}. The first step is to compute $\abs{\psi}^2$ for a fixed $\veck$ mode by linearizing the product of two Legendre polynomials that appear in the partial-wave expansion. Next is to use the Clebsch--Gordan relation,
\bea
\label{eq:Two_Pl_CG_main}
P_{l}(\cos\theta)\, P_{l'}(\cos\theta) = \sum_{L=\abs{l-l'}}^{l+l'} \mathcal{C}_{l \, l' \, L} P_L(\cos\theta)\,,
\eea
where the coefficients $\mathcal{C}_{l l' L}$ are listed in \Eq{linear_coefficient_appx} and $\cos\theta=\hat\veck\cdot\hat\vecr$. Note that $l$ labels the partial waves for a fixed $\veck$ while $L$ is for the ensemble-averaged expansion. 
We apply \Eq{Two_Pl_CG_main} to \Eq{psi_out_monochromatic_main},
integrate over the azimuthal coordinate $\phi_\veck$ and
after some algebra, acquire the multipole expansion in \Eq{averaged_multipole_main}. 
Specifically, in the infinite cutoff limit $k_\esc/\sigma_k \rightarrow \infty$, we have
\bea
\label{eq:aL_from_cL_int_simple}
\left.
\begin{aligned}
a_L(r;\,k_0,\mMearth) & = \left(\frac{1}{2\pi \sigmak^2}\right)^{3/2} 4\pi \int_0^{\infty} \dd k \, k^2 \exp\left[-\frac{k^2+k_0^2}{2 \sigmak^2}\right] c_L(r;\,k,\mMearth) \, i_L\left(\frac{k k_0}{\sigmak^2}\right)\\
\frac{\dd a_L(r;\,k_0,\mMearth)}{\dd r} & = \left(\frac{1}{2\pi \sigmak^2}\right)^{3/2} 4\pi \int_0^{\infty} \dd k \, k^2 \exp\left[-\frac{k^2+k_0^2}{2 \sigmak^2}\right] \frac{\dd c_L(r;\,k,\mMearth)}{\dd r} \, i_L\left(\frac{k k_0}{\sigmak^2}\right)
\end{aligned}
\right. \,.
\eea
Here, $i_L(x) = \sqrt{\frac{\pi}{2\,x}} I_{L+\frac{1}{2}}(x)$ is the spherical modified Bessel function of the first kind. 
We present the detailed derivation of the above multipole expansion and the generic forms of $a_L$ and $c_L$ for arbitrary $L$ in \Appx{partial_wave_appx}.

A particularly instructive component is the monopole coefficient $a_0$:
\bea
\label{eq:a0_specific}
a_0 = \left(\frac{1}{2\pi \sigmak^2}\right)^{1/2} \frac{1}{k_0}\int_0^\infty \dd k \, k \, \left\{ \exp\left[-\frac{(k-k_0)^2}{2\sigmak^2}\right] - \exp\left[-\frac{(k+k_0)^2}{2\sigmak^2}\right] \right\} \times \sum_{l=0}^{\infty} (2l+1)\,\abs{\calR_l}^2 \,.
\eea
$a_0$ corresponds to the angular average 
$\langle \langle \abs{\psi}^2\rangle\rangle_{\Omega_\vecr}/\abs{\psi_0}^2=a_0$, where
$\langle \cdots \rangle_{\Omega_\vecr}\equiv (1/4\pi) \int \dd\phi_\vecr \dd\theta_\vecr\,\sin\theta_\vecr\,(\cdots)$. \Eq{a0_specific} has been used as a benchmark computation in Refs.~\cite{Banerjee:2025dlo,Gan:2025nlu,Bouley:2025qtq}. In the limit $k_0/\sigma_k\rightarrow 0$, \Eq{a0_specific} reproduces the result for an isotropic phase-space distribution. In the low-momentum regime, we have
\bea
a_0 \simeq \psi_\sph^2/\abs{\psi_0}^2\,, ~~~~~\text{for}\,\,~k_0 R_\oplus \ll 1\,.
\eea
In this regime, the background-induced potential is dominated by $a_0-1 \simeq \psi_\sph^2/\abs{\psi_0}^2- 1$, and the force is dominated by $\dd a_0/\dd r \simeq \dd \big(\psi_\sph^2/\abs{\psi_0}^2\big)/\dd r$. In the high-momentum regime, $k_0R_\oplus \gtrsim 10$, the $L=0$ monopole contributions from $\dd a_0/\dd r$ are strongly suppressed, and the dominant ULDM signal instead comes from higher multipoles from $\dd a_L/\dd r$ with $L\geq1$.
Equivalently, angular averaging over $\Omega_\vecr$ corresponds to averaging over observer orientations, so the observer effectively experiences the DM wind from all directions, which leads to a suppression similar to the isotropic case. 
As we will see below, in the region $k_0 R_\oplus \gg 10$, corresponding to region (D), the MICROSCOPE constraint is dominated by the non-monopole multipoles, $L=1,2,\cdots$, rather than by the monopole contribution with $L=0$.

\Fig{hard_sphere_a_series} and~\Fig{hard_sphere_a_series_derivatives} show the multipole coefficients $a_L$ and their radial derivatives for the hard-sphere limit ($\mMearth \gtrsim k_0$ and $\mMearth \gtrsim R_\oplus^{-1}$) at $r=1.1\,R_\oplus$, which corresponds to the MICROSCOPE orbit. 
Shown are the calculations with a finite and infinite momentum cutoff, and the results are in excellent agreement. 
Both $a_L$ and $\dd a_L/\dd(r/R_\oplus)$ decrease rapidly with $L$ and become negligible for $L \gtrsim 5$. 
In the spherically symmetric regime, $k_0R_\oplus \ll 1$, which corresponds to region (C), the monopole terms $a_0$ and $\dd a_0/\dd(r/R_\oplus)$ dominate.\footnote{Note the same behavior holds for region (A), where $\mMearth R_\oplus \lesssim 1$ and $k_0 R_\oplus \lesssim 1$.} However, as $k_0R_\oplus$ increases and we enter region (D), the dipole contribution $a_1$ becomes comparable to the monopole, while the monopole derivative is suppressed and the force becomes dominated by the derivatives of the higher multipoles, particularly the dipole and quadrupole terms. 
This demonstrates that the spherically symmetric ansatz is no longer valid in region (D).
We also find that both $a_L$ and $\dd a_L/\dd(r/R_\oplus)$ approach constant values for $k_0R_\oplus \gtrsim 100$, signaling the onset of the optical limit. In this regime, where the scalar de Broglie wavelength is much smaller than $R_\oplus$, the field behaves as a collection of classical particles and the spatial distribution of $\langle \abs{\psi}^2\rangle$, or equivalently the scalar density $\rho_\phi$, is determined primarily by geometric parameters such as $r/R_\oplus$ and $\sigma_k/k_0$. In \Appx{geometric_optics}, we give detailed steps on computing the multipole coefficients $a_L$ using geometric optics, which provides a simple way to compute $a_L$ in the regime $k_0 R_\oplus \gg 1$ and $k_0 h \gg 1$. The latter condition ensures that the satellite altitude is much larger than the de Broglie wavelength. 
We have numerically verified that the resulting geometric-optics calculation agrees with the full partial-wave and phase-space-integrated calculation.

\subsection{Final Expressions}

Now that we have $\langle \phi^2\rangle (=\langle \abs{\psi}^2 \rangle/2)$, we can compute the background-induced potential and the corresponding force. To start with, we review the motion of the test mass $\testmass$ driven by the scalar background following Refs.~\cite{Hees:2018fpg,Banerjee:2022sqg,Gan:2025nlu}. In the ULDM mass range considered here, the scalar oscillation frequency is much larger than the relevant experimental response frequencies. The test-mass action can therefore be written as $S_\testmass = \int \dd t \, L_\testmass = - \int \dd t \, M_\testmass(\langle \phi^2 \rangle) \sqrt{g_{\mu\nu} (\dd x^{\mu}/\dd t) (\dd x^{\nu}/\dd t)}$, where $t$ is the coordinate time and $L_\testmass$ is the corresponding Lagrangian. Because the test mass motion is non-relativistic, and the test mass variation in terms of $\phi^2$ is given in \Eq{macroscopic_mass_variation}, the leading order Lagrangian is
\bea
\label{eq:L_testmass_expand}
L_\testmass =  M_\testmass \frac{\vecv_\testmass^2}{2} - \frac{\dd M_\testmass}{\dd \langle \phi^2 \rangle}\bigg|_{\phi=0} \langle \phi^2 \rangle+ \cdots\,, 
\eea
where the second term containing $\langle \phi^2 \rangle$ acts as the scalar-induced potential for the test mass. Using the Euler--Lagrange equation for \Eq{L_testmass_expand}, the scalar-induced acceleration of the test mass is $\veca_\testmass = - \vecnabla M_\testmass/M_\testmass $. Since $\vecF_\bg = M_\testmass \, \veca_\testmass = - \vecnabla V_\bg$, the background-induced potential can be identified with the scalar-induced mass shift, $V_\bg=\Delta M_\testmass$. Using \Eq{macroscopic_mass_variation}, we obtain
\bea
\label{eq:Vbg_from_geodesic}
V_\bg =  \alpha^{(2)}_\testmass \, \frac{4\pi \, M_\testmass }{\Mpl^2} \frac{\langle \phi^2 \rangle}{2} + \const \,\,\,\,\,\,\quad \Longleftrightarrow \,\,\,\,\quad \underbrace{V_\bg = \frac{\mMtest^2 \mathcal{V}_\testmass}{4} \big( \langle\abs{\psi}^2 \rangle - \abs{\psi_0}^2\big)}_{\text{This work}}\,.
\eea
To see the equivalence between the two expressions in \Eq{Vbg_from_geodesic}, we recall that the effective mass in the test mass medium is given by $\mMtest^2 = \alpha_\testmass^{(2)}(4\pi \, \rho_\testmass/\Mpl^2)$ based on \Eq{mM_def}, $\langle \phi^2 \rangle = \langle \abs{\psi}^2 \rangle/2$ from $\langle \cos^2(m_\phi t) \rangle=1/2$, and the test mass $M_\testmass = \rho_\testmass \mathcal{V}_\testmass$. Although the two representations of $V_\bg$ in \Eq{Vbg_from_geodesic} are mathematically equivalent, we use the form on the right-hand side as the default convention in this work. The reason is that the $\mMearth$-dependence is contained in $\langle |\psi|^2\rangle$, allowing $V_\bg$ to be written in a more symmetric Newtonian form with the form factor $\formfactorV$, as in \Eq{background_potential}. Note that the right-hand side of \Eq{Vbg_from_geodesic} fixes the additive constant such that $V_\bg \propto \langle \abs{\psi}^2 \rangle - \abs{\psi_0}^2$, following similar convention to Refs.~\cite{Huang:2024tog,Kalia:2024xeq,Gruzinov:2024ciz,VanTilburg:2024xib,Gan:2025nlu}. This choice is convenient because it implies
\begin{equation}
V_\bg \propto \langle |\psi|^2 \rangle/|\psi_0|^2 - 1 \propto \frac{\Delta \rho_\phi}{\rho_\phi}\,,
\end{equation}
since only spatial variations in the ULDM density contribute to the background-induced force.

\begin{figure}[t!]
\centering
\includegraphics[width=0.49\linewidth]{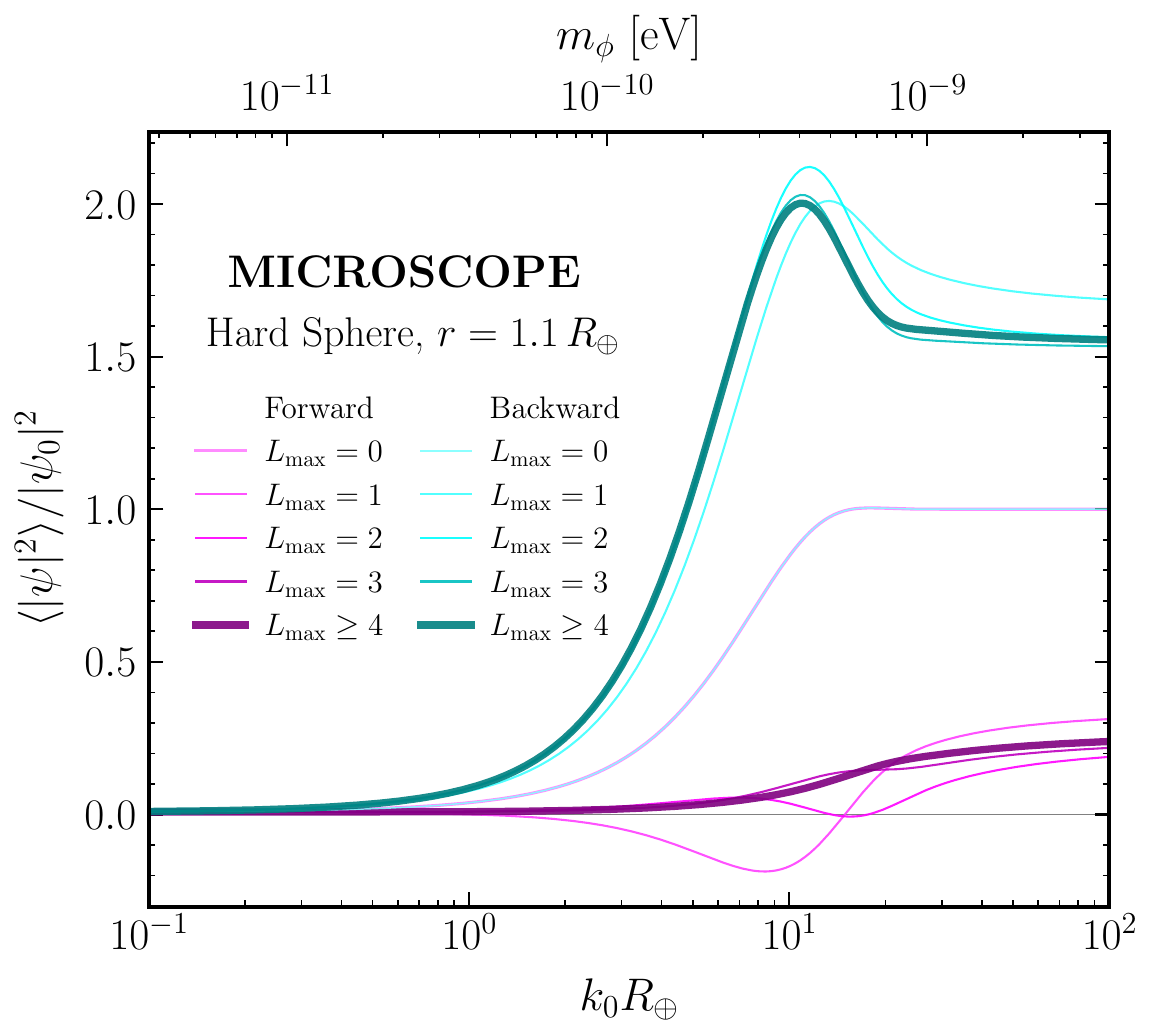}
\includegraphics[width=0.49\linewidth]{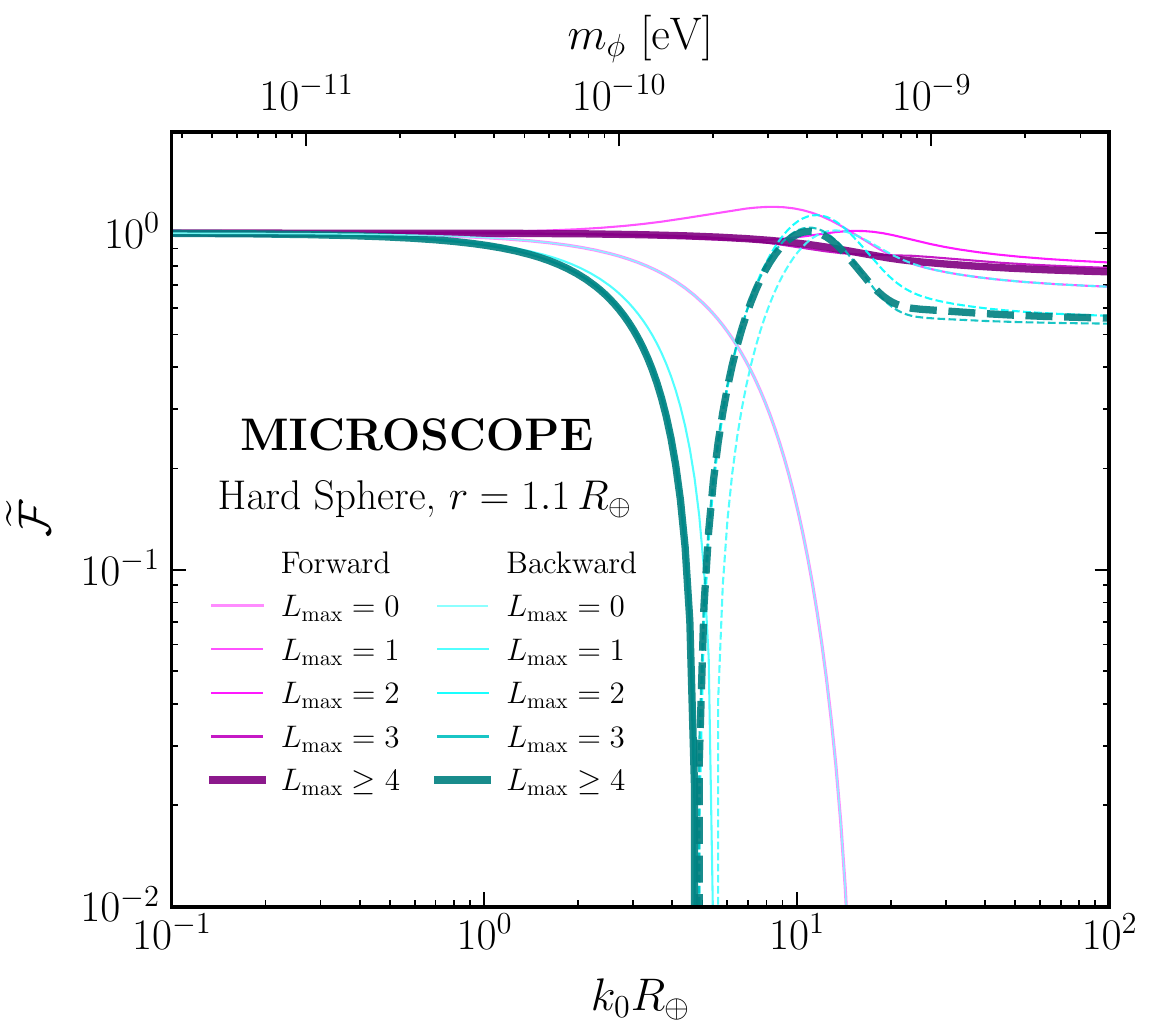}
\caption{
Dependence of the scalar configuration $\langle \abs{\psi}^2 \rangle/\abs{\psi_0}^2$ ({\bf left}) and the reduced force form factor $\widetilde{\mathcal{F}}$ ({\bf right}) on $k_0 R_\oplus$ for MICROSCOPE~($r=1.1\,R_\oplus$). The results are shown in the hard-sphere regions~(C) and~(D), as defined in \Fig{classification}. {\bf Cyan} denotes the backward direction, $\theta_\vecr=\pi$. {\bf Magenta} denotes the forward direction, $\theta_\vecr=0$. The color intensity indicates the truncation order $L_{\max}$ in the multipole expansion of \Eq{averaged_multipole_main}, with darker curves corresponding to larger $L_{\max}$. For $L_{\max}\gtrsim 4$, the multipole expansion in \Eq{averaged_multipole_main} provides a converged description of the scalar configuration. 
}
\label{fig:nonpeturb_config_formfactor_plt}
\end{figure}

Using the ensemble-averaged multipole expansion, we can now express the background-induced potential and the corresponding force in terms of the coefficients $a_L$:
\bea
\label{eq:Vbg_multipole}
V_\bg(\vecr,\veck_0) = - \frac{\mMtest^2 \mathcal{V}_\testmass}{4} \, \abs{\psi_0}^2 \, \bigg[ 1 - \sum_{L=0}^\infty a_L(r;\,k_0,\mMearth)\, P_L(\cos\theta_\vecr) \bigg]\,.
\eea
To compare with the spherically symmetric scenario discussed in Sec.~\Ref{subsec:sph_symmetric_main}, we note that $\rho_\phi = \frac{1}{2}m_\phi^2 |\phi_0|^2=\frac{1}{2}m_\phi^2 |\psi_0|^2$  and write
\bea
\label{eq:potential_formfactor}
V_\bg(\vecr,\veck_0) &=
-\frac{|\psi_0|^2}{2}\,\frac{(\mMtest^2 \mathcal{V}_\testmass)(\mMearth^2\mathcal{V}_\oplus)}{4\pi r}\times \formfactorV_\sph(r)\times \formfactorreduce(\vecr,\veck_0)\, ,
\eea
where the reduced form factor
\bea
\label{eq:reduced_form_factor_def}
\formfactorreduce(\vecr,\veck_0, \mMearth)&\equiv \frac{\formfactorV(\vecr,\veck_0, \mMearth)}{\formfactorV_\sph(r,\mMearth)}\, ,
\eea quantifies the deviation from the spherically symmetric ansatz. 
Comparing \Eq{Vbg_multipole} and \Eq{potential_formfactor}, we see that
\bea
\formfactorreduce(\vecr,\veck_0,\mMearth)&=\left( \frac{2 \pi r}{\mMearth^2 \mathcal{V}_\oplus} \times \frac{1}{\formfactorV_\sph(r)} \right) \times \bigg[1 - \sum_{L=0}^{\infty} a_L(r;\,k_0,\mMearth) \, P_L(\cos\theta_\vecr) \bigg]\, .
\eea
In the limit $k_0 R_\oplus \ll 1$, we have 
\bea
\label{eq:one_minus_aseries_sph}
1 - \sum_{L=0}^{+\infty} a_L(r;\,k_0,\mMearth) \, P_L(\cos\theta_\vecr) \simeq 1 - \frac{\psi_\sph^2}{\abs{\psi_0}^2} \quad \quad \text{for}\,\,~k_0 R_\oplus \ll 1\, .
\eea
Therefore, when $k_0 R_\oplus \ll 1$, substituting \Eq{psi_sph_config} into \Eq{one_minus_aseries_sph} and combining it with \Eq{formfactorV_sph} gives $\formfactorreduce\simeq1$, as expected from the spherically symmetric limit. However, when $k_0 R_\oplus \gtrsim 1$, this reduced form factor describes the deviation of the background-induced potential, $V_\bg \propto \langle\abs{\psi}^2\rangle-\abs{\psi_0}^2$, from the spherically symmetric ansatz. In \Fig{nonpeturb_config_formfactor_plt}, we show the dependence of the scalar configuration $\langle \abs{\psi}^2 \rangle/\abs{\psi_0}^2$ and $\widetilde{\mathcal{F}}$ on $k_0 R_\oplus$, or equivalently on $m_\phi$, for the MICROSCOPE satellite at $r=1.1\,R_\oplus$ in the forward~($\theta_\vecr=0$) and backward~($\theta_\vecr=\pi$) directions. 
\Fig{nonpeturb_config_formfactor_plt} shows that the scalar configuration and form factor behave qualitatively differently in these two directions. In the forward direction, $\langle \abs{\psi}^2 \rangle < \abs{\psi_0}^2$ throughout the range of $k_0R_\oplus$ shown, indicating persistent screening of the ULDM density at the MICROSCOPE orbit, and $\formfactorreduce$ is always positive. 
In contrast, the backward direction exhibits a density enhancement for $k_0R_\oplus \gtrsim 5$, reaching $\langle \abs{\psi}^2 \rangle \simeq 2$ near $k_0R_\oplus \sim 10$ due to the accumulation of ULDM behind the Earth. 
$\formfactorreduce$ goes from positive to negative at $k_0R_\oplus \gtrsim 5$, indicating that  $\langle\abs{\psi^2}\rangle > \abs{\psi_0}^2$ here. As expected from the spherically symmetric limit, $\formfactorreduce\simeq 1$ when $k_0R_\oplus\ll 1$.

Moving on, we define the rescaled background-induced force
\begin{align}
\label{eq:rescaled_force}
\widetilde\vecF_\text{bg}&\equiv\frac{\vecF_\text{bg}}{\abs{\psi_0}^2 \, \mMtest^2 \,\calV_\testmass/4} = - \vecnabla \left[ \langle \abs{\psi}^2 \rangle/\abs{\psi_0}^2\right]\, ,
\end{align}
which factors out the the test-mass dependence and isolates the Earth-induced contribution.
 Substituting the ensemble-averaged multipole expansion of $\langle \abs{\psi}^2\rangle$ in \Eq{averaged_multipole_main} and evaluating the gradient in spherical coordinates, we obtain
\bea
\label{eq:multipole_force_dimless}
\widetilde\vecF_\text{bg} =\underbrace{-\hat\vecr\sum_L^\infty \frac{\dd a_L}{\dd r}P_L(\cos\theta_\vecr)}_{\text{Radial derivative}}+ \underbrace{\frac{1}{r}(\sin\theta_\vecr \, \hat{\vectheta}_\vecr)\sum_{L=0}^\infty a_L(r)P_L'(\cos\theta_\vecr)}_{\text{Angular derivative}}\,,
\eea
where $\hat{\vecr} = \vecr/\abs{\vecr}$ is the unit vector in radial direction and $\hat{\boldsymbol{\theta}}_\vecr = \hat{\vecr} \times (\hat{\vecr} \times \hat{\veck}_0)/|\hat{\vecr} \times (\hat{\vecr} \times \hat{\veck}_0)|$ is the unit vector in polar angle direction, and $P_L'(\cos\theta_\vecr)= \dd P_L/\dd\cos\theta_\vecr$. Thus, the background-induced force is fully determined by $a_L$ and $\dd a_L/\dd r$, which characterize the angular and radial derivative of $V_\bg$, respectively.

%%%%%%%
\begin{figure}[t!]
\centering
\includegraphics[width=0.49\linewidth]{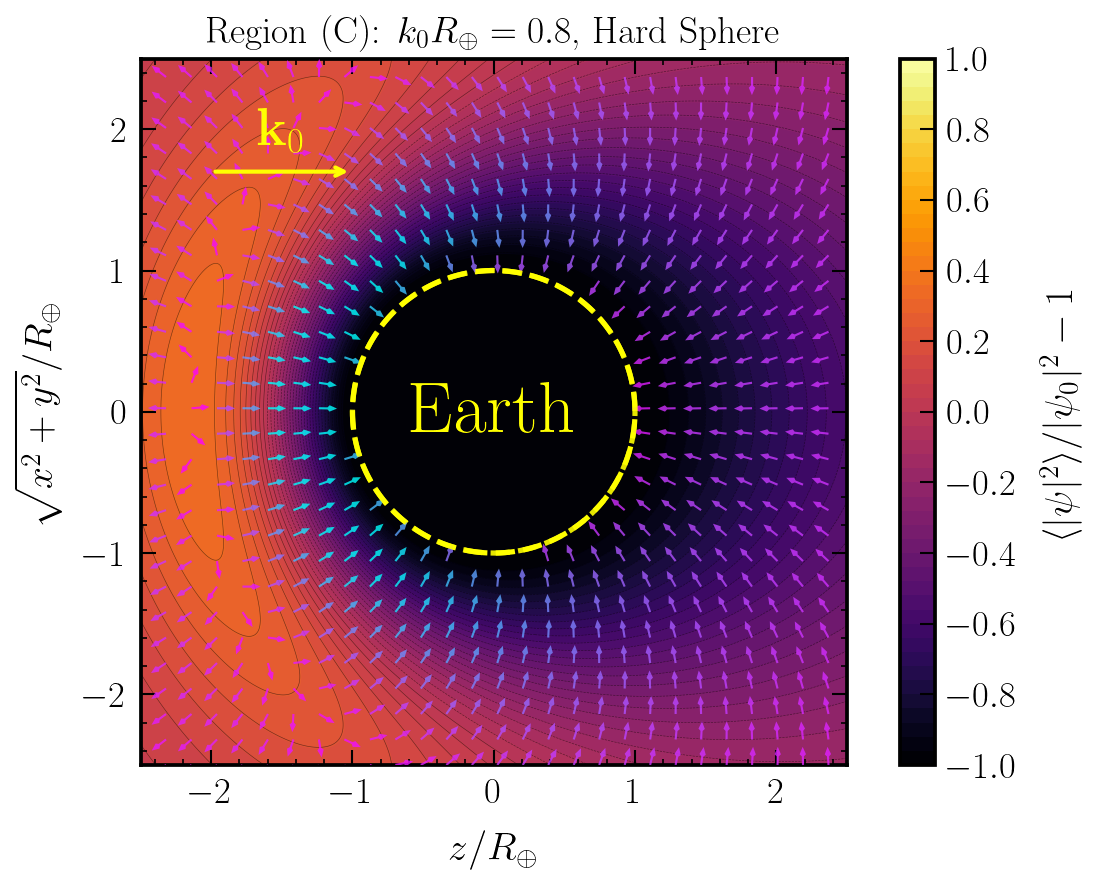}
\includegraphics[width=0.49\linewidth]{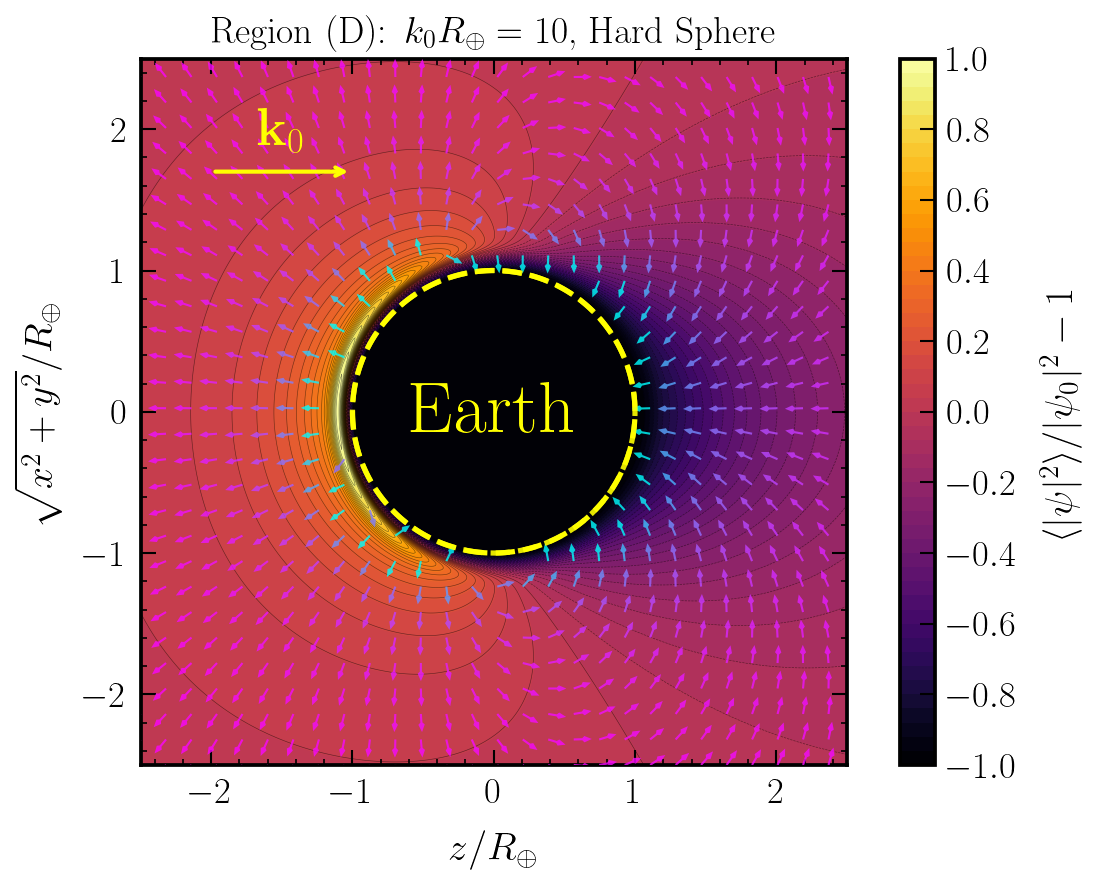}
\caption{Illustration of how the direction of the background-induced force changes as $k_0 R_\oplus$ increases. {\bf Left}. $k_0 R_\oplus = 0.8$. The scalar profile starts to accumulate at a finite distance from the Earth. Near the Earth's surface, the background-induced force still points toward the Earth, acting as an attractive force. However, the scalar configuration starts to deviate from spherical symmetry. {\bf Right}. $k_0 R_\oplus = 10$. The system becomes strongly anisotropic. The DM wind compresses the scalar profile near the Earth's surface in the backward hemisphere~($\cos\theta_\vecr \lesssim 0$). In this case, the background-induced force in the backward hemisphere reverses direction and points away from the Earth, acting as a repulsive force. By contrast, the force in the forward hemisphere~($\cos\theta_\vecr \gtrsim 0$) still points toward the Earth, acting as an attractive force. This strongly anisotropic force distribution at low altitude, generated by the Earth's scattering of the high-momentum ULDM wind, is the key origin of the unique band-split signal in the MICROSCOPE frequency space discussed below.}
\label{fig:nonpeturb_force_nonsph_plt}
\end{figure}
%%%%%%%%

Before proceeding, we highlight one of the unique characteristics of the background-induced force, compared with the force obtained in the spherically symmetric approximation in the hard-sphere limit, as demonstrated in \Fig{nonpeturb_force_nonsph_plt}. 
In the left panel, where $k_0 R_\oplus = 0.8$, the scalar profile already starts to accumulate near the Earth. 
However, the background-induced force, denoted by the small arrows, remains attractive and points toward the Earth's center. 
In the right panel, where $k_0 R_\oplus = 10$, the incident scalar momentum is large enough for the DM wind to compress the scalar profile near the Earth's surface. 
In this case, the background-induced force in the backward hemisphere~($\cos \theta_\vecr \lesssim 0$) reverses direction and becomes repulsive, while the force in the forward hemisphere~($\cos \theta_\vecr \gtrsim 0$) remains attractive. 
This anisotropy is a key feature leading to the band-split signal in MICROSCOPE and future EP experiments, providing a unique smoking-gun signature. 
We will discuss this smoking-gun signature in satellite EP tests in the following section.

\section{Equivalence Principle Tests}\label{sec:microscope}

The acceleration caused by the background-induced potential, given by \Eq{multipole_force_dimless}, can be measured in EP tests (see Table~\ref{tab:ep_test_altitudes}). As a representative example, we use the setup of the MICROSCOPE space mission~\cite{Touboul:2012ui}, which operated from April 2016 to October 2018 and tested the universality of free fall using pairs of Ti- and Pt-alloy test masses onboard the satellite. 
The satellite was in a nearly circular orbit around the Earth at 710 km of altitude. At this altitude, the Earth's gravitational acceleration is about 7.9 m/s$^2$. The two test masses are co-axial hollow cylinders. The electrostatic accelerometers measure the acceleration of each test mass. EP violation would manifest itself as a differential acceleration between the two test masses. We define the longitudinal axis of the test-mass cylinders as $\hat{\bf X}$, along which MICROSCOPE has its greatest acceleration sensitivity.

\subsection{Kinematics}\label{sec:micro_kinematics}

\begin{figure}[h!]
\centering
\includegraphics[width=0.6\linewidth]{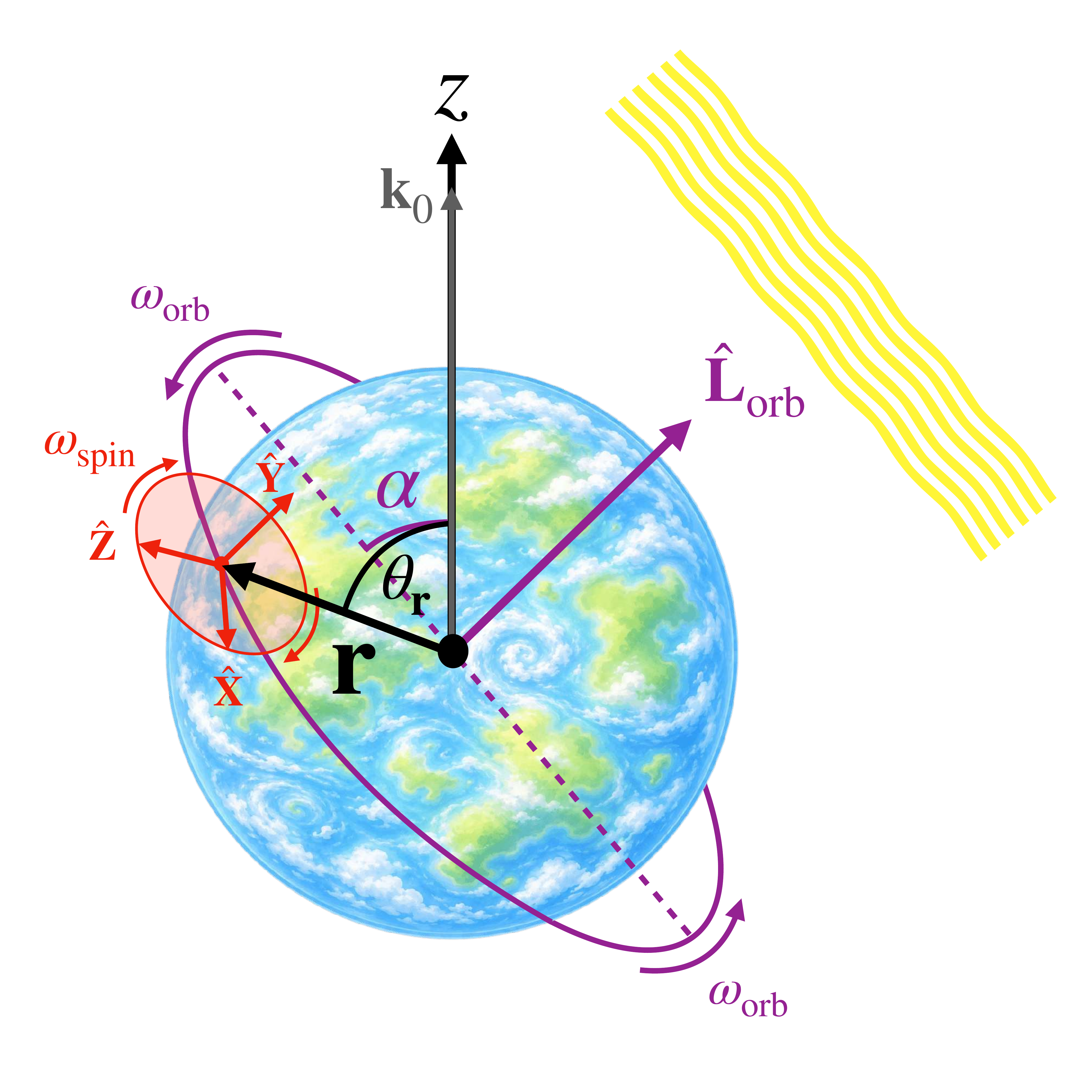}
\caption{The schematic plot of the coordinate system for the orbital motion of MICROSCOPE satellite. $\alpha$ is the angle between the MICROSCOPE orbital plane ({\bf purple}) and the DM momentum $\veck_0$.
$\theta_\vecr$ is the angle between $\vecr$ and $\veck_0$, where $\vecr$ is the Earth-centered satellite position vector. 
The satellite orbits the Earth with angular frequency $\omega_\orb$. 
For DDSSO, the unit vector of the orbital angular momentum, $\hat{\vecL}_\orb = \vecL_\orb/\abs{\vecL_\orb}$, approximately tracks the Sun direction. 
In the spin mode, MICROSCOPE instrument frame ({\bf red}) rotates with the satellite: the $\vecY$ axis is fixed parallel to the orbital angular momentum, $\vecY \equiv -\, \hat{\boldsymbol{\omega}}_\spin \simeq \hat{\vecL}_\orb$, while the $\vecX$ and $\vecZ$ axes lie in the orbital plane and rotate about $\vecY$. Since the instrument-frame rotation is opposite to the orbital rotation, the EP signal frequency for spherically symmetric central force is $\omega_\EP=\omega_\spin+\omega_\orb$.}
\label{fig:micro_orbit}
\end{figure}

The MICROSCOPE satellite operates in a dawn--dusk Sun-synchronous orbit (DDSSO); \Fig{micro_orbit} shows its orbit and the orientation and rotation of the instrument axes $\vecX$, $\vecY$, $\vecZ$,\footnote{The MICROSCOPE literature~\cite{Robert:2020ddm,Rodrigues:2022hmp,Touboul:2022yrw} uses both the instrument and satellite frames, related approximately by $\vecX_\inst \simeq -\vecZ_\sat$, $\vecY_\inst \simeq \vecX_\sat$, and $\vecZ_\inst \simeq -\vecY_\sat$. Throughout this work, $\vecX$, $\vecY$, and $\vecZ$ denote $\vecX_\inst$, $\vecY_\inst$, and $\vecZ_\inst$, respectively.} with the Sun direction indicating the DDSSO configuration. Its orbital motion and the annual precession of the orbital plane lead to characteristic frequency-space signatures on the background-induced force. To describe these signals, we first review the DDSSO kinematics and compute the annual modulation of the angle $\alpha(t)$ between the orbital plane and the mean DM momentum $\veck_0$~\footnote{We emphasize that the term annual modulation in our work refers to a different physical mechanism from the conventional annual modulation discussed in dark matter direct-detection experiments~(see Ref.~\cite{Freese:2012xd} for example) which arises from the Earth's orbital motion around the Sun. In our case, the annual modulation originates from the yearly precession of the satellite's DDSSO, which causes the signal to vary on an annual timescale.}. We then determine the rotation of the MICROSCOPE axes $\vecX$, $\vecY$, and $\vecZ$, and derive the geometric projection factors needed to express the background-induced force in the instrument frame.

MICROSCOPE has a local time of ascending node~(LTAN) of $18{:}00$, which corresponds to the local time at which the satellite crosses the equatorial plane from south to north. 
Because MICROSCOPE operated in a DDSSO, its orbital angular momentum $\hat{\vecL}_{\rm orb}$ approximately tracked the Sun. As a result, the right ascension of the ascending node (RAAN), $\Omega$, precessed at $\dot{\Omega}\simeq \omega_\odot = 360^\circ/\yr$, where $\omega_\odot$ is the annual solar angular frequency.
~\footnote{We use radians and degrees interchangeably when appropriate. In discussions of satellite kinematics and astronomical coordinates, we often express angles in degrees, denoted by ${}^\circ$, following the common convention in the relevant literature.} Therefore, we have
\bea
\label{eq:Omega_approx_fit}
\Omega - 90^\circ \simeq \omega_\odot \, t + \delta_\orb\, .
\eea
Here, we choose $t=0$ at the spring equinox. The small phase offset $\delta_\orb\simeq -2.7^\circ$ is determined by comparing this analytic parametrization with satellite orbital data~\cite{spacetrack}. Such a positive nodal precession requires a retrograde orbit, with inclination $90^\circ<i_\orb<180^\circ$, where $i_\orb$ denotes the inclination of the satellite orbit with respect to the equatorial plane. Using the nodal-precession formula~\cite{vallado2001fundamentals}, we obtain
\bea
\label{eq:Nodal_precession_Rate}
\cos i_\orb \simeq - \frac{2 \, \dot{\Omega}}{3 \, J_2} \left(\frac{R_\oplus+h}{R_\oplus}\right)^2 \frac{1}{\omega_\orb} \quad \Longrightarrow \quad i_\orb \simeq 98.2^\circ \,.
\eea
Here $\omega_\orb=2\pi/T_\orb$ is the orbital angular frequency, $T_\orb=5946\,{\rm s}$ is the orbital period~\cite{Robert:2020ddm,Rodrigues:2022hmp}, and $J_2 = 1.0826 \times 10^{-3}$ is the Earth's dynamical form factor, which quantifies the deviation of the Earth's gravitational potential from spherical symmetry~\cite{moritz2000geodetic,luzum2011iau}. Since MICROSCOPE followed an almost circular orbit, we neglect the orbital eccentricity. For $i_\orb>90^\circ$, the torque from the Earth's oblateness gives a positive nodal precession, i.e., $\dot{\Omega} > 0$, so that the orbital angular momentum $\vecL_\orb$ precesses around the Earth's spin axis, approximately toward the north celestial pole~(NCP). The orbital angular-momentum direction $\hat{\vecL}_\orb=\vecL_\orb/\abs{\vecL_\orb}$ in the ECL frame can then be written as
\bea
\label{eq:vL_ECL}
(\hat{\vecL}_\orb)_\ECL 
& = 
\begin{pmatrix}
1  &    0                & 0                     \\
0  &    \cos\varepsilon  & \sin\varepsilon \\
0  &   -\sin\varepsilon  & \cos\varepsilon
\end{pmatrix}
\begin{pmatrix}
\sin i_\orb  \cos(\Omega - 90^\circ) \\
\sin i_\orb \sin(\Omega - 90^\circ) \\
\cos i_\orb
\end{pmatrix}\,.
\eea
where $\varepsilon=23.44^\circ$ is the obliquity of the ecliptic, i.e. the angle between the ecliptic and equatorial planes~\cite{moritz2000geodetic,luzum2011iau}. Here the matrix on the right-hand side of \Eq{vL_ECL} is the rotation matrix from the EQU frame to the ECL frame, while the vector multiplying it is $(\hat{\vecL}_\orb)_\EQU$, parametrized by the inclination $i_\orb$ and the equatorial longitude $\Omega-90^\circ$ of the direction of $\vecL_\orb$.

In the ECL coordinate system, the mean DM momentum direction $\hat{\veck}_0$ is fixed and points opposite to the Cygnus~(DM wind apex) direction $\hat{\vecr}_\cyg$, as given by
\bea
\label{eq:k0_cyg_main}
\hat{\veck}_0 = - \hat{\vecr}_\cyg\,.
\eea
Specifically, the ECL components of $\hat{\vecr}_\cyg$ are $\hat{\vecr}_\cyg
= \big(\cos\beta_\cyg\cos\lambda_\cyg,\; \cos\beta_\cyg\sin\lambda_\cyg,\; \sin\beta_\cyg\big)_\ECL$ with $(\lambda_\cyg,\beta_\cyg) =  (341.6^\circ, 60.6^\circ)$, where  $\lambda_\cyg$ and $\beta_\cyg$ are the ecliptic longitude and ecliptic latitude, respectively. The values of $(\lambda_\cyg,\beta_\cyg)$ are obtained by identifying the DM wind apex direction with the solar velocity direction in the Galactic frame,
$\hat{\vecr}_\cyg=\vecv_\odot/\abs{\vecv_\odot}$,
using the velocity parameters in Ref.~\cite{Baxter:2021pqo}, and then transforming this direction to the ECL frame at equinox J2000 using Ref.~\cite{LAMBDA_CoordConv}.
 
\begin{figure}[t!]
\centering
\includegraphics[width=0.83\linewidth]{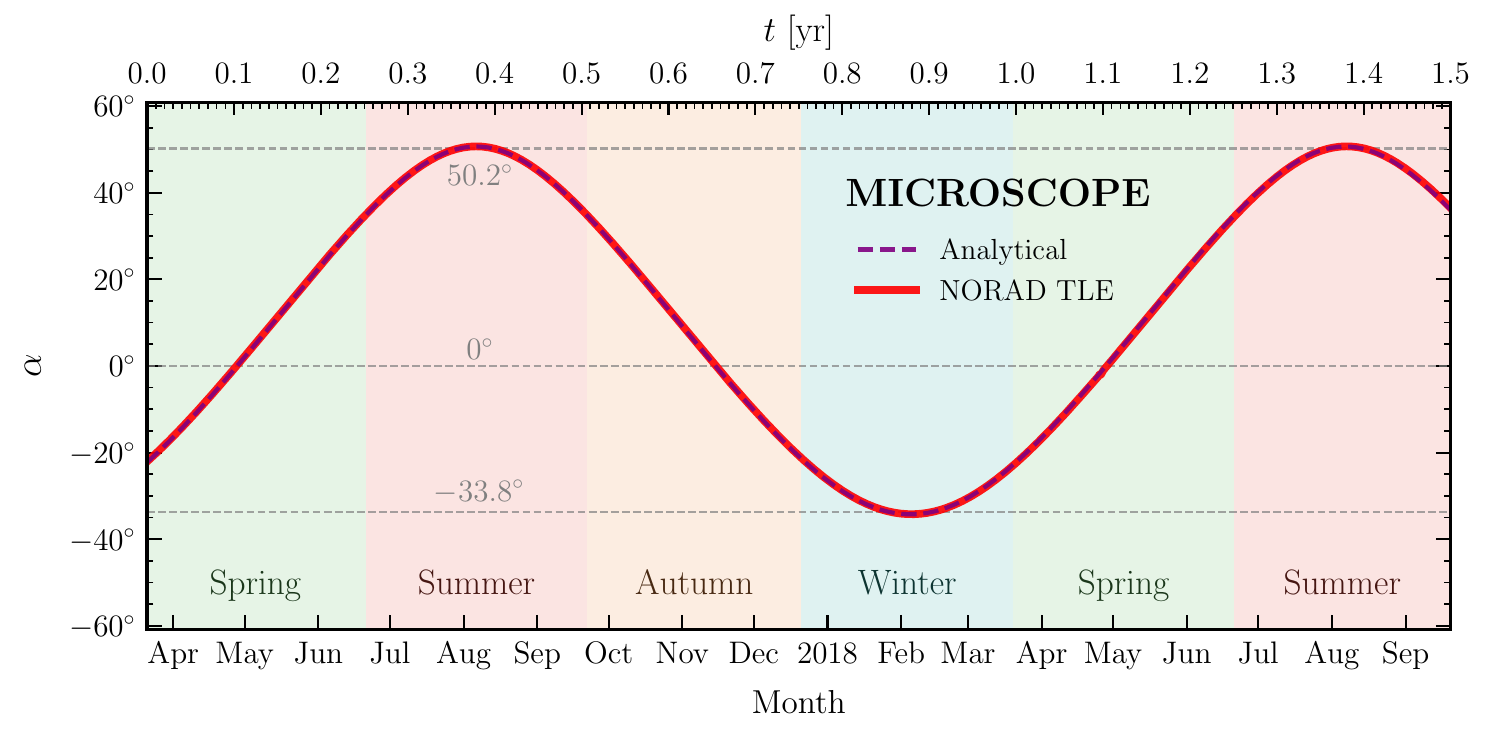}
\caption{Annual modulation of the angle $\alpha(t)$ between the MICROSCOPE orbital plane and the mean DM momentum $\veck_0$ as the Earth orbits the Sun. The {\bf thick red} line is computed using the real-time TLE data from NORAD~\cite{spacetrack}. The {\bf purple dashed} line shows the analytic result, which agrees well with the TLE-based computation. Here $t=0$ is chosen at the spring equinox in 2017.}
\label{fig:alpha_t}
\end{figure}

The annual modulation of the MICROSCOPE orbit is characterized by the angle between the orbital plane and $\hat\veck_0$~\footnote{Here we approximate $\veck_0 \simeq m_\phi \mathbf{v}_\odot$,
neglecting the Earth's orbital velocity, which satisfies
$|\mathbf{v}_\oplus| \ll |\mathbf{v}_\odot|$. Here, $\mathbf{v}_\oplus$ denotes the Earth's orbital velocity around the Sun. We have verified numerically that including the Earth's orbital velocity leads to negligible impact on the resulting
MICROSCOPE constraints.},
\bea
\label{eq:alpha_Lorb_k0}
\alpha(t) = \arcsin\,(\hat{\vecL}_\orb \cdot \hat{\veck}_0)\, .
\eea
This angle exhibits an annual modulation due to the Earth's revolution around the Sun. 
In \Fig{alpha_t}, we show the annual modulation of $\alpha(t)$ during the MICROSCOPE mission. 
To our knowledge, this modulation has not been explicitly discussed in previous MICROSCOPE EP-test analyses, which mainly focused on central-force signals. 
The thick red solid line is computed using satellite two-line element~(TLE) data from the North American Aerospace Defense Command~(NORAD). 
The purple dashed line shows the analytic result obtained by substituting \Eq{Omega_approx_fit}, \Eq{vL_ECL}, and \Eq{k0_cyg_main} into \Eq{alpha_Lorb_k0}, and agrees well with the TLE-based result. 
Moreover, $\alpha(t)$ varies between $-33.8^\circ$ and $50.2^\circ$, where the sign specifies on which side of the orbital plane $\hat{\veck}_0$ lies. 
Thus, the MICROSCOPE orbital plane is never perpendicular to $\hat{\veck}_0$. 
This nonzero projection is important for the frequency band splitting in the $\vecX$-axis signal in the regime $k_0R_\oplus \gtrsim 1$. 
Since MICROSCOPE data are collected in several-day-long sessions at different times during the mission, the annual variation of $\alpha(t)$ shown in \Fig{alpha_t} allows us to include the corresponding angular information in the data analysis, as discussed later.

%%%%
%%%%

We now describe the satellite orbital motion and the spin of the instrument frame. For convenience of illustration, we perform the computation in the orbital frame. We start with the position vector $\vecr$, the DM mean momentum $\veck_0$, and instrument axes $\vecX$, $\vecY$, $\vecZ$, as given by
\bea
\label{eq:r_k0_XYZ}
\hat{\vecr} = 
\begin{pmatrix}
\cos\Theta_\orb \\
\sin\Theta_\orb \\
0
\end{pmatrix}_\orb, 
\,\,\,
\hat{\veck}_0 = 
\begin{pmatrix}
0 \\
-\cos\alpha \\
\sin\alpha
\end{pmatrix}_\orb,
\,\,
\vecX = 
\begin{pmatrix}
\cos(-\Theta_\spin) \\
\sin(-\Theta_\spin) \\
0
\end{pmatrix}_\orb, 
\, 
\vecY = 
\begin{pmatrix}
 0 \\
 0 \\
1
\end{pmatrix}_\orb, 
\, 
\vecZ = 
\begin{pmatrix}
\cos (-\frac{\pi}{2} - \Theta_\spin) \\
\sin (-\frac{\pi}{2} - \Theta_\spin) \\
0
\end{pmatrix}_\orb,
\eea
where the subscript ``$\orb$'' denotes components in the orbital frame, shown in purple in \Fig{micro_orbit}. For the orbital frame, the $z$-axis is given by $\hat{\vecz}_\orb = \hat{\vecL}_\orb$ and the $x$-axis of the orbital frame is given by $\hat{\vecx}_\orb = \vecL_\orb \times \veck_0/|\vecL_\orb \times \veck_0|$. Here, $\Theta_\orb$ and $\Theta_\spin$ denote the orbital and spin phases, respectively, given by
\bea
\label{eq:Theta_orb_spin_def}
\Theta_\orb = \omega_\orb \, t + \Theta_{\orb,\,0}\,, \quad \quad \, \Theta_\spin = \omega_\spin \, t + \Theta_{\spin,\,0}\,,
\eea
where $\omega_\orb$ and $\omega_\spin$ are the orbital and spin angular frequencies, and $\Theta_{\orb,0}$ and $\Theta_{\spin,0}$ are the corresponding initial phases at $t=0$. \Eq{r_k0_XYZ} shows that $\vecZ$ is obtained from $\vecX$ by the substitution $\Theta_\spin \rightarrow \Theta_\spin + \pi/2$. Note that the MICROSCOPE satellite has multiple modes~\cite{Robert:2020ddm,Rodrigues:2022hmp}: inertial mode, V2 mode, and V3 mode, where the first is used for calibration and the latter two for the EP tests. Specifically, we have
\bea
\label{eq:spin_mode}
\underbrace{\omega_{\spin, {\rm inertial}} = 0}_{\text{Calibration}}\,, \quad \quad \quad \underbrace{\omega_{\spin, 2} = \frac{9}{2} \omega_\orb, \quad \,\, \omega_{\spin, 3} = \frac{35}{2} \omega_\orb}_{\text{EP Tests}}\,. 
\eea
For a conventional fifth-force search with a spherically symmetric central force, the signal is searched for at $\omega_\EP = \omega_\spin + \omega_\orb$, 
where the plus sign arises because the instrument-frame rotation is opposite to the orbital rotation. In the nonspherical background-induced force search as we will discuss later, the EP-violating signal contains multiple frequency bands around $\omega_\EP$. For consistency with previous literature, we also define $\omega_\EP=\omega_\spin + \omega_\orb$.

The angle between $\hat{\vecr}$ and $\veck_0$ is given by
\begin{equation}
\label{eq:cos_theta_r_in_orb}
\cos\theta_\vecr = \hat{\veck}_0 \cdot \hat{\vecr} = - \cos\alpha \sin \Theta_\orb\,,
\end{equation}
which describes the relative angle between the satellite position vector and the mean momentum $\veck_0$. Since $\alpha(t)$ varies only on the annual timescale, the time dependence of $\theta_\vecr$ within a given MICROSCOPE session is dominated by the orbital period $T_\orb$.

\subsection{Force Projection and Frequency Band Structure}\label{subsec:force_proj_band_structure}

In this section, we project the background-induced force in \Eq{multipole_force_dimless} onto the instrument-frame axes $\vecX$, $\vecY$, and $\vecZ$, to find $\vecX\cdot\widetilde{\vecF}_\bg$, $\vecY\cdot\widetilde{\vecF}_\bg$, and $\vecZ\cdot\widetilde{\vecF}_\bg$, which determine the signal measured by the instrument. 
Using the kinematics discussed in \Sec{micro_kinematics}, we first compute the geometric factors required for the projection, namely the inner products of the instrument-frame axes with $\hat{\vecr}$ and $\hat{\vectheta}_\vecr$. 
We then compute the projected force and express it in both the time and frequency domains. Next, we extract the $\omega_\EP$ component along the sensitive axis $\vecX$, which determines the main-band E\"otv\"os parameter used in the MICROSCOPE data analysis. We further find that the main-band amplitude carries an annual modulation through its dependence on $\alpha$, an effect we incorporate into the MICROSCOPE data analysis. Finally, we discuss the main band and the orbital sidebands generated by the orbital motion. As we discuss below, these bands provide a distinctive smoking-gun signature in frequency space and can improve the detection sensitivity in a full spectral analysis.

We begin by computing the geometric factors, namely the inner products of the instrument-frame axes with $\hat{\vecr}$ and $\hat{\vectheta}_\vecr$. This is most conveniently done in the $\veck_0$-frame, where the scalar profile is axially symmetric. Using the definition of $\hat{\vectheta}_\vecr$ and the equation $|\hat{\vecr}\times(\hat{\vecr}\times\hat{\veck}_0)|=\sin\theta_\vecr$, we have
\bea
\sin\theta_\vecr \, \hat{\boldsymbol{\theta}}_\vecr = \hat{\vecr} \times (\hat{\vecr} \times \hat{\veck}_0) = \cos\theta_\vecr\,\hat{\vecr}-\hat{\veck}_0\,.
\eea
Using the expressions for $\hat{\vecr}$ and $\hat{\veck}_0$ in \Eq{r_k0_XYZ}, we obtain $\hat{\vectheta}_\vecr$ in the orbital frame. The geometric factors are then
\bea
\label{eq:XYZ_base_proj}
\begin{aligned}
\hat{\vecr} \cdot \vecX & = \cos\Theta_\EP\\
\sin\theta_\vecr \, \thetarhat \cdot \vecX & = - \cos\alpha \, \cos\Theta_\orb \sin \Theta_\EP
\end{aligned}
, 
\quad \,\,
\begin{aligned}
\hat{\vecr} \cdot \vecY & = 0 \\
\sin\theta_\vecr \, \thetarhat \cdot \vecY & = - \sin\alpha
\end{aligned}
,
\quad \,\,
\begin{aligned}
\hat{\vecr} \cdot \vecZ & = - \sin \Theta_\EP\\
\sin\theta_\vecr \, \thetarhat \cdot \vecZ & = - \cos\alpha \cos\Theta_\orb \cos \Theta_\EP 
\end{aligned}.
\eea
Here, $\Theta_\EP=\Theta_\spin + \Theta_\orb$.
As shown in \Eq{XYZ_base_proj}, the geometric factors for the $\vecZ$ axis can be obtained from those for the $\vecX$ axis by the substitution $\Theta_\EP\rightarrow\Theta_\EP+\pi/2$. This equivalence through substitution can be seen directly from \Eq{r_k0_XYZ}, as  $\vecZ$ can be acquired from $\vecX$ through $\Theta_\spin \rightarrow \Theta_\spin + \pi/2$, while $\Theta_\orb$ remains the same.

Substituting \Eq{XYZ_base_proj} into \Eq{multipole_force_dimless}, we obtain the projections of the background-induced force onto the three instrument axes:
\bea\label{eq:FXYZ_projection}
\vecX \cdot \widetilde{\vecF}_\bg &= \underbrace{ (- \cos\Theta_\EP) \sum_{L=0}^\infty \frac{\dd a_L}{\dd r} P_L(\cos \theta_\vecr)}_{(\vecX \cdot \widetilde{\vecF}_\bg)_{\vecr}}  + \underbrace{(- \cos\alpha \, \cos\Theta_\orb \sin\Theta_\EP) \sum_{L=0}^\infty \frac{a_L(r)}{r} P_L'(\cos\theta_\vecr)}_{ (\vecX \cdot \widetilde{\vecF}_\bg)_{\thetarhat} }\, ,\\
\vecY \cdot \widetilde{\vecF}_\bg &= \underbrace{(-\sin \alpha) \sum_{L=0}^\infty \frac{a_L(r)}{r} P'_L(\cos\theta_\vecr)}_{ (\vecY \cdot \widetilde{\vecF}_\bg)_{\thetarhat} }\, ,\\
\vecZ \cdot \widetilde{\vecF}_\bg &= \underbrace{\sin\Theta_\EP  \sum_{L=0}^\infty \frac{\dd a_L}{\dd r} P_L(\cos\theta_\vecr)}_{(\vecZ \cdot \widetilde{\vecF}_\bg)_{\vecr}} + \underbrace{ (-\cos\alpha \cos\Theta_\orb \cos\Theta_\EP) \sum_{L=0}^\infty \frac{a_L(r)}{r} P'_L(\cos\theta_\vecr)}_{(\vecZ \cdot \widetilde{\vecF}_\bg)_{\thetarhat}}\, .
\eea
Since the $\vecX$ and $\vecZ$ lie in and rotate within the orbital plane, their projections of $\widetilde{\vecF}_\bg$ receive both radial and angular contributions, while the $\vecY$ projection, which is perpendicular to the plane, receives only the angular contribution.

\begin{figure}[t!]
\centering
\includegraphics[width=1\linewidth]{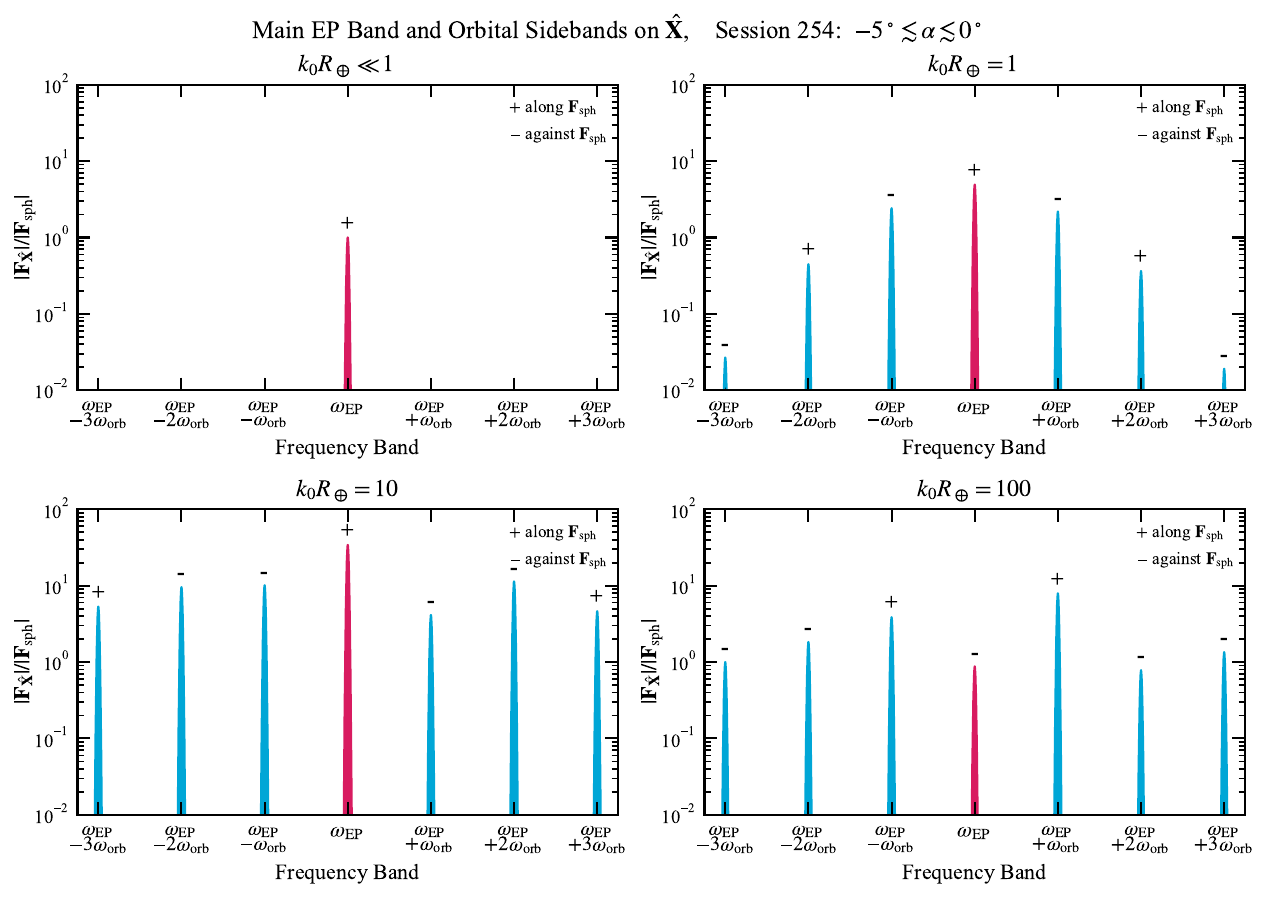}
\caption{Illustration of the main band and orbital sidebands of the $\vecX$-projected background-induced force $\vecF_{\vecX}$ for MICROSCOPE Session 254. The vertical axis shows the relative force amplitude, $|\vecF_{\vecX}|/|\vecF_{\sph}|$, where $|\vecF_{\sph}|$ is the magnitude of the background-induced force in the spherically symmetric ansatz, given in \Eq{F_bg_sph}. The {\bf magenta} peak denotes the main band at $\omega_\EP$ and the {\bf cyan} peaks denote the sidebands at $\omega_\EP \pm \omega_\orb$, $\omega_\EP \pm 2\,\omega_\orb$, and $\omega_\EP \pm 3\,\omega_\orb$. We assign a $+/-$ sign to each peak when the corresponding $\vecF_{\vecX}$ component points parallel/anti-parallel to the direction of $\vecF_{\sph}$, which points toward the Earth as shown in \Fig{nonpeturb_sph_plt}. 
When $k_0 R_\oplus \ll 1$, the background-induced force is dominated by the monopole contribution. In this limit, the signal contains a single peak at $\omega_\EP$, and the force has the same direction and magnitude as $\vecF_{\sph}$. By contrast, when $k_0 R_\oplus \gtrsim 1$, higher multipoles become important and generate the sideband structure. Including these sidebands can improve the sensitivity, especially when $k_0 R_\oplus \gtrsim 100$, corresponding to $m_\phi \gtrsim 4\times10^{-9}\,\eV$.}
\label{fig:X_band}
\end{figure}

To make the frequency band structure in \Eq{FXYZ_projection} explicit, we expand
\begin{equation}\label{eq:PL_cos_theta_r_expansion}
P_L(\cos \theta_\vecr) = \frac{1}{2^L}\sum_{k=0}^{\left \lfloor{L/2}\right \rfloor}(-1)^k \binom{L}{k}\binom{2L - 2k}{L}\, (\cos \theta_\vecr)^{L - 2k}\,,
\end{equation}
where $\left \lfloor{\cdots}\right \rfloor$ represents the floor function, and use \Eq{cos_theta_r_in_orb}. 
This expansion shows that the $\vecX$ and $\vecZ$ projections contain terms like $\cos\Theta_\EP(\sin\Theta_\orb)^\mathscr{N}$ or $\sin\Theta_\EP(\sin\Theta_\orb)^\mathscr{N}$, while the $\vecY$ projection contains terms like $(\sin\Theta_\orb)^\mathscr{N}$ where 
$\mathscr{N}$ is a non-negative integer. 
This structure makes explicit that the $\vecX$ and $\vecZ$ projections contain a main band at $\omega_\EP$ and orbital sidebands at $\omega_\EP\pm n\,\omega_\orb$, while the $\vecY$ projection has a frequency-independent (DC) main band and sidebands $n\,\omega_\orb$ ($n\geq1$). 
Note that for the background-induced force described in the spherically symmetric case~\cite{Hees:2018fpg,Berezhiani:2018oxf,Banerjee:2022sqg}, or the Yukawa-type central force~\cite{Moody:1984ba,Feinberg:1989ps,Ferrer:2000hm,Damour:2002mi,Damour:2010rp,Damour:2010rm,Berge:2017ovy,Fichet:2017bng,Bauer:2023czj,Grossman:2025jub}, the force projection in $\vecX$ and $\vecZ$ only has the main $n=0$ band and no $\vecY$-axis signal, and so previous MICROSCOPE analyses focused on the main $\omega_\EP$ band~\cite{Touboul:2017grn,MICROSCOPE:2019jix,MICROSCOPE:2022doy,Rodrigues:2022hmp,Touboul:2022yrw}.

We now discuss the $\vecX$ projection in detail, since MICROSCOPE is most sensitive along this axis. Combining \Eq{cos_theta_r_in_orb}, \Eq{PL_cos_theta_r_expansion} and \Eq{FXYZ_projection}, we decompose the $\vecX$ projection as
\bea
\label{eq:X_Fbg_band}
\vecX \cdot \widetilde{\vecF}_\bg
& = R_0^{X} \cos\Theta_{\rm EP}\\
& \,\, +\sum_{n=1}^{\infty}\bigg[ (R_n^{X}+T_n^{X})\cos\left(\Theta_{\rm EP}+n\,\Theta_{\rm orb}+\frac{n\pi}{2}\right)
+
(R_n^{X}-T_n^{X})\cos\left(\Theta_{\rm EP}-n\,\Theta_{\rm orb}-\frac{n\pi}{2}\right)
\bigg]\,.
\eea
Here $R_n^X$ denotes the contribution from the radial component of the force, while $T_n^X$ denotes the contribution from the angular component. The general expressions for $R_n^X$ and $T_n^X$, valid to arbitrary multipole order, are given in \Eq{RnX_TnX_appx} of \Appx{frequency_band_full_appx}. In the following, we list only the leading-order contributions to the main band and the first few sidebands for illustration.

The MICROSCOPE measurements focus on the acceleration along the $\vecX$ direction with the frequency of $\omega_{\rm EP}$, which corresponds to the main-band coefficient $R_0^X$:
\bea
\label{eq:R0X_force}
\text{Main band in $\vecX$~($\omega_\EP$):}\quad \quad 
R_0^{X}
=&-\frac{\dd a_0}{\dd r}
+\left(\frac{1}{2}-\frac{3}{4}\cos^2\alpha\right)\frac{\dd a_2}{\dd r}
+ ... \,.
\eea

When $k_0 R_\oplus \ll 1$, $\dd a_2/\dd r \ll \dd a_0/\dd r$, and so $R_0^X \simeq - \dd a_0/\dd \, r \simeq - \dd\big(\psi_\sph^2/\abs{\psi_0}^2\big)/\dd r$, which reduces to the analytical result in  Refs.~\cite{Hees:2018fpg,Berezhiani:2018oxf,Banerjee:2022sqg}. 
However, as shown in \Fig{hard_sphere_a_series_derivatives}, $\dd a_0/\dd r$ deviates from $\dd\big(\psi_\sph^2/\abs{\psi_0}^2\big)/\dd r$ when $k_0 R_\oplus \gtrsim 1$.  
Furthermore, when $k_0 R_\oplus \gtrsim 10$, $\dd a_0/\dd r$ vanishes, while $\dd a_2/\dd r$ dominates. 
This indicates that in this regime, the MICROSCOPE constraint from the $\omega_\EP$-band still holds, but is dominated by the quadrupole contribution, $\dd a_2/\dd r$, rather than the monopole contribution, $\dd a_0/\dd r$. 
Moreover, the coefficient of $\dd a_2/\dd r$ in \Eq{R0X_force} depends on $\alpha$, and is therefore annually modulated. 
Therefore, when treating the MICROSCOPE data in different segments given by Table~\ref{tab:microscope_segments},  we can incorporate the annual orbit-angle modulation shown in \Fig{alpha_t}.

We now turn to the remaining Fourier components at $\omega_\EP\pm n\,\omega_\orb$ with $n=1,2,3,\cdots$. Organizing them by the sideband number $n$, we obtain the leading terms
\bea
\label{eq:X_1st_sideband_coefficient}
\text{First sideband in $\vecX$~($\pm\,\omega_\orb$):} \quad \quad \left\{
\begin{aligned}
R_1^X
=&\cos\alpha\left[
-\frac{1}{2}\frac{\dd a_1}{\dd r}
+\left(\frac{3}{4}-\frac{15}{16}\cos^2\alpha\right)\frac{\dd a_3}{\dd r}
+\ldots
\right]\,,
\\
T_1^X
=&\cos\alpha\left[
\frac{1}{2}\frac{a_1}{r}
+\left(-\frac{3}{4}+\frac{15}{16}\cos^2\alpha\right)\frac{a_3}{r}
+\ldots
\right]\,.
\end{aligned}
\right.
\eea
\bea
\text{Second sideband in $\vecX$~($\pm \,2\,\omega_\orb$)}: \quad \quad \left\{\begin{aligned}
R_2^X
=&\cos^2\alpha\left[
-\frac{3}{8}\frac{\dd a_2}{\dd r}
+\left(\frac{15}{16}-\frac{35}{32}\cos^2\alpha\right)\frac{\dd a_4}{\dd r}
+ \ldots
\right]\,,
\\
T_2^X
=&\cos^2\alpha\left[
\frac{3}{4}\frac{a_2}{r}
+\left(-\frac{15}{8}+\frac{35}{16}\cos^2\alpha\right)\frac{a_4}{r}
+ \ldots
\right]\,.
\end{aligned}\right.
\eea
\bea
\text{Third sideband in $\vecX$ ($\pm \,3\,\omega_\orb$):}\quad \quad \left\{
\begin{aligned}
R_3^X
=&\cos^3\alpha\left[
-\frac{5}{16}\frac{\dd a_3}{\dd r}
+\left(\frac{35}{32}-\frac{315}{256}\cos^2\alpha\right)\frac{\dd a_5}{\dd r}
+ \ldots
\right]\,,
\\
T_3^X
=&\cos^3\alpha\left[
\frac{15}{16}\frac{a_3}{r}
+\left(-\frac{105}{32}+\frac{945}{256}\cos^2\alpha\right)\frac{a_5}{r}
+ \ldots
\right]\,.
\end{aligned}
\right. 
\eea
\begin{figure}[t!]
\centering
\includegraphics[width=1\linewidth]{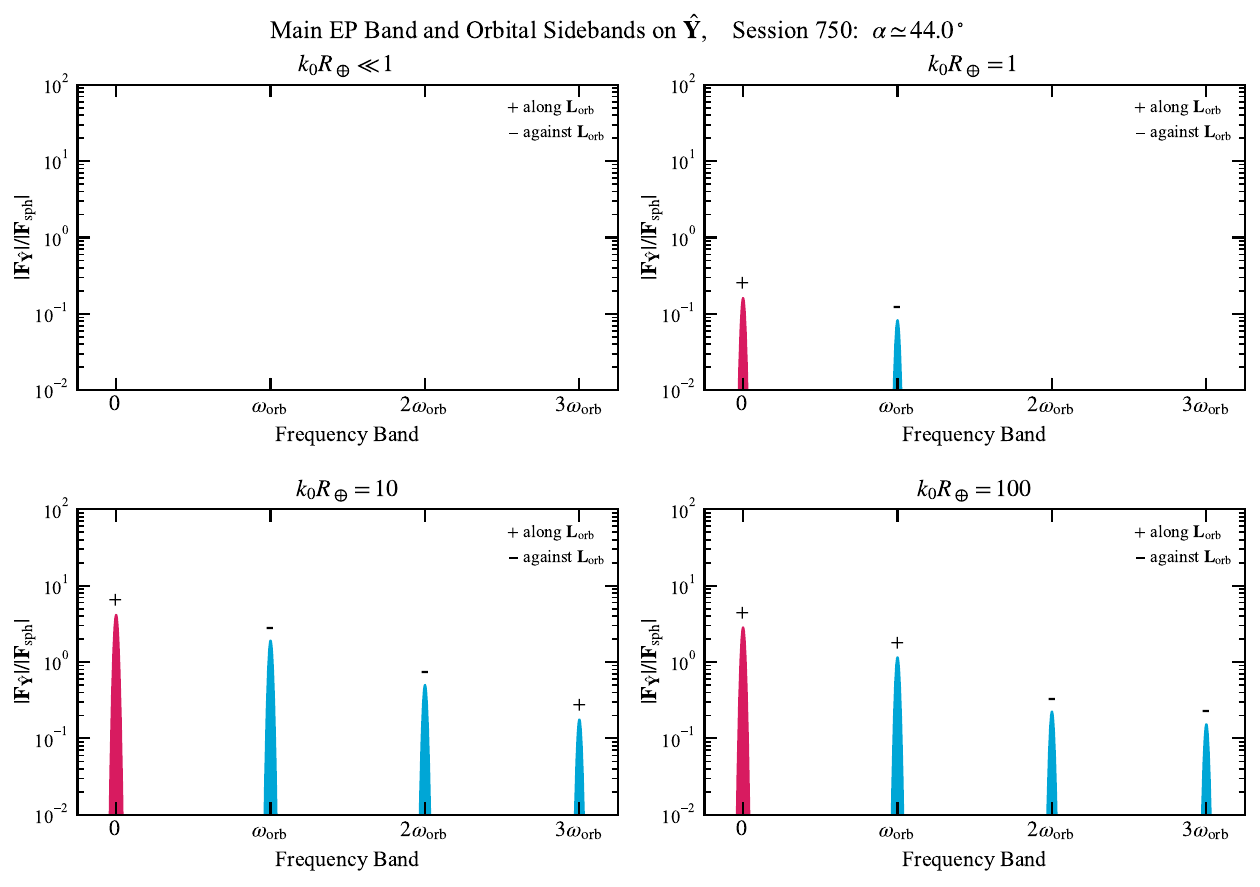}
\caption{Illustration of the main DC component and orbital sidebands of the $\vecY$-projected background-induced force $\vecF_{\vecY}$ induced by the MICROSCOPE orbital motion for Session 750. The {\bf magenta} peak denotes the main band, which gives the DC signal, where $\omega=0$. The vertical axis shows the normalized force amplitude, $|\vecF_{\vecY}|/|\vecF_{\sph}|$, where $|\vecF_{\sph}|$ is the magnitude of the background-induced force in the spherically symmetric ansatz, given in \Eq{F_bg_sph}. The {\bf cyan} peaks denote the sidebands at $ \omega_\orb$, $ 2\,\omega_\orb$, and $ 3\,\omega_\orb$, etc.
We use +/- for each peak if the corresponding $\vecF_{\vecY}$ is parallel/anti-parallel to the direction of $\hat{\vecL}_{\orb}$.
Note that there is no signal when $k_0 R_\oplus\ll 1$ since there is no force on the tangential direction in spherically symmetric case.}
\label{fig:Y_band}
\end{figure}
%
% The full frequency-band decomposition of the background-induced force along the $\vecX$ axis is given in \Appx{band_split_calc_appx}.
In \Fig{X_band}, we show the relative amplitudes of the main band and orbital sidebands. 
We find that in the low-momentum regime ($k_0 R_\oplus \ll 1$), the main band dominates, because the DM field profile is nearly spherically symmetric around the Earth and $a_0$ dominates the series expansion. 
When increasing $m_\phi$ toward the high momentum region, $k_0 R_\oplus \gg 1$, the incident DM waves start to resolve the finite size of the Earth, and the contribution from $a_L$ and $\dd a_L/\dd r$ for $L \geq 1$ become substantial, and could dominate over the spherically symmetrical contribution from $a_0$. 
Physically speaking, for the high momentum region, the DM field would develop a non-spherical field profile $\langle \abs{\psi}^2 \rangle$ around the Earth, with sizable dipole, quadrupole, and higher-multipole components. 
As a result, the signals probed by the satellite experience an extra ``orbital modulation'' beyond $\omega_{\rm EP}$, and it is the reason why the sideband of $\omega_{\rm EP} \pm \omega_{\rm orb}$ becomes comparable or even dominates over the main band when $k_0 R_\oplus \gg 1$. As shown in \Fig{X_band}, when $k_0R_\oplus\gtrsim100$, corresponding to $m_\phi\gtrsim4\times10^{-9}\,\eV$, the first sidebands at $\omega_\EP\pm\omega_\orb$ are about one order of magnitude larger than the main band. Assuming that the noise power spectrum at the sideband frequencies is not qualitatively different from that at the main band, including the sidebands can considerably improve the sensitivity. 
We explore this possibility further in \Sec{future}.
Since \Eq{X_Fbg_band} gives the time-domain evolution in terms of $R_n^X$ and $T_n^X$, the same formalism can also be applied to a full time-domain analysis using MICROSCOPE time-series data.

The same mechanism populates the $\vecY$ projection: the higher multipoles ($L\geq1$) that grow at large $k_0 R_\oplus$ generate components at the orbital harmonics $n\,\omega_\orb$. 
The key difference is that for the $\vecY$ projection, the leading component has zero frequency and therefore corresponds to a DC signal. 
For completeness, we give the $\vecY$ projection:
\bea
\label{eq:Y_Fbg_band}
\vecY \cdot \widetilde{{\bf F}}_{\rm bg} = T_0^Y + \sum_{n=1}^\infty T_n^Y \, \cos(n \,\Theta_\orb + n \frac{\pi}{2}).
\eea
Here $T_0^Y$ denotes the DC component, while $T_n^Y$ with $n \geq 1$ denote the orbital-harmonic coefficients. The complete expressions for the coefficients $T_n^Y$, valid at arbitrary multipole order, are given in \Eq{TnY_appx} of \Appx{frequency_band_full_appx}. For illustration, we present below only the leading contributions to the DC component and the first few orbital harmonics:
\bea
\text{Main band in $\vecY$~(DC Component):}\quad \quad T_0^Y = - \sin \alpha \left[ \frac{a_1}{r} + \left(-\frac{3}{2} + \frac{15}{4} \cos^2\alpha \right)\frac{a_3}{r} + \cdots \right],
\eea
\bea
\text{First sideband in $\vecY$~($\omega_\orb$):}\quad \quad T_1^Y  = - \sin \alpha \cos\alpha \left[ 3 \, \frac{a_2}{r} + \left(-\frac{15}{2} + \frac{105}{8} \cos^2\alpha \right) \frac{a_4}{r} + \cdots\right],
\eea
\bea
\text{Second sideband in $\vecY$~($2\,\omega_\orb$):}\quad \quad T_2^Y = - \sin \alpha \cos^2 \alpha \left[ \frac{15}{4}\frac{a_3}{r} + \left(-\frac{105}{8} + \frac{315}{16} \cos^2\alpha \right) \frac{a_5}{r} + \cdots\right],
\eea
\bea
\text{Third sideband in $\vecY$~($3\,\omega_\orb$):}\quad \quad T_3^Y = - \sin \alpha \cos^3 \alpha \left[ \frac{35}{8}\frac{a_4}{r} + \left(-\frac{315}{16} + \frac{3465}{128} \cos^2\alpha \right) \frac{a_6}{r} + \cdots\right].
\eea
$T_{1}^Y, T_{2}^Y, T_{3}^Y,\cdots$ are the one-sided sideband components, as shown in \Fig{Y_band}.
% \TY{The full harmonic decomposition of the background-induced force along the $\vecY$ axis is given in \Appx{band_split_calc_appx}. }
Note that the force signal in the $\vecY$-axis has more apparent annual modulation, given that there is a universal factor $\sin\alpha$ at the front of all coefficients in \Eq{Y_Fbg_band}.

We emphasize that the frequency-band structure shown in \Fig{X_band} and \Fig{Y_band} is a distinctive smoking-gun signature. For a force described by a spherically symmetric ansatz~\cite{Hees:2018fpg,Berezhiani:2018oxf,Banerjee:2022sqg}, or for a Yukawa-type central force~\cite{Moody:1984ba,Feinberg:1989ps,Ferrer:2000hm,Damour:2002mi,Damour:2010rp,Damour:2010rm,Berge:2017ovy,Fichet:2017bng,Bauer:2023czj,Grossman:2025jub}, the force has only a radial component. It therefore produces only the $\omega_\EP$ main-band signal along the $\vecX$ axis, while the $\vecY$ projection vanishes. Although the relative amplitudes depend on the unknown parameter $m_\phi$ through $k_0 R_\oplus$, the frequency spacing is fixed by the orbital frequency $\omega_\orb$. Moreover, such an orbital-sideband structure is resolvable by
MICROSCOPE~\cite{Touboul:2017grn,MICROSCOPE:2019jix,MICROSCOPE:2022doy,Rodrigues:2022hmp,Touboul:2022yrw}, or the future Galileo Galilei experiment~\cite{Nobili:2000bzv,Nobili:2012uj,Nobili:2017cxu,Nobili:2018eym}. 
Specifically, for these satellites, $\omega_\orb\sim1\,{\rm mHz}$ is much larger than the Fourier resolution $\Delta\omega=2\pi/T_{\rm session}\sim0.01\,{\rm mHz}$ for a several-day session. 
The annual modulation of $\alpha$ can in principle generate additional yearly sidebands around each component, separated by $\omega_\odot$. 
However, this annual splitting, $\omega_\odot\sim1\,{\rm yr}^{-1}\sim10^{-5}\,{\rm mHz}$, is much smaller than both the session-level frequency resolution and the orbital-sideband spacing. We therefore neglect this effect in \Fig{X_band} and \Fig{Y_band}.

%%%%%%%%%%%%%%%%%%%%%%%%%
\subsection{Revised MICROSCOPE Constraints}\label{subsec:microscope_revisit}

In this section, we rederive the MICROSCOPE constraints on quadratically coupled ULDM using the force formalism developed above. 
The constraints from MICROSCOPE found in 2022 a limit on the E\"otvos parameter~\cite{MICROSCOPE:2022doy},
\bea
\label{eq:eotvos_def}
\eta(\ti,\pt) = 2 \frac{a_{\ti,\oplus}-a_{\pt,\oplus}}{a_{\ti,\oplus}+a_{\pt,\oplus}}=[-1.5\pm 2.3\,({\rm stat})\pm 1.5\,({\rm syst})]\times 10^{-15}\,.
\eea
This MICROSCOPE constraint currently provides one of the leading sensitivities in space-based EP tests, and has been used to constrain a wide range of models, including long-range forces~\cite{Fayet:2017pdp,Fayet:2018cjy, Amaral:2024tjg}, axions~\cite{Grossman:2025cov,Gan:2025nlu,Gue:2024onx,Gan:2025icr}, DM-nuclear scattering~\cite{Day:2023mkb,Liu:2025jlx,Fu:2026atp}, and scalar ULDM~\cite{Hees:2018fpg,Banerjee:2022sqg,Gan:2025icr, Gue:2025nxq,Delaunay:2025pho,Gan:2025nlu,Bouley:2025qtq}.
In our analysis, we go beyond the E\"otv\"os parameter and use the segment-level MICROSCOPE data~\cite{Touboul:2022yrw}, presented in Table~\ref{tab:microscope_segments}, 
which precisely takes into account of the satellite position and orientation during each segment.

Before presenting our analysis, we first review previous studies of MICROSCOPE constraints on quadratically coupled ULDM. 
The MICROSCOPE constraint on repulsive quadratically coupled ULDM was first explored in Ref.~\cite{Hees:2018fpg}, using the spherically symmetric ansatz and recasting the first MICROSCOPE constraint on the E\"otv\"os parameter~\cite{Berge:2017ovy}. This constraint was later updated in Ref.~\cite{Banerjee:2022sqg} using the same ansatz and the final MICROSCOPE constraint on the E\"otv\"os parameter~\cite{MICROSCOPE:2022doy}. 
The attractive case was considered in Refs.~\cite{Gue:2025nxq,Delaunay:2025pho}, where only the $s$-wave component was included when deriving the constraint. 
Ref.~\cite{Gan:2025nlu} went beyond the $s$-wave approximation by including the full partial-wave expansion and phase-space integration. However, its MICROSCOPE recast was intended only as a simplified illustration: it showed that the constraint does not vanish in the high-$k_0$ regime, due to  the interplay between descreening and incoherence, which is clarified in this work as the nonzero higher-multipole contributions. However, it avoided the full numerical treatment by considering only the forward and backward directions. 
Ref.~\cite{Bouley:2025qtq} also considered the full partial-wave expansion and phase-space integration.
However, it performed an angular average, which is equivalent to retaining only the $\dd a_0/\dd r$ component, and therefore obtained a conservative constraint. 
Including higher multipoles with $L\geq 1$ can further strengthen the constraint, since these contributions do not vanish in the high-momentum regime. 
Moreover, none of the above works included the MICROSCOPE satellite orbital motion, the instrument-frame rotation, or the annual modulation of the angle between the orbital plane and $\veck_0$, all of which induce distinctive smoking-gun signal features. 
With these prior developments in mind, we revisit the MICROSCOPE constraint using the toolbox developed in this work, together with previous efforts, especially Refs.~\cite{Gan:2025nlu,Bouley:2025qtq,Brzeminski:2026rox}.

\begin{table*}[t!]
\centering
\setlength{\tabcolsep}{6pt}
\renewcommand{\arraystretch}{1.5}
\begin{tabular}{|c|c|c|c|cc|c|ccc|}
\hline
\makecell{Segment}\rule[-3.3ex]{0pt}{7.9ex}
% \makecell{Segment}
& \makecell{Spin}
& \makecell{Date\\[2.7pt](Start)}
& \makecell{Duration\\[2.7pt](orbits)}
& \makecell{$\alpha_{\rm start}$}
& \makecell{$\alpha_{\rm end}$}
& \makecell{$\eta_{\obs}$\\[3.9pt]$(10^{-15})$}
& \makecell{$\sigma_{\rm stat}$\\[3.9pt]$(10^{-15})$}
& \makecell{$\sigma_{\rm syst}$\\[3.9pt]$(10^{-15})$}
& \makecell{$\sigma_\tot$\\[3.9pt]$(10^{-15})$} \\
\hline
210   & V3 & 2017-02-14 & 50  & $-34^\circ$ & $-33^\circ$ & $-29.2$ & $13.1$ & $1.8$ & $13.2$ \\
212   & V3 & 2017-02-18 & 60  & $-33^\circ$ & $-32^\circ$ & $9.5$   & $11.9$ & $1.0$ & $11.9$ \\
218   & V3 & 2017-02-28 & 120 & $-31^\circ$ & $-28^\circ$ & $6.7$   & $8.1$  & $1.1$ & $8.2$ \\
234   & V3 & 2017-03-15 & 92  & $-25^\circ$ & $-22^\circ$ & $5.9$   & $8.3$  & $1.0$ & $8.4$ \\
236   & V3 & 2017-03-21 & 120 & $-22^\circ$ & $-18^\circ$ & $2.6$   & $6.6$  & $1.2$ & $6.7$ \\
238   & V3 & 2017-03-29 & 120 & $-18^\circ$ & $-13^\circ$ & $5.8$   & $6.4$  & $1.2$ & $6.5$ \\
252   & V3 & 2017-04-13 & 106 & $-9^\circ$  & $-5^\circ$  & $-14.9$ & $7.3$  & $1.1$ & $7.4$ \\
254   & V3 & 2017-04-20 & 120 & $-5^\circ$  & $0^\circ$   & $-14.1$ & $7.0$  & $1.5$ & $7.2$ \\
256   & V3 & 2017-04-29 & 120 & $1^\circ$   & $6^\circ$   & $-5.3$  & $7.4$  & $1.1$ & $7.5$ \\
326--1 & V3 & 2017-09-27 & 66  & $32^\circ$  & $29^\circ$  & $-16.3$ & $9.6$  & $1.6$ & $9.7$ \\
326--2 & V3 & 2017-10-01 & 34  & $29^\circ$  & $28^\circ$  & $-10.4$ & $13.5$ & $1.6$ & $13.6$ \\
358   & V3 & 2017-10-14 & 92  & $21^\circ$  & $17^\circ$  & $15.8$  & $10.9$ & $1.1$ & $11.0$ \\
402   & V2 & 2017-12-06 & 18  & $-13^\circ$ & $-14^\circ$ & $28.4$  & $43.6$ & $7.3$ & $44.2$ \\
404   & V3 & 2017-12-07 & 120 & $-14^\circ$ & $-18^\circ$ & $4.7$   & $6.7$  & $1.0$ & $6.8$ \\
406   & V3 & 2017-12-16 & 20  & $-19^\circ$ & $-19^\circ$ & $5.9$   & $14.9$ & $3.2$ & $15.2$ \\
438   & V2 & 2018-01-16 & 32  & $-31^\circ$ & $-32^\circ$ & $-23.4$ & $24.6$ & $5.5$ & $25.2$ \\
442   & V2 & 2018-01-22 & 40  & $-33^\circ$ & $-33^\circ$ & $-1.5$  & $19.1$ & $7.3$ & $20.4$ \\
748   & V2 & 2018-09-03 & 24  & $45^\circ$  & $44^\circ$  & $-23.4$ & $24.6$ & $7.3$ & $25.6$ \\
750   & V3 & 2018-09-05 & 8   & $44^\circ$  & $44^\circ$  & $66.9$  & $38.4$ & $7.3$ & $39.1$ \\
\hline
\end{tabular}
\caption{MICROSCOPE SUEP data for the 19 science segments from Ref.~\cite{Touboul:2022yrw} used in this work. The labels V2 and V3 denote the satellite spin modes, with $\omega_{\spin,2}=(9/2)\,\omega_\orb$ and $\omega_{\spin,3}=(35/2)\,\omega_\orb$, respectively. The orbital period is $T_\orb=5946\,{\rm s}$~\cite{Robert:2020ddm,Rodrigues:2022hmp}. The angles $\alpha_{\rm start}$ and $\alpha_{\rm end}$ denote the angle between the MICROSCOPE orbital plane and the mean DM momentum direction $\veck_0$ at the beginning and end of each segment, as given by \Fig{alpha_t}. The per-session estimate of the E\"otv\"os parameter along the $\hat{{\bf X}}$ axis, $\eta_\obs$, and its uncertainties, $\sigma_{\rm stat}$, $\sigma_{\rm syst}$, $\sigma_\tot$, are quoted in units of $10^{-15}$.}
\label{tab:microscope_segments}
\end{table*}

\begin{figure}[t!]
\centering
\includegraphics[width=0.60\linewidth]{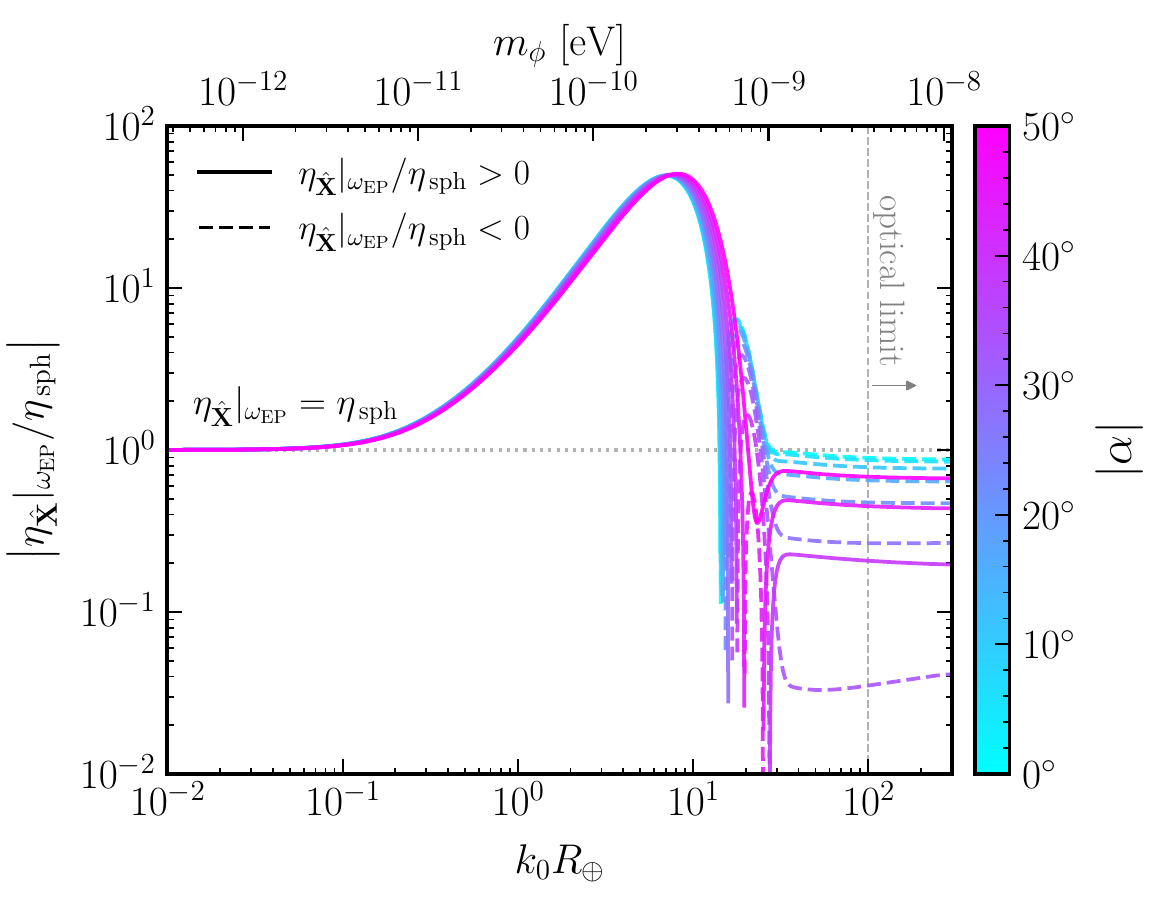}
\caption{Ratio between the E\"otv\"os parameter $\eta_{\vecX}|_{\omega_\EP}$ along the sensitive axis $\vecX$ in the main $\omega_\EP$ band, defined in \Eq{eotvos_X_EPband}, and the corresponding E\"otv\"os parameter obtained in the spherically symmetric ansatz of Refs.~\cite{Hees:2018fpg,Banerjee:2022sqg}. We show ten curves with $\abs{\alpha}$ uniformly sampled from $0^\circ$ to $50^\circ$, with colors varying from cyan to magenta.}
\label{fig:eotvos_ratio_EP}
\end{figure}

In this work, we use the MICROSCOPE data reported in Ref.~\cite{Touboul:2022yrw} which, along with the corresponding $\alpha(t)$ for each segment computed in this work, are summarized in Table~\ref{tab:microscope_segments}.
Importantly, this data is for the main frequency band $\omega_{\rm EP}$.
The analysis includes all 19 SUEP~(Sensor Unit for the Equivalence Principle test) data segments, which measure the E\"otv\"os parameter $\eta_{\vecX}|_{\omega_\EP}$. From \Eq{eotvos_def} we can compute the E\"otv\"os parameter induced by the background-induced force. Because the constraint in Ref.~\cite{Touboul:2022yrw} for each segment is for the main band $\omega = \omega_\EP$ projected on the $\vecX$-axis, 
we have for the numerator
\bea
\label{eq:acc_diff_EP_X}
a_{\ti, \oplus}-a_{\pt, \oplus} = \abs{\psi_0}^2 \frac{\mMti^2 \calV_\ti}{4 M_\ti} R_0^X - \abs{\psi_0}^2 \frac{\mMpt^2 \calV_\pt}{4 M_\pt}\,R_0^X \,.
\eea
Here \(R_0^X\) is the signed coefficient of the
\(\cos\Theta_\EP\) term in the decomposition of
$\vecX\cdot\widetilde{\vecF}_\bg$ in \Eq{X_Fbg_band}. \Eq{mM_def} gives $\mMtest^2 \, \calV_\testmass/4 M_\testmass = \pi \, \alpha_\testmass^{(2)}/\Mpl^2$, where $\testmass =\ti,\pt$, and $\abs{\psi_0}^2 = 2 \rho_\phi/m_\phi^2$.
The denominator of \Eq{eotvos_def}, $a_{\pt, \oplus}+a_{\ti, \oplus} \simeq 2 \, g$, where $g = (1/\Mpl^2)\,M_\oplus/(R_\oplus+h)^2 \simeq 7.9\,\text{m}/\text{s}^2$ is the gravitational acceleration at the MICROSCOPE orbit.
Therefore, we have
\bea
\label{eq:eotvos_X_EPband}
\eta_{\vecX}|_{\omega_\EP} = -\frac{2 \rho_\phi}{m_\phi^2} \, \frac{\pi \, (\alpha_\ti^{(2)}- \alpha_\pt^{(2)})\,(R_\oplus+h)^2}{ M_\oplus  \, R_\oplus } \left[ \frac{\dd a_0}{\dd (r/R_\oplus)}
- \left(\frac{1}{2}-\frac{3}{4}\cos^2\alpha\right)\frac{\dd a_2}{\dd (r/R_\oplus)} + \cdots \right].
\eea
In the above equation, $\alpha_\testmass^{(2)}$ is the effective coupling between the scalar and the testmass $\testmass = \ti, \pt$, as defined in \Eq{macroscopic_mass_variation}. Since $\alpha_\testmass^{(2)}$ is a function of $d_i^{(2)}$, we can recast the MICROSCOPE constraint on the $m_\phi-d_i^{(2)}$ plane afterwards. Comparing with the spherically symmetric ansatz in Refs.~\cite{Hees:2018fpg,Banerjee:2022sqg}, we have
\bea
\label{eq:eta_ratio}
\frac{\eta_{\vecX}|_{\omega_\EP}}{\eta_\sph} = \frac{1}{\frac{\dd \big(\psi_\sph^2/\abs{\psi_0}^2\big)}{\dd (r/R_\oplus)}} \,\, \left[ \frac{\dd a_0}{\dd (r/R_\oplus)}
- \left(\frac{1}{2}-\frac{3}{4}\cos^2\alpha\right)\frac{\dd a_2}{\dd (r/R_\oplus)} + \cdots \right]\, ,
\eea
which we show in ten curves uniformly sampled over
$0^\circ \lesssim \alpha \lesssim 50^\circ$ in \Fig{eotvos_ratio_EP}. 
For $k_0 R_\oplus \ll 1$, where
$a_0 \simeq \psi_{\rm sph}^2/\abs{\psi_0}^2$, the E\"otv\"os parameter from
the background-induced force reproduces the result of
Refs.~\cite{Hees:2018fpg,Banerjee:2022sqg}. As $k_0 R_\oplus \gtrsim 0.1$,
however, the $a_0$-derivative departs from the spherically symmetric ansatz,
enhancing $\eta_{\vecX}|_{\omega_{\rm EP}}$ by up to a factor of 30.
For $k_0 R_\oplus \gtrsim 10$, the $a_2$-derivative dominates, highlighting the importance of performing the full calculation beyond spherical symmetry.  
For $k_0 R_\oplus \gtrsim 100$, the systems enters the optical-limit regime and \Eq{eta_ratio} becomes approximately constant. 
In this regime, throughout the range of $\alpha$ covered by the MICROSCOPE science
 sessions, $\abs{\eta_{\vecX}|_{\omega_\EP}}<\abs{\eta_\sph}$. 

The acceleration difference in \Eq{acc_diff_EP_X}, the E\"otv\"os parameter in \Eq{eotvos_X_EPband}, and the ratio in \Eq{eta_ratio} do not yet include screening by the MICROSCOPE satellite surrounding the test masses $\testmass = \ti, \pt$. We now account for this satellite-screening effect discussed by Refs.~\cite{Burrage:2025grx,Burrage:2026loe} in the context of determining the screening-limited upper boundary of the MICROSCOPE exclusion region. Thereafter this method is adopted in both our work and the concurrent study of Ref.~\cite{Brzeminski:2026pgz}. As reported in Refs.~\cite{Robert:2020ddm,eoPortal:MICROSCOPE}, in its folded configuration, the MICROSCOPE satellite has an approximate volume $\mathcal{V}_\sat \simeq 1.4\,{\rm m} \times 1.0\,{\rm m} \times 1.6\,{\rm m} \simeq 2.2 \, {\rm m}^3$ and a mass $M_\sat \simeq 301.6\,\kg$. 
The satellite structure is composed primarily of aluminum~\cite{Robert:2020ddm,eoPortal:MICROSCOPE}, and we therefore approximate its dilaton charges by those of aluminum, $(Q_\sat)_i \simeq (Q_{\rm Al})_i$,
\begin{equation}
\label{eq:sat_dilaton_charge}
\begin{array}{@{}lll@{}}
\quad \quad \quad (Q_\sat)_g = 1, 
& \quad \quad \quad (Q_\sat)_e \simeq 1.8 \times 10^{-3},
& \quad \quad \quad (Q_\sat)_{m_e} \simeq 2.6 \times 10^{-4}, \\[6pt]
\quad \quad \quad (Q_\sat)_{\hat{m}} \simeq 9.7 \times 10^{-2},
& \quad \quad \quad (Q_\sat)_{\delta m} \simeq 4.2 \times 10^{-5},
& \quad \quad \quad (Q_\sat)_{H} \simeq 0.3.
\end{array}
\end{equation}
Using $M_\sat$ and $\mathcal{V}_\sat$ mentioned above, we take its average density to be $\rho_\sat \simeq M_\sat /\mathcal{V}_\sat \simeq 0.13\,{\rm g}/{\rm cm}^3$. A precise calculation of satellite screening would require a detailed model of the satellite geometry and material distribution. For tractability, we approximate the satellite as a uniform sphere with mass $M_\sat$ and radius $R_\sat \simeq (3 \mathcal{V}_\sat/4\pi)^{1/3}$, and place the test masses at its center. Within this approximation, the scalar field at the center is $\psi_\sat= \psi_{\rm no \, sat}\times 1/\cosh(\mMsat R_\sat)$ and $(\vecnabla \psi)_\sat = (\vecnabla \psi)_{\rm no \, sat} \times \mMsat R_\sat/\sinh(\mMsat R_\sat)$~\cite{Burrage:2025grx}, where the subscripts ``sat'' and ``no sat'' denote the quantities in the presence and absence of the surrounding satellite. Therefore, the satellite-screening factor multiplying the force signal is
\begin{equation}
\label{eq:F_sat_asymp}
\mathcal{F}_\sat \simeq \frac{2\,\mMsat \, R_\sat}{\sinh(2\mMsat R_\sat)} \simeq 
\begin{cases}
1\,, & \quad \quad \mMsat R_\sat \ll 1\,, \\[4pt]
4 \, \mMsat R_\sat \, e^{-2\,\mMsat R_\sat}\,, & \quad \quad \mMsat R_\sat \gg 1\,.
\end{cases}
\end{equation}
where the exponential suppression in the strongly screened regime is controlled by the penetration depth of the scalar field inside the satellite. Satellite screening therefore becomes important once $\mMsat R_\sat \sim 1$ and exponentially suppresses the force signal at larger couplings. Specifically, we have
\bea
\alpha^{(2)}_\sat \simeq 10^{24} \quad \quad \Longleftrightarrow \quad \quad \mMsat R_\sat \simeq 1 \quad \quad \quad \text{(Satellite Screening)}.
\eea
Here, $\alpha_\sat^{(2)}$ labels the critical coupling at which satellite screening becomes important. This condition can be mapped onto the $m_\phi-d^{(2)}_i$ plane through $d^{(2)}_{\sat, i} \sim \alpha^{(2)}_\sat/(Q_\sat)_i$, using the satellite dilaton charges given in \Eq{sat_dilaton_charge}. The uniform-sphere model is a simplifying approximation. Alternatively, the satellite may be modeled as a finite-thickness shell surrounding the test masses. For a shell thickness $\Delta L_\sat < R_\sat$, the relevant screening exponent scales as $\mMsat \,\Delta L_\sat \propto (d_i^{(2)} \Delta L_\sat/R_\sat)^{1/2}$, implying that a thinner shell produces weaker satellite screening. For fixed satellite mass and outer radius, this scaling implies that the uniform-sphere model produces stronger screening than a thin-shell model. We therefore adopt the uniform sphere as our fiducial model to obtain a conservative constraint in the analysis below. Another effect that may modify the constraint in the strongly coupled regime is the matter effect of the atmosphere, which effectively increases the screening radius of the Earth. In \Appx{atm_matter_effect}, we show that this additional effect is negligible throughout most of the parameter space of interest, except in the ultra-strong-coupling regime.

To calculate the bound on $d_i^{(2)}$ from MICROSCOPE, we perform a $\chi^2$ test on the data in Table~\ref{tab:microscope_segments}. The statistical uncertainty is obtained by MICROSCOPE~\cite{Touboul:2022yrw} from their data-analysis framework, and it is also verified that the statistical error in each segment is compatible with the Gaussian distribution. 
The systematic uncertainty is separately reported by MICROSCOPE by combining the error budgets. 
Following the common convention, we evaluate the total error $\sigma_\tot$ by adding the statistical and systematic errors in quadrature, i.e., $\sigma_\tot = \sqrt{\sigma_{\rm stat}^2 + \sigma_{\rm syst}^2}$. From now we use $\eta_{\obs,I}$ and $\sigma_{\tot,I}$ to indicate the observed E\"otv\"os parameter and total error, respectively, for the $I-$th SUEP segment given in Table~\Ref{tab:microscope_segments}. 
We treat the MICROSCOPE data in each segment as Gaussian distributed around the theoretical prediction, with the standard deviation given by $\sigma_{\tot, I}$. We follow the likelihood analysis method in Ref.~\cite{Cowan:2010js} and use the $\chi^2$ statistic without nuisance parameters or correlated noise between segments. Possible off-diagonal contributions arising from correlated systematic uncertainties between segments are neglected, as their impact is expected to be subdominant given $\sigma_{{\rm syst},I} \ll \sigma_{{\rm stat},I}$. For the 19 scientific segments and for a certain $m_\phi$, the $\chi^2$ statistic with respect to the coupling $d^{(2)}_i$ is,
\bea
\label{eq:chisq_MICRO}
\chi^2(d^{(2)}_i;m_\phi) = \sum_{I=1}^{N_{\rm seg}} \bigg[ \frac{ \eta_{\obs,I} - \mathcal{F}_\sat \, \eta_{\vecX}|_{\omega_\EP}(d^{(2)}_i;m_\phi, \alpha_I)}{\sigma_{\tot,I}} \bigg]^2 \,,
\eea
where $\eta_{\vecX}|_{\omega_\EP}(d^{(2)}_i;m_\phi, \alpha_I)$ is the contribution from ULDM $\phi$ with mass $m_\phi$ and coupling $d^{(2)}_i$, estimated at the $I-$th segment with DM wind direction $\alpha_I$, and $N_{\rm seg}=19$ is the number of SUEP scientific segments as given in Table~\ref{tab:microscope_segments}. Here, we take $\alpha_I = (\alpha_{{\rm start},I} + \alpha_{{\rm end},I})/2$, the average of its values at the beginning and end of each session, for our numerical calculations\footnote{We have verified that choosing either $\alpha_I=\alpha_{{\rm start},I}$ or $\alpha_I=\alpha_{{\rm end},I}$ does not noticeably change our numerical results. This is because each session lasts only a few days, whereas $\alpha(t)$ varies on an annual timescale, leading to corrections to the resulting constraints at approximately the percent level.}. Here, $\eta_{\vecX}|_{\omega_\EP}(d^{(2)}_i;m_\phi, \alpha_I)$ denotes the unscreened prediction given in \Eq{eotvos_X_EPband}, while satellite screening is incorporated separately through the satellite screening factor $\mathcal{F}_\sat$. We restrict our analysis to scenarios in which only a single coupling $d^{(2)}_i$ is nonzero. Here our $\chi^2$ statistic \Eq{chisq_MICRO} is equivalent to the log-likelihood for the Gaussian distributed errors
\bea
L(d_i^{(2)};m_\phi)=
\prod_{I=1}^{N_{\rm seg}}\frac{1}{\sqrt{2\pi}\,\sigma_{\tot,I}}\exp\left[-\frac{1}{2}\left (\frac{ \eta_{\obs,I} -  \mathcal{F}_\sat \, \eta_{\vecX}|_{\omega_\EP}(d^{(2)}_i;m_\phi, \alpha_I)}{\sigma_{\tot,I}} \right)^2\right]\,.
\eea
Therefore $\chi^2 = -2\log L + \text{const}$.
To generate the upper limit of the coupling $d^{(2)}_i$, we use the test statistic $\Delta \chi^2$ defined as
\begin{equation}
\label{eq:delta_chisq_MICRO}
\Delta \chi^2(d_i^{(2)};m_\phi) = \chi^2(d_i^{(2)};m_\phi) - \min_{d_i^{(2)}} \chi^2(d_i^{(2)};m_\phi)\,,
\end{equation}
where the second term minimizes $\chi^2(d^{(2)}_i; m_\phi)$ in terms of $d_i^{(2)}$ for a fixed $m_\phi$. It is equivalent to the log-likelihood ratio of $L(d_i^{(2)};m_\phi)$ and the maximized likelihood $\max_{d_i^{(2)}}L(d_i^{(2)};m_\phi)$ as $\Delta \chi^2 = -2 \log(L/\max_{d_i^{(2)}} L)$. Based on Wilks' theorem, $\Delta \chi^2$ follows a $\chi^2-$distribution with one degree-of-freedom. We set a 95\% C.L. one-sided upper limit on $d^{(2)}_i$ by requiring $\Delta \chi^2(d^{(2)}_i;m_\phi) = 2.71$. For comparison, when deriving the MICROSCOPE constraint using the spherically symmetric ansatz, we set $\mathcal{F}_\sat=1$, thereby switching off the satellite-screening to reproduce the treatment adopted in Refs.~\cite{Hees:2018fpg,Banerjee:2022sqg}. We will return to this comparison in the subsequent discussion.

\begin{figure}[t!]
\centering
\includegraphics[width=0.6\linewidth]{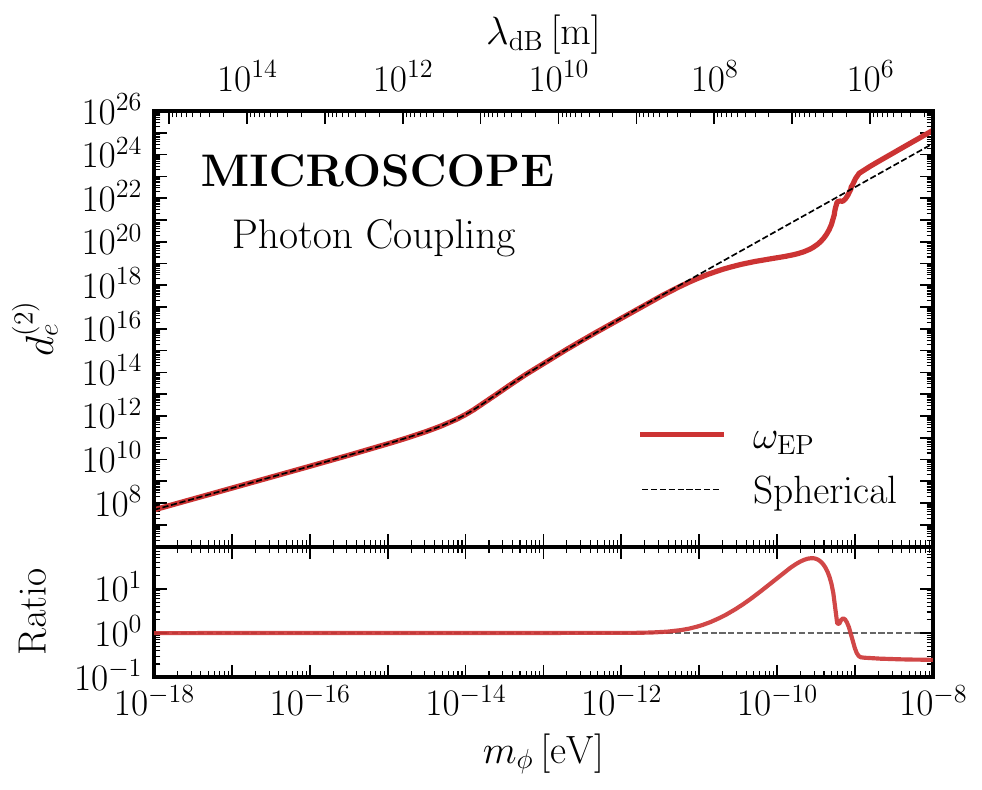}
\caption{$\chi^2$ limit on $d_e^{(2)}$ from MICROSCOPE, for the lower-coupling boundary where satellite and atmospheric screening are negligible. The {\bf red solid} line is derived from the EP-band signal $\eta_{\vecX}|_{\omega_\EP}$ in \Eq{eotvos_X_EPband}, while the {\bf black dashed} line is obtained using the previous spherically symmetric ansatz~\cite{Hees:2018fpg,Banerjee:2022sqg}. The lower panel shows the ratio of the lower-coupling boundary of the limit on $d_e^{(2)}$ obtained using the previous spherically symmetric ansatz to that obtained using the EP-band signal $\eta_{\vecX}|_{\omega_\EP}$ in \Eq{eotvos_X_EPband}. Note that this ratio does not depend on the specific scalar--SM coupling considered, as long as the coupling stays below the strongly coupled satellite-screening regime.}
\label{fig:micro_compare}
\end{figure}

Compared with previous studies that assumed a spherically symmetric ansatz, we find notable modifications to the resulting constraints. We find that for $m_\phi \sim 10^{-11}$ eV, the constraint receives an enhancement by over an order of magnitude, but weakens at higher masses, for all values of $\alpha$ considered (see \Fig{eotvos_ratio_EP}). Both of these effects are consequences of deviating from spherical symmetry. The combined result is shown in \Fig{micro_compare}. Here, we compare the upper bound on $d_e^{(2)}$ obtained from two $\chi^2$ analyses -- one using the previous spherically symmetric ansatz~\cite{Hees:2018fpg,Banerjee:2022sqg}, and the other using the EP-band signal $\eta_{\vecX}|_{\omega_\EP}$ in \Eq{eotvos_X_EPband} -- together with the ratio between the upper bounds.
We find similar results for the other couplings, which we show in \Fig{microscope_constraint}.

Note that when imposing the MICROSCOPE constraint in this section using $\chi^2$-analysis, we only consider the main EP band with $\omega = \omega_\EP$, i.e., the magenta band in \Fig{X_band}. As we have discussed in the previous section, performing an analysis with the full frequency band structure can further enhance the constraint. We quantify this enhancement in the next section. 

%%%%%%%%%%%%%%%%%%%%%%%%%
\subsection{Future Experimental Probes and Analyses}
\label{sec:future}
%%%%%%%%%%%%%%%%%%%%%%%%%
In the previous section, we derived the MICROSCOPE constraint using our force formalism and the published segment-level measurements in the main $\omega_\EP$ band~\cite{Touboul:2022yrw}, as given in Table~\ref{tab:microscope_segments}. The same force formalism also predicts orbital sidebands at $\omega_\EP\pm n\,\omega_\orb$. In principle, these sidebands can
provide additional sensitivity, because they contain independent coherent components of the same background-induced force signal. A full frequency-band analysis of MICROSCOPE, however, would require the frequency-dependent uncertainties at each sideband, which are not directly available from Ref.~\cite{Touboul:2022yrw}. 
We therefore leave a dedicated full-band MICROSCOPE analysis to future work. 
In this section, we instead formulate the general $\Delta \chi^2$ including the sideband contributions and illustrate its impact with next-generation EP-tests (see Table~\ref{tab:ep_test_altitudes}). The benchmark parameters are inspired by the Galileo Galilei proposal~\cite{Nobili:2000bzv,Nobili:2012uj,Nobili:2017cxu,Nobili:2018eym}, but the result should be interpreted as a generic next-generation projection.

For the benchmark projection, we consider a rapidly spinning space-based EP test with $\omega_\spin\simeq 2\pi\times1\,{\rm Hz}$, an observing time $T_\obs\simeq1\,{\rm day}$, and an effective E\"otv\"os sensitivity $|\eta_{\rm NG}| \lesssim 10^{-17}$ representative of a next-generation target. For the satellite-screening estimate, we take $\mathcal{V}_\sat \simeq \pi \times 0.75\,{\rm m}^2 \times 1.5\,{\rm m}$ and $M_\sat \simeq 437.9\,{\rm kg}$~\cite{ThalesAleniaSpace:2009GG}. Motivated by the reference geometry
of the MICROSCOPE mission~\cite{MICROSCOPE:2022doy,Rodrigues:2022hmp,Touboul:2022yrw} and the proposed Galileo Galilei design~\cite{Nobili:2000bzv,Nobili:2012uj,Nobili:2017cxu,Nobili:2018eym}, we adopt a configuration where the spin axis is perpendicular to the orbital plane and the force-sensitive plane coincides with the orbital plane. We do not attempt to model the mission-long evolution of the spin--orbit geometry or the associated systematic-errors. 
With this
benchmark geometry, the $\vecX$-projected force decomposition in \Eq{X_Fbg_band} directly gives the multiband E\"otv\"os signal,
\bea
\label{eq:eta_X_band}
\eta_{\vecX}(t)
& =
\eta_{\vecX}|_{\omega_\EP} \cos\Theta_{\rm EP}\\
& \,\, +\sum_{n=1}^{\infty}\left[
\eta_{\vecX}|_{\omega_\EP + n\, \omega_\orb}\cos\left(\Theta_{\rm EP}+n\,\Theta_{\rm orb}+\frac{n\pi}{2}\right)
+
\eta_{\vecX}|_{\omega_\EP - n\, \omega_\orb}\cos\left(\Theta_{\rm EP}-n\,\Theta_{\rm orb}-\frac{n\pi}{2}\right)
\right]\,.
\eea
Here, the coefficients of the E\"otv\"os parameter in the frequency domain are given by
\bea
\label{eq:eotvos_X_general_band}
\eta_{\vecX}|_{\omega_\EP \pm n\, \omega_\orb} = \frac{2 \rho_\phi}{m_\phi^2} \, \frac{\pi \, (\alpha_\ti^{(2)}- \alpha_\pt^{(2)})\,(R_\oplus+h)^2}{ M_\oplus  \, R_\oplus } \times (R_n^X \pm T_n^X),
\eea
where $R_n^X$ and $T_n^X$ are given in \Subsec{force_proj_band_structure}. The coefficients $\eta_{\vecX}|_{\omega_\EP \pm n\, \omega_\orb}$ in \Eq{eta_X_band} and \Eq{eotvos_X_general_band} correspond to the unscreened signal, while satellite screening is incorporated through the multiplicative factor $\mathcal{F}_\sat$ in the construction of $\Delta\chi^2$ in \Eq{chisq_nextg_generic}. Note that when $n=0$, where $T_0^X=0$, \Eq{eotvos_X_general_band} simplifies to the expression for the main band \Eq{eotvos_X_EPband}. We first compute the autocorrelation of \Eq{eotvos_X_general_band} using $C_\eta(\tau) = \langle  \eta_{\vecX}(t) \, \eta_{\vecX}(t+\tau) \rangle_t$, and then use the Wiener--Khinchin theorem to obtain the two-sided power spectral density~(PSD) $S_\eta^{(2)}(f) = \int_{-\infty}^{\infty} \, d\tau \, e^{-2 i \pi f \tau} C_\eta(\tau)$. Since $f\geq0$, $2\,S_\eta^{(2)} \rightarrow S_\eta$ and we have the one-sided PSD
\bea
S_{\eta}(f) = \underbrace{\frac{1}{2}\,\Big[\eta_{\vecX}|_{\omega_\EP}\Big]^2 \delta\left(f-f_\EP\right)}_{\text{Main band}} + \underbrace{\sum_{n=1}^\infty \frac{1}{2}\,\Big[\eta_{\vecX}|_{\omega_\EP \pm n\, \omega_\orb}\Big]^2 \delta\left(f-(f_\EP \pm n\,f_\orb)\right)}_{\text{Sidebands}}\, . 
\eea
The main band and orbital sidebands are spectrally resolved because $f_\orb \gg 1/T_\obs$. Using a matched-filter statistic for the coherent oscillatory signal induced by the Earth's background-induced force~\cite{Maggiore:2007ulw}, we have
\bea
\text{SNR}^2 =2 \, \Nsens \, T_\obs \int^\infty_{0} \dd f \, \frac{S_{\eta}(f)}{S_{n}(f)},
\eea
where $S_{n}(f)$ is the one-sided noise power spectrum. $\Nsens$ is the number of independent force-sensitive axes: $\Nsens=1$ for
MICROSCOPE (sensitive along $\vecX$ only) and $\Nsens=2$ for Galileo Galilei which has a 2D force-sensitive plane~\cite{Nobili:2012uj,Nobili:2017cxu,Nobili:2018eym}. Because $\omega_\spin \gg \omega_\orb$, we approximately have the noise power spectrum $S_{n}(f_\EP \pm n\,f_\orb) \simeq S_{n}(f_\EP)$ in the first few sidebands, which gives the dominant contribution to the signal. Here we assume that the noise PSD varies smoothly. Narrow instrumental spectral features may invalidate this approximation, requiring the actual noise PSD at each sideband frequency for a reliable sensitivity estimate. Therefore, from the previous SNR analysis, we have $\Delta \chi^2 = \text{SNR}^2$, which gives
\bea
\label{eq:chisq_nextg_generic}
\Delta \chi^2(d^{(2)}_i;m_\phi) 
& = \underbrace{ \Nsens \, \bigg[ \frac{ \mathcal{F}_\sat \, \eta_{\vecX}|_{\omega_\EP}(d^{(2)}_i;m_\phi, \alpha)}{\sigma_{\tot}} \bigg]^2}_{\text{Main band}}\\
& \quad + \underbrace{ \Nsens \,\sum_{n=1}^\infty \left\{ \bigg[ \frac{ \mathcal{F}_\sat\, \eta_{\vecX}|_{\omega_\EP + n\omega_\orb}(d^{(2)}_i;m_\phi, \alpha)}{\sigma_{\tot}} \bigg]^2 + \bigg[ \frac{ \mathcal{F}_\sat \, \eta_{\vecX}|_{\omega_\EP - n\omega_\orb}(d^{(2)}_i;m_\phi, \alpha)}{\sigma_{\tot}} \bigg]^2 \right\} }_{\text{Sidebands}},
\eea
where $\sigma_\tot \simeq \sqrt{S_n(f_\EP)/T_\obs}$ for the main band and adjacent sidebands, and $\mathcal{F}_\sat$ is the satellite-screening factor given in \Eq{F_sat_asymp}, evaluated using the benchmark satellite mass and volume specified above in this section~\cite{ThalesAleniaSpace:2009GG}. Compared with \Eq{chisq_MICRO}, the additional terms in the second line account for the sidebands $\omega_\EP \pm n\,\omega_\orb$, with $n=1,2,\ldots$.

\begin{figure}[t!]
\centering
\includegraphics[width=0.6\linewidth]{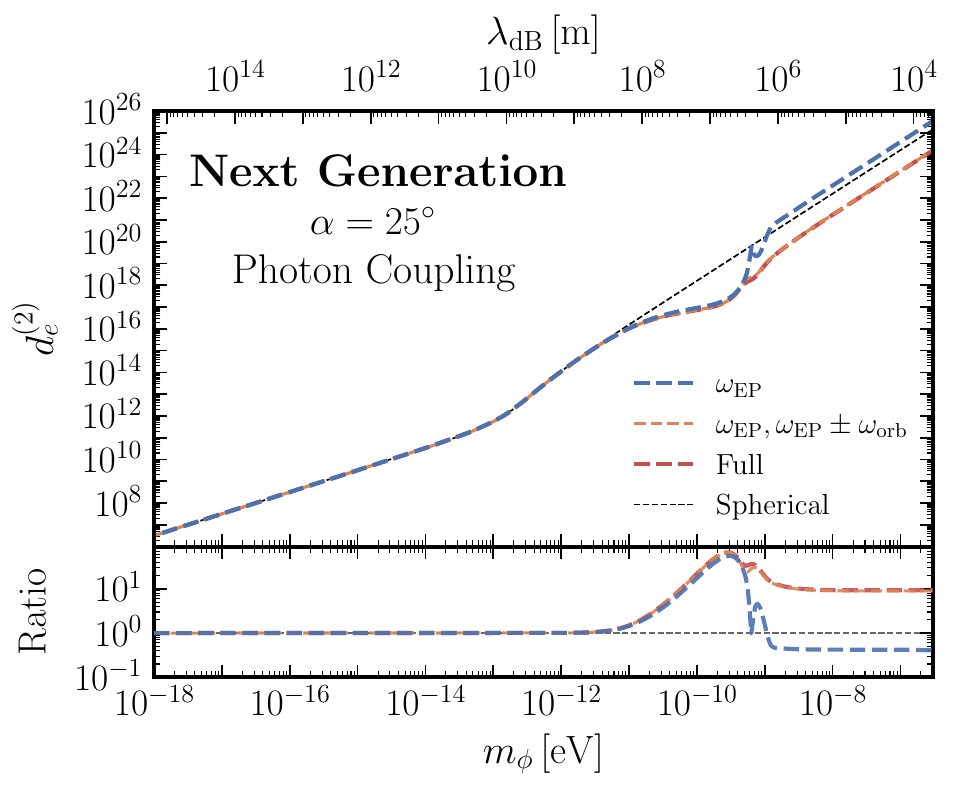}
\caption{Next-generation EP test sensitivity to $d_e^{(2)}$ with the benchmark parameters inspired by the Galileo Galilei proposal~\cite{Nobili:2000bzv,Nobili:2012uj,Nobili:2017cxu,Nobili:2018eym}, for the lower-coupling boundary where satellite and atmospheric screening are negligible. The {\bf blue dashed} is the EP-band signal $\eta_{\vecX}|_{\omega_\EP}$ in \Eq{eotvos_X_EPband}. The {\bf red dashed} and {\bf orange dashed} lines correspond to adding the first sideband $\omega_\EP \pm \omega_\orb$ and all sidebands, respectively. 
The {\bf black dashed} line comes from the spherically symmetric ansatz~\cite{Hees:2018fpg,Banerjee:2022sqg}. 
The lower panel shows the ratio of $d_e^{(2)}$ obtained using the spherically symmetric ansatz over that obtained using the band information listed above. 
Including the sideband information can strengthen the constraint by roughly one order of magnitude for $m_\phi\gtrsim 10^{-9}$ eV. 
This ratio is independent of the choice of scalar--SM coupling, as long as the coupling stays below the strongly coupled satellite-screening regime. Varying $\alpha$ changes the ratio by ${\cal O}(1)$ but not the qualitative conclusion.}
\label{fig:ng_full_compare}
\end{figure}

To illustrate the impact of the sidebands on the sensitivity, we consider a single observation window at $\alpha=25^\circ$, which is a typical value of the angle between the orbital plane and $\veck_0$, as shown in Table~\ref{tab:microscope_segments}. 
As before, we take the 95\% C.L. one-sided upper limit $\Delta \chi^2 = 2.71$ on $d^{(2)}_i$. We estimate the uncertainty as $\sigma_\tot \simeq |\eta_{\rm NG}|/\sqrt{2.71} \simeq 10^{-17}/\sqrt{2.71}$. In \Fig{ng_full_compare}, we show the sensitivity lines for the next-generation EP-tests from considering the only main band at $\omega_\EP$ and from including the sidebands $\omega_\EP\pm n\,\omega_\orb$. We find that including the sidebands can improve the sensitivity by approximately an order of magnitude relative to the main band for $m_\phi\gtrsim 10^{-9}\,\text{eV}$. This enhancement can be understood from \Fig{X_band}, which shows that the sideband force becomes comparable to, and can even exceed, the main-band force at large $k_0R_\oplus$. Furthermore, we find that the $n=1$ sideband dominates over the higher orbital sidebands, so including modes with $n>1$ provides little additional gain in sensitivity.

\begin{figure}[h!]
\centering
\includegraphics[width=0.85\linewidth]{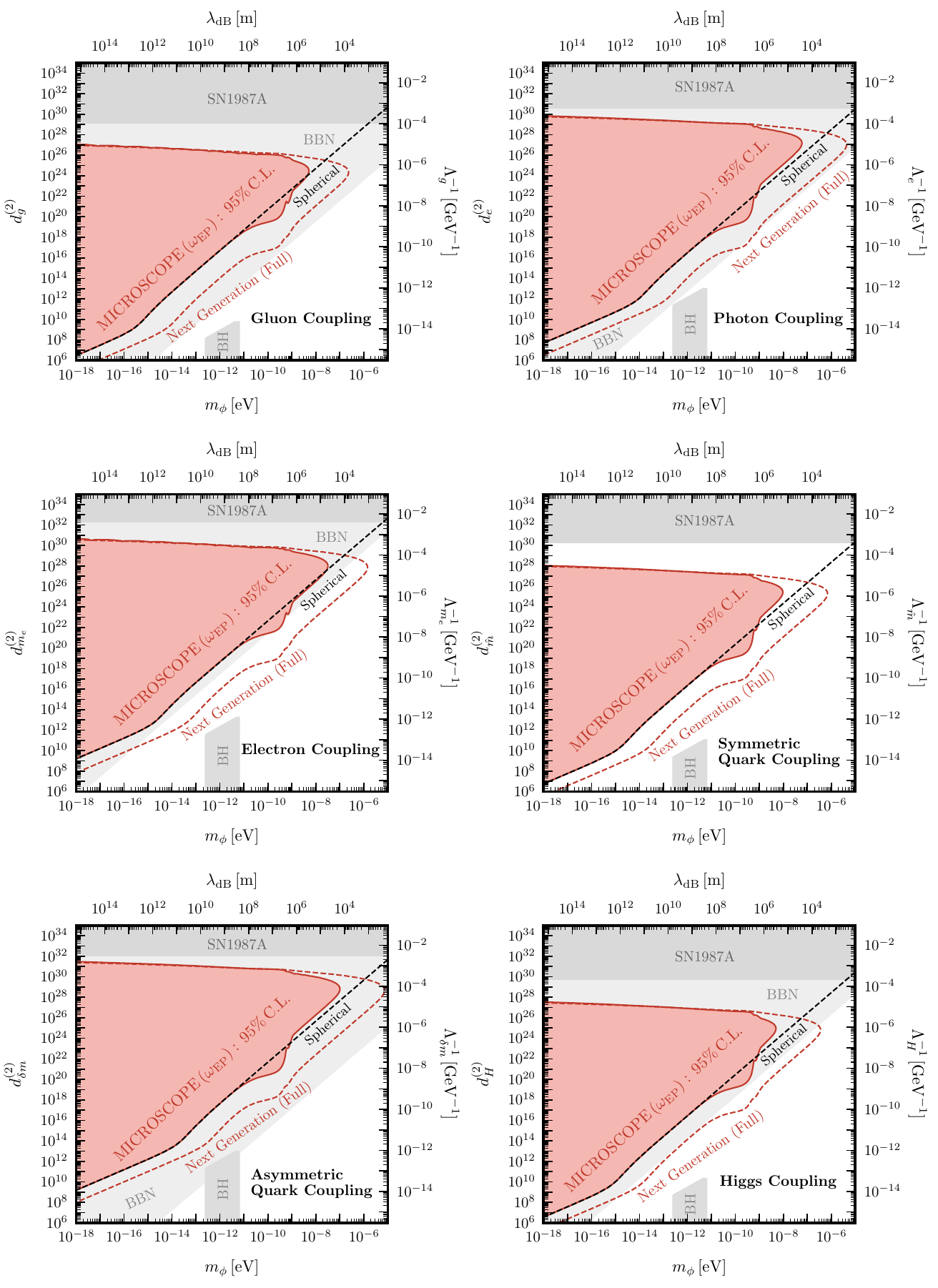}
\vspace{-1.2em}
\caption{MICROSCOPE $95\%\,\mathrm{C.L.}$ constraints from the EP-band signal ({\bf red solid}) and the spherically symmetric ansatz ({\bf black dashed}), and the full-band $95\%\,\mathrm{C.L.}$ projections from a next-generation experiment ({\bf red dashed}) for $d_i^{(2)}$. For large couplings, satellite screening suppresses the signal while atmospheric screening introduces an $\mathcal{O}(1)$ correction, with the precise effects depending on the specific satellite/atmosphere model (see Secs.~\ref{subsec:microscope_revisit} and App.~\ref{appx:atm_matter_effect}). Also shown in {\bf gray} are the constraints from BBN~\cite{Bouley:2022eer}, black hole superradiance~(BH)~\cite{Davoudiasl:2019nlo,Baryakhtar:2020gao}, and SN1987A~\cite{Raffelt:1990yz, Olive:2007aj,Day:2023mkb,Gan:2025nlu}. Note that the BBN constraints do not apply to the symmetric quark coupling $d^{(2)}_{\hat m}$.}
\label{fig:microscope_constraint}
\end{figure}

Note that the ratio shown in \Fig{ng_full_compare} depends only on the geometric parameter $\alpha$ and is independent of $d_i^{(2)}$. 
While varying $\alpha$ changes the numerical value of the ratio, it modifies the full-band sensitivity by at most an $\mathcal{O}(1)$ factor and does not change the conclusion that including the sidebands improves the sensitivity. 
This is because, for $k_0 R_\oplus \lesssim 10$, the sensitivity is dominated by the $\alpha$-independent $d a_0/dr$ contribution to $\eta_{\vecX}|_{\omega_\EP}$. 
For $k_0 R_\oplus \gtrsim 10$, the sensitivity is instead dominated by the first sidebands at $\omega_\EP \pm \omega_\orb$, whose amplitudes are proportional to $\cos\alpha$ (see \Eq{X_1st_sideband_coefficient}). 
Since $\alpha$ remains acute throughout the orbit (see \Fig{alpha_t}), the resulting variation remains an $\mathcal{O}(1)$ effect.
%%%%%%%%%%%%
%%%%%%%%%
We summarize the MICROSCOPE constraints and next-generation experiment sensitivities for the full suite of dilatonic couplings in \Fig{microscope_constraint}. For comparison, the constraints from BBN~\cite{Bouley:2022eer} and black hole superradiance~\cite{Davoudiasl:2019nlo,Baryakhtar:2020gao} are shown in gray. As shown in Ref.~\cite{Bouley:2022eer}, the BBN constraints do not apply to the symmetric quark coupling $d_{\hat m}^{(2)}$. Therefore, for this coupling, MICROSCOPE provides the strongest existing constraint. More generally, for scalar--SM couplings to which BBN constraints apply, next-generation EP tests such as Galileo Galilei and STE-QUEST can reach sensitivities comparable to, or even stronger than, the BBN constraints.

\begin{table}[t]
\centering
\renewcommand{\arraystretch}{1.7}
\setlength{\tabcolsep}{7.0pt}
\begin{tabular}{|c|l|c|c|c|}
\hline
Category & Experiment & Altitude $h$ [km] & E\"otv\"os $\eta$ & Status \\
\hline
\multirow{6}{*}{Satellite}
& MICROSCOPE~\cite{MICROSCOPE:2022doy,Rodrigues:2022hmp,Touboul:2022yrw}
& $710$
& $[-1.5\pm2.3\pm1.5]\times 10^{-15}$
& Completed \\
& CSS~\cite{he2023space,Zhang:2026lzu}
& $390$
& $[-3.1 \pm 4.6]\times 10^{-7}$
& Completed \\
& Galileo Galilei~\cite{Nobili:2000bzv,Nobili:2012uj,Nobili:2017cxu,Nobili:2018eym}
& $600$
& $(\sim 10^{-17})$
& Proposed \\
& STE-QUEST~\cite{STE-QUEST:2022eww,Struckmann:2023ybg}
& $1400$
& $(\sim 10^{-17})$
& Proposed \\
& STEP~\cite{Sumner:2006vex,Overduin:2012uk}
& $550$
& $(\sim 10^{-18})$
& Proposed \\
& QTEST~\cite{Williams:2015ima}
& $400$
& $(5\times 10^{-16})$
& Proposed \\
\hline
Sounding Rocket
& SR-POEM~\cite{Reasenberg:2010iz}
& $\geq 800$
& $(\leq 10^{-16})$
& Proposed \\
\hline
Torsion-Balance
& E\"ot-Wash~\cite{Schlamminger:2007ht}
& $0.07$
& $[0.3\pm 1.8]\times 10^{-13}$
& Completed \\
\hline
\end{tabular}
\caption{Representative EP-test experiments. We list their categories, approximate altitudes, E\"otv\"os sensitivities, and status. Values in parentheses denote projected sensitivities for proposed experiments. Among the completed experiments, MICROSCOPE provides the leading sensitivity to quadratically coupled ultralight dark matter
in the parameter space considered here. Ground-based E\"ot-Wash experiments, for example, are more strongly affected by Earth's screening due to their proximity to the Earth's surface.}
\label{tab:ep_test_altitudes}
\end{table}

%%%%%%%%%%%%%%%%%%%%%%%%%

% \subsection{Results}

%%%%%%%%%%%%%%%%%%%%%%%%%
\section{Discussion and Conclusion}
\label{sec:conclusion}
%%%%%%%%%%%%%%%%%%%%%%%%%

Quadratically coupled ultralight scalar dark matter remains a comparatively unexplored dark-matter candidate despite its rich phenomenology and the strong experimental sensitivity of equivalence-principle (EP) tests. A central challenge in interpreting these searches is that the coupling to ordinary matter modifies not only fundamental parameters but also the scalar field configuration itself, giving rise to screening effects and a background-induced force. Accurately predicting observable signals therefore requires a self-consistent treatment of the scalar distribution in the presence of matter.

In this work, we developed a general framework that incorporates phase-space averaging for calculating the background-induced force beyond the spherically symmetric approximation. We showed that the commonly used spherically symmetric ansatz breaks down once the ULDM de Broglie wavelength becomes comparable to the Earth's radius, $m_\phi \gtrsim 10^{-11},\mathrm{eV}$, and that the resulting modifications can substantially alter the predicted experimental signal. Applying our formalism to the MICROSCOPE mission, we derived updated constraints on the dilatonic couplings that differ significantly from previous results in this regime, by more than an order of magnitude at some masses.

A key result of this work is the identification of a novel frequency-band structure generated by Earth screening. 
We found that the orbital motion of satellite EP tests lead to characteristic orbital sidebands in the sensitive axis ($\vecX$) signal and generate nonzero signals along the spin axis ($\vecY$). 
This results in a signal that exhibits a nontrivial structure in frequency space, suggesting that a full-band analysis could substantially improve experimental sensitivity.
Using the proposed Galileo Galilei mission as a benchmark, we find that a full-band analysis improves the reach by more than an order of magnitude, and we expect comparable gains for STE-QUEST.
These findings also motivate a dedicated reanalysis of the MICROSCOPE data using the full frequency-band information.

We further demonstrated that the high-mass and strongly coupled regime, corresponding to region (D), admits a simple physical interpretation. In this region, the scalar field approaches the optical-limit configuration and be accurately described using geometric optics. The agreement between the geometric-optics and partial-wave descriptions provides a unified understanding of the scalar distribution across a wide range of parameters.

The methods developed here are readily applicable beyond the scenario considered in this work, including scattering between scalar field and the test mass, as well as attractive scalar interactions, axions, long-range neutrino-mediated forces, and vector models such as $B-L$ gauge bosons. More broadly, our results highlight the importance of matter-induced distortions of ULDM fields and demonstrate that they can generate qualitatively new experimental signatures. 
The physical insights obtained in this work may help guide the design of future experiments and analyses aimed at probing ULDM and other weakly coupled new physics.

\vspace{3mm}
{\it Note Added: While this work was being completed, we
became aware of overlapping work in preparation from~\cite{Brzeminski:2026pgz}.}

\section*{Acknowledgement}

We thank Dawid Brzeminski, Yifan Chen, Shao-Feng Ge, Aaron Pierce, G\'eraldine Servant, Lian-Tao Wang, and Yue Zhao for helpful discussions. The work of X.G. is supported by the Deutsche Forschungsgemeinschaft under Germany’s Excellence Strategy - EXC
2121 “Quantum Universe” - 390833306. H.X. is supported by the Siyuan Postdoctoral (Overseas Talent Recruitment) Program of Shanghai Jiao Tong University and the Shanghai ``Super Postdoc'' Incentive Program. T-T.Y. is supported by the U.S. Department of Energy under Grant Number DE-SC0011640.

\appendix
\counterwithin{figure}{section}
\counterwithin{table}{section}

\section{Dilaton Charges}\label{appx:dilaton_charge_appx}

The dimensionless scalar--SM coupling $\alpha^{(2)}_\mathcal{A}$, defined as the fractional change in the mass of body $\mathcal{A}$ induced by the background scalar field, governs the interaction between the scalar and SM. Its composition dependence is encoded in the material-dependent dilaton charges $(Q_\calA)_{i}$, which are determined by hadronic, nuclear, and atomic physics.

Following Refs.~\cite{Damour:2010rp,Damour:2010rm}, we present a pedagogical and self-contained derivation of the scalar and dilaton charges, and give closed-form analytical expressions in terms of the mass and atomic numbers $(A,\,Z)$, the QCD nucleon matrix elements $(\sigma_{\pi N},\,\Delta\sigma)$, the electromagnetic nucleon-mass parameters $(C_p,\,C_n)$, the electron mass $m_e$, and the nuclear binding-energy parameters entering the semi-empirical mass formula~(SEMF). 
Previous literature~\cite{Damour:2010rp,Damour:2010rm,Hees:2018fpg,Bouley:2022eer} has quoted numerical values for these charges. The analytical forms derived here make the composition dependence explicit and are 
updated with modern hadronic inputs. The dilaton charges for the effective scalar--SM coupling are discussed in \Subsec{effective_coupling} and have been widely studied in Refs.~\cite{Damour:2010rp,Damour:2010rm,Hees:2018fpg,Bouley:2022eer}. For completeness, we also provide the corresponding expressions for the Higgs portal in \Subsec{higgs_portal} and for the universal coupling and light QCD axion in \Subsec{other_coupling}. Although we take the quadratic coupling as the benchmark example throughout this work, the discussion in this section generalizes straightforwardly to scalar--SM couplings with other powers of the scalar field.

\subsection{Scalar--SM Coupling}\label{subsec:scalar_sm_coupling_appx} 

In this subsection, we derive the scalar--SM coupling $\alpha^{(2)}_\calA$
 and express it in terms of the dilaton charges $(Q_\calA)_i$, which characterize the sensitivity of a test mass to variations of fundamental constants. We first consider a single-component object and discuss the multicomponent generalization at the end of the subsection.

We begin by defining the dimensionless scalar--SM coupling: 
\bea
\label{eq:scalar_charge_def_appx}
\frac{\Delta M_\calA}{M_\calA} = \frac{\Delta m_\calA}{m_\calA} = \alpha^{(2)}_\calA \times \Phi^{(2)}\,, \quad \quad \text{where} \quad \Phi^{(2)}=\frac{2\pi \phi^2}{\Mpl^2}\,.
\eea
In the above equation, ``$\calA$'' is the subscript labeling the object, and ``$(2)$'' indicates a quadratic coupling.
Here, $\Phi^{(2)}$ is the dimensionless representation of the quadratic scalar.~\footnote{The discussion in this section extends directly to an $\mathscr{N}$-th power scalar coupling to the SM sector, defined by $\Phi^{(\mathscr{N})} = (\sqrt{4\pi}\phi/\Mpl)^\mathscr{N}/\mathscr{N}\,!$, where $\mathscr{N}\,!$ is the symmetry factor. The corresponding theories are obtained from the effective scalar--SM interaction~(\Subsec{effective_coupling}), the Higgs portal~(\Subsec{higgs_portal}), and the universal coupling~(\Subsec{other_coupling}) upon the replacement $\Phi^{(2)} \rightarrow \Phi^{(\mathscr{N})}$. However, the interaction of the QCD axion remains quadratic, as ensured by the $\mathbb{Z}_2$ CP symmetry~\cite{GrillidiCortona:2015jxo,Kim:2023pvt,Beadle:2023flm,Gan:2025nlu}. The linear case in Ref.~\cite{Damour:2010rp} is recovered by taking $\Phi^{(1)} = \sqrt{4\pi}\phi/\Mpl$. In this work, we focus on the quadratic case, $\Phi^{(2)}$.} 
$M_\calA = N_\calA \times m_{\calA}$ is the mass of macroscopic object, while $m_{\calA}$ is the atomic mass and $N_\calA$ is the total number of atoms.

The atomic mass receives three contributions that motivate its parametrization: 
\begin{itemize}
    \item {\it Nucleon masses --} In the combined chiral limit $m_u,m_d\to 0$ and electromagnetic-decoupling limit $\alpha_{\rm em}\to 0$, $\Lambda_{\rm QCD}$, generated through dimensional transmutation in the QCD sector, is the only scale determining ordinary hadron masses. This reflects the fact that the dominant chirally symmetric contribution to hadron masses is set by gluodynamics~\cite{Shifman:1978zn,Wilczek:2012sb}. Once the non-zero chiral-symmetry-breaking parameters $\hat m/\Lambda_{\rm QCD}$ and $\delta m/\Lambda_{\rm QCD}$, as well as a non-zero $\alpha_{\rm em}$, are restored, hadron masses~(such as the proton and neutron masses) are correspondingly shifted.
    \item {\it Nuclear binding energy --} The nuclear binding energy is controlled by $\Lambda_{\rm QCD}$, $\hat m/\Lambda_{\rm QCD}$, and $\alphaem$.
    \item {\it Electron mass --} The electron rest mass, which arises from the Higgs mechanism independently of QCD, contributes directly to the atomic mass.
\end{itemize}   
Together, these three contributions, lead to the following parametrization~\cite{Damour:2010rp}:
\bea
\label{eq:m_A_appx_general}
m_\calA = \Lambda_\text{QCD} \times \overline{m}_\calA\Big(\alphaem, \frac{m_e}{\Lambda_\text{QCD}}, \frac{\hat{m}}{\Lambda_\text{QCD}}, \frac{\delta m}{\Lambda_\text{QCD}}\Big)\,.
\eea
Here, $\Lambda_{\rm QCD}$ is chosen as the reference scale, since it dominates the hadronic contributions to atomic mass. The reduced atomic mass $\bar m_{\mathcal A}$ is therefore dimensionless and depends only on the dimensionless variables $\alpha_{\rm em}$, $m_e/\Lambda_{\rm QCD}$, $\hat m/\Lambda_{\rm QCD}$, $\delta m/\Lambda_{\rm QCD}$. We expand on the atomic mass discussion in \Subsec{atom_mass}.

Differentiating \Eq{scalar_charge_def_appx} with respect to $\Phi^{(2)}$, we have 
$\alpha^{(2)}_\calA = \partial \log m_\calA / \partial \, \Phi^{(2)}$. 
Substituting \Eq{m_A_appx_general} and applying the chain rule, we obtain
\bea
\label{eq:coupling_dilaton_charge_relation_appx}
\alpha^{(2)}_\calA = (Q_\calA)_{g} \, d^{(2)}_g + (Q_\calA)_e \, d^{(2)}_e + (Q_\calA)_{m_e} \, (d^{(2)}_{m_e}-d^{(2)}_g) + (Q_\calA)_{\hat{m}} \, (d^{(2)}_{\hat{m}}-d^{(2)}_g)+ (Q_\calA)_{\delta m} \, (d^{(2)}_{\delta m}-d^{(2)}_g)\,,
\eea
where $d^{(2)}_i$ is the dimensionless couplings varying the fundamental constants as given by \Eq{vary_const}. Then we have the dilaton charges defined as
\bea
\label{eq:dilaton_charge_def_appx}
& (Q_\calA)_{g} = 1\,, \quad \quad \quad (Q_\calA)_{e} =\frac{\partial \log \overline{m}_\calA}{\partial \, \log \alphaem}\,, \quad \quad \quad  (Q_\calA)_{m_e} = \frac{\partial \, \log \overline{m}_\calA}{\partial \, \log\,(m_e/\Lambda_\text{QCD})}\,,\\
& \quad \quad (Q_\calA)_{\hat{m}} = \frac{\partial \, \log \overline{m}_\calA}{\partial \, \log\,(\hat{m}/\Lambda_\text{QCD})}\,, \quad \quad (Q_\calA)_{\delta m} = \frac{\partial \, \log \overline{m}_\calA}{\partial \, \log\,(\delta m/\Lambda_\text{QCD})}\,,
\eea
which quantify the sensitivities of the atomic mass to variations in the fundamental constants. Note that $(Q_\calA)_g = \partial \log m_\calA/\partial \, \log \Lambda_\text{QCD} = 1$, because $\Lambda_\text{QCD}$ is chosen as the reference scale for the atomic mass, as given by \Eq{m_A_appx_general}.

We now generalize the above discussion to multicomponent objects, such as the Earth or the test masses used in EP experiments. We let $I$ label the different constituents, and write the total mass as
$M_{\calA} = \sum_I m_{\calA_I} N_{\calA_I}$. We define
$f_{\calA_I} = m_{\calA_I} N_{\calA_I}/M_{\calA}$
to be the mass fraction of constituent $I$, where $m_{\calA_I}$ and $N_{\calA_I}$ denote the atomic mass and the number of atoms of constituent $I$, respectively. Since the scalar--SM coupling is defined through the fractional mass variation in \Eq{scalar_charge_def_appx}, it follows that $\alpha^{(2)}_\calA$ is the mass-fraction-weighted average of the constituent couplings. Applying \Eq{scalar_charge_def_appx} and \Eq{coupling_dilaton_charge_relation_appx} to each constituent
then yields the corresponding dilaton charges,
\bea
\label{eq:coupling_dilaton_charge_multicomponent}
\alpha^{(2)}_\calA = \sum_I f_{\calA_I} \, \alpha^{(2)}_{\calA_I}\,, 
\quad \quad \quad 
(Q_\calA)_i = \sum_I f_{\calA_I} \, (Q_{\calA_I})_i\,,
\eea
with $\sum_I f_{\calA_I} = 1$. For a realistic object with multiple constituents, we evaluate its scalar-SM coupling and dilaton charges using \Eq{coupling_dilaton_charge_multicomponent}.

\subsection{Atomic Mass}\label{subsec:atom_mass}

In this section, we introduce the composition of atomic mass in terms of the constituent rest masses and the nuclear binding energy. We briefly explain its origin and then discuss how it depends on variations of the fundamental constants given by \Eq{vary_const}.

To start with, we decompose the atomic mass $m_\calA$ as 
\bea
\label{eq:m_atom_mass_appx}
m_{\calA} = (E_\calA)_\rmass + (E_\calA)_\bind\,,
\eea
where $(E_\calA)_\rmass$ is the rest mass of the nucleons and electrons, and $(E_\calA)_\bind$ is the nuclear binding energy. Here, the rest mass is 
\bea
\label{eq:rest_mass_appx}
(E_\calA)_\rmass = Z\,m_p + N\,m_n + Z\,m_e\,, \quad \quad \quad (N=A-Z)\,
\eea
where $m_n$ is the neutron mass, $m_p$ is the proton mass, $m_e$ is the electron mass. The electron mass arises from the Higgs mechanism, so it is free from the variation of $\alphaem$, $\Lambda_\text{QCD}$, $\hat{m}$, $\delta m$. 
The proton and neutron masses are given by
\bea
\label{eq:proton_neutron_mass_appx}
m_p = m_{N,0} + \sigmapi - \frac{1}{2} \Delta \sigma + C_p\,, \quad \quad m_n = m_{N,0} + \sigmapi + \frac{1}{2} \Delta \sigma +C_n\,.
\eea
In the above equation, $m_{N,0}$ is the nucleon mass in the chiral limit ($m_u, m_d \to 0$), arising purely from gluon dynamics; $\sigma_{\pi N} = \hat{m}\,\partial \, m_{p,n}/\partial \, \hat{m} = m_\pi^2\,\partial \, m_{p,n}/\partial \, m_\pi^2$ is the pion-nucleon sigma term; $\Delta\sigma = - 2 \, \delta m \, \partial \, m_p/\partial \,\delta m = 2 \, \delta m \, \partial \, m_n/\partial \,\delta m$ quantifies the neutron-proton mass difference arising from isospin violation ($\delta m = m_d - m_u \neq 0$) at electroweak scale; and $C_{p,n} = \alpha_\text{em}\,\partial \,m_{p,n}/\partial\,\alpha_\text{em}$ is the electromagnetic contribution to the proton and neutron masses. Specifically, we have
\bea
\label{eq:proton_neutron_mass_scaling_appx}
m_{N,0} \propto \Lambda_\text{QCD}\,, \quad \,\,\, \sigma_{\pi N} \propto \Lambda_\text{QCD} \cdot \left( \frac{\hat{m}}{\Lambda_\text{QCD}}\right)\,, \quad \,\,\, \Delta \sigma \propto \Lambda_\text{QCD} \cdot \left(\frac{\delta m}{\Lambda_\text{QCD}}\right)\,,  \quad \,\,\, C_{p,n} \propto \alphaem \cdot \Lambda_\text{QCD}\,.
\eea
The nuclear binding energy is modeled by the 
Bethe--Weizs\"acker semi-empirical mass formula~(SEMF)~\cite{Weizsacker:1935bkz,Bethe:1936zz,Blatt:1952ije,kaplan1955nuclear}, given by
\bea
\label{eq:binding_energy_five_term_appx}
(E_\calA)_\bind = \underbrace{\calE_\text{vol} \, A}_{(E_\calA)_\text{vol}} + \underbrace{\calE_\text{surf} A^{2/3}}_{(E_\calA)_\text{surf}} + \underbrace{\calE_\text{asym} \frac{(A-2Z)^2}{A}}_{(E_\calA)_\text{asym}} + \underbrace{\calE_\text{em} \frac{Z(Z-1)}{A^{1/3}}}_{(E_\calA)_\text{em}} + \cdots\,.
\eea
In the above equation, the subscripts ``$\text{vol}$'', ``$\text{surf}$'', ``$\text{asym}$'', and ``$\text{em}$'' represent the volume, surface, asymmetry, and Coulomb~(electrostatic) energy contributions, respectively. 
The ellipsis ``$\cdots$'' denotes the pairing energy, which is subdominant and therefore neglected. 
The $(Z,A)$ dependence of each term can be understood through simple physical arguments. 
The volume term is the binding energy which $A$ nucleus in the bulk contribute, so $(E_\calA)_\text{vol}<0$ and $(E_\calA)_\text{vol} \propto A$. 
The surface term is the correction to the volume term, since the nucleons at the nuclear surface are not attracted by the bulk nucleons, giving $(E_\mathcal{A})_\text{surf} > 0$ and $(E_\mathcal{A})_\text{surf} \propto A^{2/3}$. 
The asymmetric term comes from the asymmetric Pauli exclusion of the number of proton and neutron in the nuclei. When $Z \neq N$, the proton and neutron Fermi surfaces are different, giving the dominant contribution to the energy difference. In addition, $\rho$-meson exchange generates an extra but subdominant contribution when $Z \neq N$~\cite{Serot:1984ey,Serot:1997xg}. 
Therefore, we have $(E_\calA)_\text{asym} > 0$. 
The factor $[(N-Z)/A]^2$ indicates the lowest order expansion, since the linear expansion in terms of $(N-Z)/A$ vanishes because the nuclear force is invariant under the isospin symmetry $n \leftrightarrow p$~($N \leftrightarrow Z$). Multiplying the extra bulk extensivity factor $A$ gives $(E_\calA)_\text{asym} \propto A \times [(N-Z)/A]^2\propto (A-2Z)^2/A$. The Coulomb energy term describes the repulsion among protons inside the nucleus. Therefore, we have $(E_\calA)_\text{em} > 0$. The numerator $Z(Z-1)$ counts
the number of distinct proton pairs, obtained by subtracting the unphysical self-energy of each proton from the naive $Z^2$ counting. The denominator $A^{1/3}$ reflects the fact that the Coulomb energy is inversely proportional to the nuclear radius
$(R_\mathcal{A})_\text{nucleus} \propto A^{1/3}$.

We now discuss the dependence of each coefficient $\mathcal{E}$ on $\alpha_\text{em}$ and $m_\pi$, where the latter follows the Gell-Mann--Oakes--Renner relation $m_\pi^2 \propto \hat{m}$~\cite{Gell-Mann:1968hlm,Scherer:2002tk}. From Ref.~\cite{Donoghue:2006du}, the volume term $\mathcal{E}_\text{vol}$ and surface term $\mathcal{E}_\text{surf}$ both arise from the nuclear central force mediated by pion exchange, and therefore both vary with $m_\pi^2$. The asymmetry 
term $\mathcal{E}_\text{asym}$ depends on the Fermi momentum $k_F$, which is set by the nuclear saturation density $n_\text{sat}$~\cite{Serot:1984ey,Serot:1997xg}. Since the saturation condition is determined by the balance of nuclear forces governed by pion exchange, $k_F$ 
inherits a dependence on $m_\pi^2$~\cite{Donoghue:2006du}. The Coulomb term is given by $(E_\mathcal{A})_\text{em} = (3/5)\,\alpha_\text{em}\,Z(Z-1)/(R_\mathcal{A})_\text{nucleus}$~\cite{Weizsacker:1935bkz,Bethe:1936zz,Blatt:1952ije,kaplan1955nuclear}, where $(R_\mathcal{A})_\text{nucleus} = A^{1/3}\,r_0$. This follows from modeling the nucleus as a uniformly charged liquid drop, with $r_0$ the empirical nucleon radius. This gives $\mathcal{E}_\text{em} = 
(3/5)\,\alpha_\text{em}/r_0 \propto \alphaem$. Furthermore, the relation $r_0\,k_F = (9\pi/8)^{1/3}$ follows from combining the nuclear saturation density $n_\text{sat} = 1/(\frac{4 \pi}{3} \,r_0^3)$ with the degenerate Fermi gas expression $n_\text{sat} = \gamma_N \int_{|\veck| \leq k_F} \frac{\dd^3 \veck}{(2\pi)^3}$, where $\gamma_N = 2 \times 2$ accounts for the spin of both proton and neutron. Since $k_F$ is governed by $m_\pi^2$ through the saturation condition, $r_0$ inherits the same dependence, and consequently $\mathcal{E}_\text{em}$ acquires an implicit dependence on $m_\pi^2$.

\subsection{Dilaton Charge: Analytical Expressions}\label{subsec:dilaton_charge}

In this section, we derive the dilaton charge for the coupling scenarios discussed in this work: the effective scalar--SM coupling, the Higgs portal, the universal coupling, and the light QCD axion. 
The analytical results for the effective scalar--SM coupling are consistent with Ref.~\cite{Damour:2010rp}, but is expressed in a more compact form that makes individual contributions explicit and readily accommodates updated QCD and nuclear inputs.
We also define $F_\mathcal{A} \equiv
A\,m_\text{amu}/m_\mathcal{A} \simeq 1$,~\footnote{This ratio accounts for the deviation of the atomic mass from the naive estimate $A\,m_\text{amu}$ due to nuclear binding, neutron-proton mass differences, and electron masses with ordering $\mathcal{O}(\mev)$. The deviation from unity is therefore suppressed as $F_\mathcal{A} - 1 \sim \mathcal{O}(10^{-3})$, and can be safely set to $F_\mathcal{A} \simeq 1$ in the computation of the dilaton charge.} where $m_\text{amu} \equiv m_{{}^{12}\text{C}}/12 \simeq
931.49~\text{MeV}$~\cite{ParticleDataGroup:2022pth}.

We begin with the effective scalar--SM coupling introduced in \Eq{Ldamour} of \Subsec{effective_coupling}. Combining \Eq{m_atom_mass_appx}, \Eq{rest_mass_appx}, and \Eq{binding_energy_five_term_appx}, we obtain the atomic mass $m_\calA$. Substituting $m_\calA$ into \Eq{dilaton_charge_def_appx} and using the scaling relations given in \Eq{proton_neutron_mass_appx}, we find
\bea
\label{eq:dilaton_charge_scalar_SM_appx}
& (Q_\calA)_{e} = F_\calA \frac{1}{m_\text{amu}} \left[ C_n +  (C_p-C_n)\frac{Z}{A} + \calE_\text{em} \frac{Z(Z-1)}{A^{4/3}} \right] \,, \quad \quad \quad \quad \quad \text{(Effective Scalar-SM, \Subsec{effective_coupling})}\,,\\
& (Q_\calA)_{m_e} = F_\calA \frac{m_e}{m_\text{amu}} \frac{Z}{A}\,,\\
& (Q_\calA)_{\hat{m}} =  F_\calA \frac{1}{m_\text{amu}} \left[  \sigma_{\pi N} + \frac{\partial \, \calE_\text{vol}}{\partial \, \log m_\pi^2} + \frac{\partial \, \calE_\text{surf}}{\partial \, \log m_\pi^2} \frac{1}{A^{1/3}} + \frac{\partial \, \calE_\text{asym}}{\partial \, \log m_\pi^2} \frac{(A-2Z)^2}{A^2} + \frac{\partial \, \calE_\text{em}}{\partial \log m_\pi^2} \frac{Z(Z-1)}{A^{4/3}} \right]\,,\\
& (Q_\calA)_{\delta m} = F_\calA \frac{\Delta \sigma}{2\,m_\text{amu}} \frac{A-2Z}{A} \,.
\eea
The dilaton charge $(Q_\mathcal{A})_e$ for the variation of $\alpha_\text{em}$ receives two contributions: the electromagnetic self-energies of the proton and neutron (the first two terms), and the Coulomb energy in the nuclear binding energy (the third term). The dilaton charge 
$(Q_\mathcal{A})_{m_e}$ for the electron mass variation is proportional to the electron number fraction $Z/A$ in ordinary matter. This dilaton charge is actually equivalent to doing replacement $\bar{\psi}_e \psi_e \rightarrow  \langle \bar{\psi}_e \psi_e \rangle = n_e$ to include the matter effect in the Lagrangian directly. The dilaton charge $(Q_\mathcal{A})_{\hat{m}}$ for the symmetric quark mass variation has two main contributions: the pion-mass dependence of the nucleon mass, encoded in the $\sigma_{\pi N}$ term, and the pion-mass dependence of the nuclear binding energy coefficients, encoded in the remaining terms. Finally, the dilaton charge $(Q_\mathcal{A})_{\delta m}$ for the asymmetric quark mass variation arises solely from isospin violation ($\delta m = m_d - m_u \neq 0$). Here, nuclear binding contributes no additional term here because the nuclear force maintains isospin symmetry. After acquiring the dilaton charge given by \Eq{dilaton_charge_scalar_SM_appx}, one can compute the total effective scalar-SM coupling $\alpha_\calA^{(2)}$ using 
\Eq{coupling_dilaton_charge_relation_appx}.

Secondly, we discuss the dilaton charge for the Higgs portal of \Subsec{higgs_portal}, where the scalar-SM interaction $\alpha^{(2)}_\calA$ is determined by the single UV parameter $d_H^{(2)}$. Substituting \Eq{low_energy_dh} into \Eq{coupling_dilaton_charge_relation_appx}, then we have
\bea
\label{eq:Higgs_Dilaton_Charge_appx}
\alpha_\calA^{(2)} = (Q_{\calA})_H\, d_H^{(2)}\,, \quad   (Q_{\calA})_H = \frac{2}{9} + \frac{\alphaem}{\pi} (Q_{\calA})_e + \frac{7}{9}
\left[ (Q_{\calA})_{m_e}  + (Q_{\calA})_{\hat{m}} + (Q_{\calA})_{\delta m} \right] \,\,\, (\text{Higgs Portal, \Subsec{higgs_portal}})\,,
\eea
where $(Q_\calA)_H$ is the dilaton charge associated with the Higgs portal. The factor $2/9$ is universal for all materials and therefore cancels in the differential dilaton charge between two test masses in EP tests. However, this factor still contributes to the dilaton charge of the source mass and dominates over the other contributions. We also see that the contribution from $(Q_{\calA})_e$ is suppressed by the one-loop factor $\alphaem/\pi$ arising from the $h\gamma\gamma$ vertex, as discussed in \Subsec{higgs_portal}, and can therefore be neglected when computing the dilaton charge.

Thirdly, we discuss the universal coupling of \Subsec{other_coupling}, in which the scalar-SM coupling originates from the $\phi$-dependent conformal rescaling of the spacetime metric. 
In this scenario, the conformal transformation is controlled by a single parameter $d_\text{univ}$. Substituting \Eq{low_energy_universal} into \Eq{coupling_dilaton_charge_relation_appx}, we obtain
\bea
\label{eq:Universal_Dilaton_Charge_appx}
\alpha_\calA^{(2)} = (Q_{\calA})_\text{univ} \,\, d_\text{univ}^{(2)}\,, \quad \,\, (Q_{\calA})_\text{univ} = 1 \quad \quad \text{(Universal Coupling, \Subsec{other_coupling})}\,.
\eea
$(Q_{\calA})_\text{univ}$ is the dilaton charge for the universal coupling, corresponding to the case in which the effective scalar--SM coupling is identical for all mass terms. Therefore, the equivalence principle is strictly preserved, as illustrated in Refs.~\cite{Fujii:1996td,Hui:2010dn,Armendariz-Picon:2011ydk,Nitti:2012ev,Sibiryakov:2020eir}.

Finally, we discuss the charge for the light QCD axion from \Subsec{other_coupling}. Based on the QCD-axion-induced quark field rotation derived in Refs.~\cite{Ubaldi:2008nf,GrillidiCortona:2015jxo,Kumamoto:2024wjd}, we have
\bea
\label{eq:axion_quark_mass_vary_appx}
\frac{\Delta m_\pi^2}{m_\pi^2} = \frac{\Delta \hat{m}}{\hat{m}} = \mathcal{F}_\phi(\theta_\phi) - 1\,, \quad \frac{\Delta \delta m}{\delta m} = \frac{ \mathcal{F}_\phi(\theta_\phi)-1}{\mathcal{F}_\phi(\theta_\phi)}\,, \quad \quad \,\, \text{where \,\, $\mathcal{F}_\phi(\theta_\phi) = \sqrt{1-\beta_{ud} \sin^2\left(\frac{\theta_\phi}{2}\right)}$}\,.
\eea
In \Eq{axion_quark_mass_vary_appx}, $\theta_\phi = \phi/f_\phi$ is the axion misalignment angle, and $\beta_{ud} = 4 \, m_u m_d/(m_u+m_d)^2 \simeq 0.88$. Substituting $\Delta m_\pi^2/m_\pi^2$ and $\Delta \delta m/\delta m$ from \Eq{axion_quark_mass_vary_appx} into \Eq{proton_neutron_mass_appx} gives the proton and neutron rest masses as functions of $\theta_\phi$. Similarly, substituting $\Delta m_\pi^2/m_\pi^2$ into
\Eq{binding_energy_five_term_appx} gives the nuclear binding energy as a function of $\theta_\phi$. The QCD axion charge and its analytic representation can then be defined in direct analogy with the dilaton case. Defining $(Q_\calA)_\text{axion} \equiv \partial \log m_\calA/\partial \, (\delta \theta^2_\phi/2)$, where $\delta \theta_\phi \equiv \delta \phi/f_\phi$ is the deviation from the extrema, $\theta_\phi=0$~(maximum) and $\theta_\phi=\pi$~(minimum), we obtain~
\bea
\label{eq:axion_charge_appx}
\begin{aligned}
& \alpha^{(2)}_\calA |_{\theta_\phi=0,\pi} = (Q_{\calA})_\text{axion}|_{\theta_\phi=0,\pi} \,\, \frac{\Mpl^2}{4\pi f_\phi^2}\,, \quad \quad \quad \quad \quad \quad \quad \quad \quad \quad \quad \quad \text{(Light QCD Axion, \Subsec{other_coupling})}\,,\\
& (Q_{\calA})_\text{axion} |_{\theta_\phi = 0} = - F_\calA \frac{\beta_{ud}}{4 \, m_\text{amu}} \bigg[ \left( \sigma_{\pi N} + \Delta \sigma \frac{A-2Z}{2A} \right) \\
& \quad \quad \quad  \quad \quad \quad \quad \,\,\, + \left( \frac{\partial \, \calE_\text{vol}}{\partial \, \log m_\pi^2} + \frac{\partial \, \calE_\text{surf}}{\partial \, \log m_\pi^2} \frac{1}{A^{1/3}} + \frac{\partial \, \calE_\text{asym}}{\partial \, \log m_\pi^2} \frac{(A-2Z)^2}{A^2} + \frac{\partial \, \calE_\text{em}}{\partial \log m_\pi^2} \frac{Z(Z-1)}{A^{4/3}}  \right) \bigg]\,,\\
& (Q_{\calA})_\text{axion} |_{\theta_\phi = \pi} = F_\calA \frac{\beta_{ud}}{4 \sqrt{1-\beta_{ud}}\, m_\text{amu}} \bigg[ \left( \sigma_{\pi N} + \frac{\Delta \sigma}{1-\beta_{ud}} \frac{A-2Z}{2A} \right) \\
& \quad \quad \quad  \quad \quad \quad \quad \,\,\, + \left( \frac{\partial \, \calE_\text{vol}}{\partial \, \log m_\pi^2} + \frac{\partial \, \calE_\text{surf}}{\partial \, \log m_\pi^2} \frac{1}{A^{1/3}} + \frac{\partial \, \calE_\text{asym}}{\partial \, \log m_\pi^2} \frac{(A-2Z)^2}{A^2} + \frac{\partial \, \calE_\text{em}}{\partial \log m_\pi^2} \frac{Z(Z-1)}{A^{4/3}}  \right) \bigg]\,.
\end{aligned}
\eea
In the equations listed above, $(Q_{\calA})_\text{axion} |_{\theta_\phi = 0}$ is the axion charge evaluated at $\theta_\phi = 0$. Since this corresponds to the hilltop of the potential, it explains the negative sign of $(Q_{\calA})_\text{axion}|_{\theta_\phi = 0}$. For the $\theta_\phi=0$ case, we verify that the contribution from the nuclear rest mass is consistent with Ref.~\cite{Kumamoto:2024wjd}, and the contribution from the nuclear binding energy is consistent with Refs.~\cite{Gue:2024onx,Bauer:2024hfv}. By contrast, $(Q_\calA)_\text{axion}|_{\theta_\phi=\pi}$ is the axion charge evaluated at $\theta_\phi = \pi$. Since this corresponds to the potential minimum, $(Q_{\calA})_\text{axion}|_{\theta_\phi = \pi}$ carries a positive sign. This quantity can be used to characterize the axion effective mass after the axion phase transition has occurred. From \Eq{axion_charge_appx}, $\sigma_{\pi N}$, despite being one of the dominant 
contributions to the absolute matter effect,~\footnote{The $\sigma_{\pi N}$ term arises from approximating the quadratic axion--SM coupling in the finite-density medium as $a^2 \langle \bar{\boldsymbol{N}} \boldsymbol{N} \rangle \simeq a^2 \, n_{\boldsymbol{N}}$~\cite{Hook:2017psm,Balkin:2022qer}, where $n_{\boldsymbol{N}}$ is the nucleon number density.} cancels universally when computing the differential acceleration of two test masses. Therefore, axion-induced EP violation cannot be properly accounted for without the isospin-violating term $\Delta \sigma$ and the nuclear binding energy terms $\varE_\text{surf}$, $\varE_\text{asym}$, $\varE_\text{em}$.

\subsection{Nuclear Physics Inputs}
\label{subsec:nuclear_inputs}

In the previous section, we derived the analytical expressions for the dilaton charges. We now combine these expressions with inputs from nuclear and hadronic physics to obtain their numerical values. Since the numerical formulas most commonly used in the literature were introduced in Refs.~\cite{Damour:2010rp,Damour:2010rm} in 2010, it is useful to identify which are limited by the underlying model assumptions and which can be updated with modern data. 

We start by introducing the dependence of nuclear binding energy on the pion mass. Because the following discussion is built on the point-like contact interactions in the model of Refs.~\cite{Damour:2010rp,Donoghue:2006du}, it cannot be further updated within this framework. From Refs.~\cite{Damour:2010rp,Donoghue:2006du}, such pion-mass dependence is given by
\bea
\label{eq:bind_mpisq_derivative_appx}
\begin{array}{rcl@{\;}l@{\qquad\qquad}rcl@{\;}l@{\;}l}
\displaystyle
\frac{\partial \varE_{\rm vol}}{\partial \log m_\pi^2}
&=&
\displaystyle
\frac{\partial \varE_{\rm vol}}{\partial \log G_S}\,
&
\displaystyle
\frac{\partial \log G_S}{\partial \log m_\pi^2}\,,
&
\displaystyle
\frac{\partial \varE_{\rm asym}}{\partial \log m_\pi^2}
&=&
\displaystyle
\varE_{\rm asym}^{\rm (SEMF)}
&
\displaystyle
\frac{\partial \log \varE_{\rm asym}}{\partial \log G_S}\,
&
\displaystyle
\frac{\partial \log G_S}{\partial \log m_\pi^2}\,,
\\[3mm]
\displaystyle
\frac{\partial \varE_{\rm surf}}{\partial \log m_\pi^2}
&=&
\displaystyle
\frac{\partial \varE_{\rm surf}}{\partial \log G_S}\,
&
\displaystyle
\frac{\partial \log G_S}{\partial \log m_\pi^2}\,,
&
\displaystyle
\frac{\partial \varE_{\rm em}}{\partial \log m_\pi^2}
&=&
\displaystyle
\varE_{\rm em}^{\rm (SEMF)}
&
\displaystyle
\frac{\partial \log k_F}{\partial \log G_S}\,
&
\displaystyle
\frac{\partial \log G_S}{\partial \log m_\pi^2}\,.
\end{array}
\eea
In the above equations, $G_S$ is the coupling strength of the four-fermion contact interaction $\bar{\boldsymbol{N}}\boldsymbol{N}\bar{\boldsymbol{N}}\boldsymbol{N}$ with $\boldsymbol{N} = (p, n)$, parametrizing the scalar channel of the nucleon--nucleon interaction. Because this effective coupling encodes pion-exchange contributions to the nuclear force, $G_S$ depends on the pion mass. The Coulomb energy $\varE_{\rm em}$ is proportional to $1/r_0 \propto k_F$, where the Fermi momentum $k_F$ increases as the nuclear interaction $G_S$ strengthens, thereby acquiring an implicit dependence on the pion mass. To evaluate the above pion-mass dependencies, we firstly use the following numerical inputs
\bea
\label{eq:GS_appx}
\frac{\partial \,\varE_\text{vol}}{\partial \log G_S} \simeq -120\,\mev\,, \quad \,\, \frac{\partial \, \varE_\text{surf}}{\partial \log G_S} \simeq  97\,\mev\,, \quad \,\, \frac{\partial \log k_F}{\partial \log G_S} \simeq 0.525\,, \quad\,\, \frac{\partial \log G_S}{\partial \log m_\pi^2}
\simeq -0.35\,.
\eea
The numerical values in \Eq{GS_appx} can be found in Refs.~\cite{Damour:2010rp,Donoghue:2006du}. To evaluate $\partial \log \varE_\text{asym}/\partial \log G_S$, we need
\bea
\label{eq:calE_asym_appx}
\calE_\text{asym} = \frac{k_F^2}{6 \sqrt{M_*^2 + k_F^2}}+\frac{G_\rho k_F^3}{12 \pi^2}\,, \quad \quad M_* = m_{N,0} + \frac{\gamma_N G_S k_F^3}{6 \pi^2}\,, \quad G_S \simeq -355.4~{\rm GeV}^{-2}\,, \,\,\, G_\rho \simeq 15\, \gev^{-2}\,,
\eea
where $M_*$ is the effective nucleon mass, and
$G_\rho = g_\rho^2/m_\rho^2$ denotes the coupling strength in the isovector
$\rho$-meson channel. Here, $g_\rho \simeq 3$ is the corresponding $\rho \boldsymbol{N}\boldsymbol{N}$ coupling strength, while $m_\rho \simeq 775\,\mev$ is the $\rho$-meson mass~\cite{ParticleDataGroup:2024cfk}.  For $\varE_\text{asym}$ given by \Eq{calE_asym_appx}, we take the derivative in terms of $G_S$ using the chain rules, then we have $\partial \log \varE_\text{asym}/\partial \log G_S \simeq 2.4$, which is consistent with Ref.~\cite{Damour:2010rp}. Note that the numerical values given in \Eq{GS_appx} and \Eq{calE_asym_appx} depends on the model assuming the nucleon-nucleon contact interaction and nucleon is point-like particle based on the effective theory arised by Serot and Walecka~\cite{Serot:1984ey,Serot:1997xg}. Therefore, there is currently no generally accepted modern determination that clearly supersedes the Damour and Donoghue~(DD 2010) values~\cite{Damour:2010rp,Damour:2010rm}.

To eventually determine the values of \Eq{bind_mpisq_derivative_appx}, we also need the following standard textbook values
\bea
\label{eq:SEMF_kF_r0_appx}
\varE^{\text{(SEMF)}}_\text{asym} \simeq 23 \,\mev\,, \quad \quad \varE^{\text{(SEMF)}}_\text{em} \simeq 0.72 \,\mev\,, \quad \quad k_F = 1.3 \,\text{fm}^{-1}\,, \quad \quad k_F \, r_0  = \left(\frac{9\pi}{8}\right)^{1/3}\,.
\eea
The coefficients $\varE_\text{asym}^\text{(SEMF)}$ and
$\varE_\text{em}^\text{(SEMF)}$ are obtained by
fitting the SEMF model to nucleus masses, whose standard
textbook values can be found in
Refs.~\cite{kaplan1955nuclear,Hodgson:1997,Damour:2010rp}. 
Although modern mass-table fits can shift $\varE_\text{asym}^\text{(SEMF)}$ and
$\varE_\text{em}^\text{(SEMF)}$ mildly, the changes
are typically at the $\mathcal{O}(10\%)$ and $\mathcal{O}(1\%)$ level for asymmetric term and Coulomb term, respectively. Moreover, the value of $k_F$ could be updated by the modern nuclear saturation density measurement through $k_F = (3\pi^2 n_\text{sat}/2)^{1/3}$, but the change of $k_F$ is typically at the $\mathcal{O}(1\%)$ level. Given the small size of these updates, we use the values quoted in \Eq{SEMF_kF_r0_appx} directly. Substituting \Eq{GS_appx}, \Eq{SEMF_kF_r0_appx}, and previously acquired $\partial \log \varE_\text{asym}/\partial \log G_S \simeq 2.4$ into \Eq{bind_mpisq_derivative_appx}, we have
\bea
\label{eq:bind_mpisq_derivative_num_appx}
\frac{\partial \varE_{\rm vol}}{\partial \log m_\pi^2} \simeq 42 \, \mev\,, \quad \frac{\partial \varE_{\rm surf}}{\partial \log m_\pi^2} \simeq -34 \, \mev\,, \quad \frac{\partial \varE_{\rm asym}}{\partial \log m_\pi^2} \simeq -20 \,\mev\,, \quad \frac{\partial \varE_{\rm em}}{\partial \log m_\pi^2} \simeq -0.13 \,\mev\,,
\eea
which is consistent with the numerical results given by Ref.~\cite{Damour:2010rp}. Based on the above discussion, we know that \Eq{bind_mpisq_derivative_num_appx} encodes the model assumption given by Serot and Walecka~\cite{Serot:1984ey,Serot:1997xg} and there are no universally accepted modern values, so we continue using these values for the nuclear binding energy derivatives in dilaton charge computation.

\begin{table}[t]
\centering
\renewcommand{\arraystretch}{1.7}
\setlength{\tabcolsep}{10pt}
\begin{tabular}{|c|c|ccc|c|}
\hline
 & $\sigma_{\pi N}\,\text{[MeV]}$ & $C_n - C_p\,\text{[MeV]}$ & $\Delta\sigma\,\text{[MeV]}$ & $m_n - m_p\,\text{[MeV]}$ & $C_n\,\text{[MeV]}$ \\
\hline
DD 2010~\cite{Damour:2010rp,Damour:2010rm}   & $45$~\cite{Gasser:1990ce} & $-0.76$~\cite{Gasser:1982ap} & $3.10$~\cite{Damour:2010rp} & $2.34$~$\cancel{\text{BBN}}$ & $-0.13$~\cite{Gasser:1982ap} \\
This Work    & $\simeq 60$~\cite{Hoferichter:2015dsa,Gupta:2021ahb,FLAG:2024oxs} & $-0.87$\,\cite{Borsanyi:2014jba,ParticleDataGroup:2024cfk} & $2.16$\,\cite{Borsanyi:2014jba,ParticleDataGroup:2024cfk} & $1.29$~\cite{ParticleDataGroup:2024cfk} & $-0.15$~(rescaled) \\
\hline
\end{tabular}
\caption{Comparison of the main hadronic inputs adopted by Damour and 
Donoghue~\cite{Damour:2010rm,Damour:2010rp}, denoted DD 2010, with those 
used in this work. For This Work, we extract $C_n - C_p$ and $\Delta\sigma$ 
by combining the QCD/QED ratio $(m_n-m_p)_{\text{QCD}}/(m_n-m_p)_{\text{QED}}$ 
reported by the BMW Collaboration~\cite{Borsanyi:2014jba} with the experimental value of $m_n-m_p$~\cite{ParticleDataGroup:2024cfk}. The 
value of $C_n$ is obtained by rescaling its DD 2010 value while keeping 
the ratio $C_n/C_p$ fixed. Note that the DD 2010 value of $m_n-m_p$ 
is inconsistent with the experimental value used in standard 
BBN, motivating the updated inputs adopted here.}
\label{tab:asymmetry_parameters}
\end{table}

Unlike the quantities above, which describe the pion-mass dependence of nuclear binding energies, several hadronic parameters can be updated, as summarized in Table~\ref{tab:asymmetry_parameters}.
We firstly use the updated $\sigma_{\pi N} \simeq 60\,\mev$~\cite{Hoferichter:2015dsa,Gupta:2021ahb,FLAG:2024oxs} compared with $45\,\mev$ used in Damour and Donoghue~(DD 2010)~\cite{Damour:2010rp,Damour:2010rm}. For the pion--nucleon sigma term $\sigma_{\pi N}$, two benchmark values, $\sigma_{\pi N}\simeq 40\,\text{MeV}$ and $\sigma_{\pi N}\simeq 60\,\text{MeV}$, are commonly quoted, reflecting a long-standing tension between direct lattice-QCD calculations and dispersive analyses of pion-nucleon scattering data. Recent reanalyses suggest that this tension could be alleviated once excited-state contamination is properly accounted for, favoring the higher value~\cite{Hoferichter:2015dsa,Gupta:2021ahb}, which we adopt here. For the isospin-breaking parameters $C_n-C_p$ and $\Delta \sigma$, we use the first-principles lattice QCD+QED result $(m_n-m_p)_\text{QCD}/(m_n-m_p)_\text{QED} = \Delta \sigma/(C_n-C_p) \simeq -2.49$ from the BMW collaboration~\cite{Borsanyi:2014jba}, together with the measured neutron--proton mass difference, $m_n-m_p = 1.29\,\text{MeV}$~\cite{ParticleDataGroup:2024cfk}~\footnote{Alternative determinations of $C_n - C_p$ via the dispersive method~\cite{Walker-Loud:2012ift,Thomas:2014dxa} differ from our adopted value by at most $\sim 30\%$. While the BMW Collaboration also provides absolute predictions for $(m_n-m_p)_\text{QED}$ and $(m_n-m_p)_\text{QCD}$~\cite{Borsanyi:2014jba}, their sum does not match the measured $m_n-m_p$ exactly. We therefore use only the BMW ratio of QCD to QED contributions, anchored by the experimental value of $m_n-m_p$~\cite{ParticleDataGroup:2024cfk}.}. By contrast, the original DD 2010 values yield $m_n - m_p \simeq 2.34\,\text{MeV}$ (marked by $\cancel{\text{BBN}}$), nearly twice the measured value. Such a large neutron--proton mass splitting would substantially distort the neutron-to-proton freeze-out ratio and the primordial helium abundance, bringing it into conflict with Big Bang Nucleosynthesis constraints~\cite{Sibiryakov:2020eir,Bouley:2022eer}. In the absence of a modern update for the individual self-energies $C_n$ and $C_p$, we rescale $C_n$ keeping the ratio $C_n/C_p$ fixed to its DD 2010 value. This choice has negligible numerical impact, since $C_n$ enters $(Q_\mathcal{A})_e$ only as a subdominant contribution and is thus insensitive to variations at the $\mathcal{O}(10\%)$ level. Using the updated numbers given by Table~\ref{tab:asymmetry_parameters}, we have the numerical form of the dilaton charges of the effective scalar-SM couplings:
\bea
\label{eq:Qe_num_appx}
(Q_{\calA})_{e} 
&= \underbrace{-1.6 \times 10^{-4} \left[\frac{C_n}{-0.15\,\mev}\right]}_{\text{DD 2010}:\,-1.4 \times 10^{-4}} 
 + \underbrace{9.4 \times 10^{-4} \left[\frac{C_n - C_p}{-0.87\,\mev}\right]}_{\text{DD 2010}:\,8.2 \times 10^{-4}} \frac{Z}{A} 
 + 7.7\times 10^{-4}\left[\frac{\varE^{\text{(SEMF)}}_\text{em}}{0.72 \,\mev}\right] \, \frac{Z(Z-1)}{A^{4/3}}\,,
\eea
\bea
\label{eq:Qme_num_appx}
 (Q_{\calA})_{m_e} 
&= 5.5 \times 10^{-4} \left[\frac{m_e}{0.511\,\mev}\right]\, \frac{Z}{A}\,,
\eea
\bea
\label{eq:Qmhat_num_appx}
(Q_{\calA})_{\mhat} 
&= \underbrace{0.064 \left[\frac{\sigma_{\pi N}}{60\,\mev}\right] + 0.045\left[ \frac{\frac{\partial \varE_{\rm vol}}{\partial \log m_\pi^2}}{42\,\mev}\right]}_{\text{DD 2010}:\,0.093} 
 - 0.036\,\left[\frac{\frac{\partial \varE_{\rm surf}}{\partial \log m_\pi^2}}{-34\,\mev}\right]\, \frac{1}{A^{1/3}} \\
 & \quad - 0.02\,\left[\frac{\frac{\partial \varE_{\rm asym}}{\partial \log m_\pi^2}}{-20 \,\mev}\right] \frac{(A - 2Z)^2}{A^2} 
 - 1.4 \times 10^{-4}\,\left[\frac{\frac{\partial \varE_{\rm em}}{\partial \log m_\pi^2}}{-0.13\,\mev}\right] \,\frac{Z(Z-1)}{A^{4/3}}\,, \\[8pt]
\eea
\bea
\label{eq:Qdeltam_num_appx}
(Q_{\calA})_{\delta m} 
&= \underbrace{0.0012 \left[\frac{\Delta\sigma}{2.16\,\mev}\right]}_{\text{DD 2010}:\,0.0017}\, \frac{A - 2Z}{A}\,.
\eea
In \Eq{Qe_num_appx}--\Eq{Qdeltam_num_appx}, unlike the purely numerical results presented in DD 2010~\cite{Damour:2010rm,Damour:2010rp}, we keep the dependence on the underlying particle- and nuclear-physics inputs explicit. This is useful because these hadronic and nuclear inputs carry uncertainties and may be further updated in future work. Using the formulas above, together with the numerical inputs in \Eq{bind_mpisq_derivative_num_appx} and Table~\ref{tab:asymmetry_parameters}, we can also compute the Higgs-portal dilaton charge defined in \Eq{Higgs_Dilaton_Charge_appx} and the axion charge defined in \Eq{axion_charge_appx}. Throughout the rest of this work, we adopt the benchmark values listed in the row labeled ``This Work'' in Table~\ref{tab:asymmetry_parameters}. The corresponding dilaton charges for the Earth and the MICROSCOPE test masses are presented in Table~\ref{tab:dilaton_charges_table}. As a consistency check, using the numerical values labeled ``DD 2010'' in Table~\ref{tab:asymmetry_parameters}, we reproduce the original dilaton charges reported in DD
2010~\cite{Damour:2010rm,Damour:2010rp}. For terms whose numerical values differ from those in DD 2010, we indicate the corresponding DD 2010 values with underbrackets for comparison.

\section{Spherically Symmetric Ansatz}\label{appx:sph_ansatz_appx}

In \Subsec{sph_symmetric_main}, we derived $\psi_\sph$ by taking the $s$-wave limit $k_0 R_\oplus \ll 1$. In this appendix, we review the spherically symmetric ansatz, as given by Refs.~\cite{Hees:2018fpg,Berezhiani:2018oxf,Banerjee:2022sqg}, and discuss the screening effect using this analytical formalism. Taking the zero-momentum limit $|\veck| \to 0$, we have $E_\eff(\veck) \to 0$. Therefore, \Eq{Schrodinger_Eq} can be written as
\bea
\label{eq:sph_phi_eq_appx}
\frac{1}{r^2} \frac{\dd}{\dd r}\left( r^2 \frac{\dd \psi_\sph}{\dd r}\right) = \mMearth^2 \, \theta(R_\oplus-r) \, \psi_\sph\,,
\eea
where $\psi_\sph$ is independent of the angular coordinates. The regular solution satisfying $\psi_\sph(r\to\infty)=|\psi_0|$ is
\bea
\label{eq:psi_sph_appx}
\psi_\sph(r) = 
\abs{\psi_0} \times 
\left\{\begin{aligned}
& 1 - \mathcal{A} \, \frac{R_\oplus}{r}  \quad &(r \geq R_\oplus)\\
& \mathcal{B} \, \frac{\sinh(\mMearth \, r)}{\mMearth \, r}  \quad &(r < R_\oplus)
\end{aligned}
\right. \,,
\eea
where $|\psi_0| = \sqrt{2 \rho_\phi}/m_\phi$. Matching $\psi_\sph$ and $\dd\psi_\sph/\dd r$ at the surface $r=R_\oplus$ determines the coefficients as follows:
\bea
\label{eq:psi_sph_coefficients_appx}
\mathcal{A} = \frac{\mMearth R_\oplus - \tanh(\mMearth R_\oplus)}{\mMearth R_\oplus} = \frac{\mMearth^2 \mathcal{V}_\oplus}{4\pi R_\oplus}  J_+(\mMearth R_\oplus)\,, \quad \quad \quad \mathcal{B} = \frac{1}{\cosh(\mMearth R_\oplus)}\,.
\eea
This reproduces the spherical configuration given by Refs.~\cite{Hees:2018fpg,Berezhiani:2018oxf,Banerjee:2022sqg}. To compare with the $s$-wave approximation given in \Eq{A0_main}, \Eq{A0_lowk0_appx}, and \Eq{B0_lowk0_appx}, we have $\mathcal{A} \simeq i A_0/(k R_\oplus)$ and $\mathcal{B} \simeq B_0$. In the exterior solution, the constant term corresponds to the incident plane wave in the zero-momentum limit, while the term proportional to $\mathcal{A}$ represents the scattered-wave contribution. Substituting $\psi_\sph$ into \Eq{phi_NR_approx}, we obtain the full time-dependent scalar configuration, $\phi_\sph(t,r) = \psi_\sph(r) \cos(m_\phi t)$. Therefore, we have $\langle \phi_\sph^2 \rangle = \psi_\sph^2(r)\, \langle \cos^2(m_\phi t) \rangle= \psi_\sph^2(r)/2$. We can compute the background-induced potential using $V_\bg = (\mMtest^2 \mathcal{V}_\testmass/4) (\psi^2_\sph - \abs{\psi_0}^2)$ following \Eq{Vbg_from_geodesic}.

For $\mMearth R_\oplus \lesssim 1$, the spherically symmetric system lies in the perturbative region~(A) of \Fig{classification}. In this limit, we have the scalar configuration
\bea
\mMearth \, R_\oplus \lesssim 1: \quad \quad \psi_\sph(r) \simeq 
\abs{\psi_0} \, \left\{
\begin{aligned}
& 1-\frac{\mMearth^2\mathcal{V}_\oplus}{4\pi r} & \quad \,\, (r\geq R_\oplus)\\
& 1-\xi(r) \, \frac{\mMearth^2\mathcal{V}_\oplus}{4\pi R_\oplus} & \quad \,\, (r < R_\oplus)
\end{aligned}
\right. \,.
\eea
This shows that the deviation of $\psi_\sph$ from its asymptotic value $\abs{\psi_0}$ remains perturbative. 
The matter-induced effective mass of the Earth, $\mMearth^2$, quantifies the coupling strength, while $\calV_\oplus$ captures the coherent contribution of the entire Earth volume to the scattered wave. Outside the Earth, the scalar perturbation decays as $1/r$. Inside the Earth, the perturbation has an extra $\mathcal O(1)$ spatial profile factor encoded in
$\xi(r)=3/2-(r/R_\oplus)^2/2$, with $\xi(0)=3/2$ and $\xi(R_\oplus)=1$. Using the exterior profile $\psi_\sph(r>R_\oplus)$, the background-induced potential and force are
\bea
\label{eq:Vbg_Fbg_sph_A_appx}
\mMearth \, R_\oplus \lesssim 1: \quad \quad V_\bg \simeq - \frac{\rho_\phi}{m_\phi^2} \frac{(\mMtest^2\mathcal{V}_\testmass)(\mMearth^2\mathcal{V}_\oplus)}{4 \pi r}\,, \quad \quad \, \vecF_\bg \simeq - \frac{\rho_\phi}{m_\phi^2} \frac{(\mMtest^2\mathcal{V}_\testmass)(\mMearth^2\mathcal{V}_\oplus)}{4\pi r^2} \, \hat{\vecr}\,.
\eea
Comparing this expression with \Eq{background_potential}, we see that it corresponds to the perturbative regime in which the form factor satisfies $\formfactorV \simeq 1$. In this regime, the background-induced potential and force are not suppressed by screening, and the full volume of the Earth contributes coherently.

For $\mMearth R_\oplus \gtrsim 1$, the system enters the non-perturbative screening region~(C) of \Fig{classification}. In this regime, the scalar configuration becomes
\bea
\mMearth \, R_\oplus \gtrsim 1: \quad \quad \psi_\sph(r) \simeq 
\abs{\psi_0} \left\{
\begin{aligned}
& 1-\frac{R_\oplus}{r} + \frac{1}{\mMearth r}
    && \quad (r > R_\oplus)\\
& \frac{1}{\mMearth R_\oplus}
    && \quad (r = R_\oplus)\\
& \frac{e^{-\mMearth(R_\oplus-r)}}{\mMearth \, r}
    && \quad (\mMearth^{-1} \lesssim r < R_\oplus)\\
& 2 e^{-\mMearth R_\oplus}
    && \quad (r < \mMearth^{-1})
\end{aligned}
\right. \,.
\eea
Away from the Earth's surface where $r>R_\oplus$, the scattered-wave contribution is then approximately controlled by the geometric factor $R_\oplus/r$, and becomes insensitive to the microscopic coupling strength $\mMearth^2$. At the surface of the Earth, one obtains the factor $1/(\mMearth \, R_\oplus)$. This shows that the scalar field is suppressed at the surface due to the screening effect induced by the potential barrier from the Earth. Inside the Earth, namely for $\mMearth^{-1} \lesssim r<R_\oplus$, the scalar field decays exponentially with the penetration depth $\mMearth^{-1}$. Using the exterior profile $\psi_\sph(r>R_\oplus)$, we have
\bea
\label{eq:Vbg_Fbg_sph_C_appx}
\mMearth \, R_\oplus \gtrsim 1: \quad \quad V_\bg \simeq - \frac{\rho_\phi}{m_\phi^2} \frac{(\mMtest^2\mathcal{V}_\testmass) R_\oplus}{r} \left(1-\frac{R_\oplus}{2r}\right), \quad \quad \, \vecF_\bg \simeq - \frac{\rho_\phi}{m_\phi^2} \frac{(\mMtest^2 \mathcal{V}_\testmass) R_\oplus}{r^2} \left(1-\frac{R_\oplus}{r}\right) \hat{\vecr}\,,
\eea
which are consistent with the results given in \Subsec{sph_symmetric_main}. From \Eq{Vbg_Fbg_sph_C_appx}, we see that both the background-induced potential and force are independent of $\mMearth$. From one perspective, this behavior follows from the fact that, once the system enters the hard-sphere regime, the scalar profile outside the Earth saturates and becomes insensitive to any further increase in the coupling strength. From another equivalent perspective, this can be understood as a manifestation of the screening effect discussed in \Subsec{scattering} and \Subsec{sph_symmetric_main}. To see this explicitly, we can compare \Eq{Vbg_Fbg_sph_A_appx} with \Eq{Vbg_Fbg_sph_C_appx}. Relative to the perturbative result in \Eq{Vbg_Fbg_sph_A_appx}, the screened potential and force are suppressed by the characteristic factor $3/(\mMearth R_\oplus)^2$, up to the geometric factors shown explicitly in \Eq{Vbg_Fbg_sph_C_appx}.

This ansatz assumes that, far from the Earth, the scalar field is a single, homogeneous, zero-momentum mode $\abs{\psi_0}$. In reality the field is part of the virialized Milky Way halo, and since the Solar System moves through it, the field carries a non-zero velocity relative to the Earth and is described by a distribution of momentum modes, e.g., a boosted Maxwell--Boltzmann distribution. For $k_0 R_\oplus \ll 1$ this distinction is inconsequential: each mode is effectively homogeneous over the Earth and has the same spherical configuration, so the ansatz remains accurate. Once $k_0 R_\oplus \gtrsim 1$, i.e., $m_\phi \gtrsim 4 \times 10^{-11}\,\eV$, it fails on two aspects---for each mode the $s$-wave approximation breaks down, so higher angular-momentum $l$-modes must be retained and the scattered profile becomes non-spherical; and different modes are now distributed differently, so the field must be averaged over its momentum distribution. The former calls for a partial-wave treatment and the latter for phase-space averaging over $\veck$. 
Both are needed to compute the constraint and the distinctive multi-band signal discussed in this work.

\section{Partial Wave Analysis}\label{appx:partial_wave_appx}
In this section, we introduce the partial wave analysis that we implement to calculate the field distribution around the scattering object. Unlike the spherically symmetric case discussed in~\Appx{sph_ansatz_appx}, our discussion here applies to both the low-momentum region of $k_0 R_\oplus \ll 1$ and the high momentum region of $k_0 R_\oplus \gtrsim 1$. We first present the partial wave solution of the field from a monochromatic incident wave, then the multipole expansion of the field amplitude. We then present the phase space integration of the field amplitude and finish this section with results in different limits to compare with previous results~\cite{Gan:2025nlu,Brzeminski:2026rox}.

\subsection{Monochromatic Wave Function}

A monochromatic incident plane wave with a specified momentum $\veck$ can be expanded in spherical coordinates as
\bea
\label{eq:inc_wave_appx}
\psiinc(\vecr;\veck) = \abs{\psi_0} e^{i\veck \cdot \vecr} = \abs{\psi_0} \,\sum^{\infty}_{l=0} (2l+1)\, i^l \, j_l(k r) \, P_l(\cos\theta).
\eea
When an incident plane wave scatters off a potential barrier, such as that induced by the Earth, the total wavefunction contains both the incident wave and an outgoing scattered component. For a system with azimuthal symmetry, the scattered wave can be written as
\bea
\label{eq:sca_wave_appx}
\psisc(\vecr;\veck) = \abs{\psi_0} \,\sum^{\infty}_{l=0} (2l+1)\, i^l \, A_l \, h_l(k r) \, P_l(\cos\theta)\,.
\eea
As shown in \Fig{frame}, $\theta$ denotes the angle between the position vector $\vecr$ and the incident momentum $\veck$. $h_l = j_l + i y_l$ are the spherical Hankel functions of the first kind, $j_l$ are the Bessel functions, and $y_l$ are the spherical Neumann functions. 
$A_l$ is the amplitude of the scattered wave, which can be written as 
\bea
\label{eq:Al_Sl_deltal_appx}
A_l =  \frac{S_l - 1}{2} = \frac{e^{2i\delta_l}-1}{2}\,.
\eea
$S_l$ is the S-matrix of the $l-$th component of the partial wave, and $\delta_l$ is the phase shift. The total wavefunction outside the scatterer is obtained by adding the incident and scattered components,
\bea
\label{eq:psiout_appx}
\psiout(\vecr;\veck) = \psiinc(\vecr;\veck) + \psisc(\vecr;\veck) = \abs{\psi_0} \sum_{l=0}^{\infty} (2l+1) \, i^l \, \calR_l(kr) \, P_l(\cos\theta) \quad \quad (r\geq R_\oplus)\,.
\eea
$\calR_l(kr)$ denotes the radial component of each partial wave, as given by
\bea
\label{eq:R_l_out_appx}
\calR_l(kr) = j_l(kr) + A_l \, h_l(kr) \quad \quad (r\geq R_\oplus)\,.
\eea
Generally speaking, the wave function outside depends on the potential and it needs to be solved numerically. However, if the scattering object is modeled as a uniform sphere, analytical expressions can be obtained for both the interior and exterior wave functions.
We start from the Schr\"odinger equation inside the object 
\bea
\label{eq:Schrodinger_Eq_int_appx}
- \frac{1}{2 m_\phi} \vecnabla^2 \psiint + V_{\eff,\oplus}\,(\vecr) \, \psiint = E_{\eff}(\veck) \, \psiint\,,
\eea
where the uniform potential and the effective kinetic energy are defined respectively as
\bea
V_{\eff, \oplus}\,(\vecr) = \frac{\mMearth^2}{2 \,m_\phi}\, \quad \text{and}\quad \,\,\, E_{\eff}(\veck) = \frac{\veck^2}{2 \,m_\phi}\,,
\eea
which gives
\bea
(\vecnabla^2 + k^2_\oplus) \psiint = 0\, \quad \text{with}\quad k^2_\oplus = k^2 - \mMearth^2\,.
\eea
Considering the convergence at $r = 0$, the solution of the differential equation in a uniform sphere can be expanded in partial waves as
\bea
\label{eq:int_wave_appx}
\psiint(\vecr;\veck) = \abs{\psi_0} \,\sum^{\infty}_{l=0} (2l+1)\, i^l \, B_l \, j_l(k_\oplus r) \, P_l(\cos\theta)\,,
\eea
where $B_l$ is the amplitude of the interior wave function. By matching the outside and interior wave function at $r = R_\oplus$, we solve the amplitudes $A_l$ and $B_l$,
\bea
\label{eq:Al_Bl_appx}
& A_l = - \frac{k \, j_l(k_\oplus R_\oplus) \, j_{l+1}(k R_\oplus) - k_\oplus \, j_l(k R_\oplus) \, j_{l+1}(k_\oplus R_\oplus) }{ k \, j_l(k_\oplus R_\oplus) \, h_{l+1}(k R_\oplus) - k_\oplus \, h_l(k R_\oplus) \, j_{l+1}(k_\oplus R_\oplus) }\,,\\
& B_l = \frac{k \, j_l(k R_\oplus) \, h_{l+1}(k R_\oplus) - k \, h_l(k R_\oplus) \, j_{l+1}(k R_\oplus) }{ k \, j_l(k_\oplus R_\oplus) \, h_{l+1}(k R_\oplus) - k_\oplus \, h_l(k R_\oplus) \, j_{l+1}(k_\oplus R_\oplus) }\,.
\eea
The phase shift is given by
\bea
\label{eq:tan_delta_l_solid_sphere_appx}
\tan \delta_l = \frac{A_l/i}{A_l+1} = \frac{k \, j_l(k_\oplus R_\oplus) \, j_{l+1}(k R_\oplus) - k_\oplus \, j_l(k R_\oplus) \, j_{l+1}(k_\oplus R_\oplus)}{ k \, j_l(k_\oplus R_\oplus) \, y_{l+1}(k R_\oplus) - k_\oplus \, y_l(k R_\oplus) \, j_{l+1}(k_\oplus R_\oplus) }\,.
\eea
In the low-momentum limit, $kR_\oplus \ll 1$, the amplitudes of the exterior and interior wavefunctions are dominated by $s$-wave component~($l=0$) and can be expanded in powers of $kR_\oplus$. Specifically, the exterior amplitude from the $s$-wave component is
\bea
\label{eq:A0_lowk0_appx}
A_0 = -i \frac{\mMearth R_\oplus - \tanh(\mMearth R_\oplus)}{\mMearth R_\oplus} \, kR_\oplus - \left(\frac{\mMearth R_\oplus - \tanh(\mMearth R_\oplus)}{\mMearth R_\oplus}\right)^2 \, (k R_\oplus)^2+ \cdots\,,
\eea
while the interior amplitude from the $s$-wave component is
\bea
\label{eq:B0_lowk0_appx}
B_0 = \frac{1}{\cosh(\mMearth R_\oplus)} + \frac{-i}{\cosh(\mMearth R_\oplus)}\left(\frac{\mMearth R_\oplus - \tanh(\mMearth R_\oplus)}{\mMearth R_\oplus}\right) (k R_\oplus) + \cdots\,.
\eea

Another useful limit is the hard sphere limit, where $\mMearth \gg k$ and $\mMearth R_\oplus \gg 1$. In this scenario, the field is exponentially suppressed inside the sphere, and the outside wave function is well approximated by the hard-wall boundary condition $\calR_l(kR_\oplus)=0$. Using \Eq{R_l_out_appx}, this gives
\bea
\label{eq:Al_Bl_hardsphere_appx}
j_l(kR_\oplus)+A_l h_l(kR_\oplus)=0\,,
\qquad
A_l=-\frac{j_l(kR_\oplus)}{h_l(kR_\oplus)}\,, \qquad B_l = 0.
\eea
Using \Eq{Al_Sl_deltal_appx}, the phase shift in this limit satisfies
\bea
\label{eq:phase_shift_hardsphere_appx}
\tan\delta_l=\frac{j_l(kR_\oplus)}{y_l(kR_\oplus)}\,.
\eea
The hard-sphere expressions for $A_l$ and $B_l$ in \Eq{Al_Bl_hardsphere_appx}, together with the phase shift $\delta_l$ in \Eq{phase_shift_hardsphere_appx}, can be recovered by taking the limit $\mMearth R_\oplus \gg 1$ in \Eq{Al_Bl_appx} and using $j_l(k_\oplus R_\oplus) \simeq i^l e^{\mMearth R_\oplus}/2 \mMearth R_\oplus$.

\subsection{Phase Space Integration}

With the general expression \Eq{psiout_appx} of the outside field, we calculate the square of the monochromatic field and integrate it with respect to the Standard Halo Model (SHM). We begin by simplifying the squared of a monochromatic wavefunction with fixed incident momentum $\veck$. Specifically, we have
\bea
\label{eq:phi_squared_appx}
\abs{\psi(\vecr;\veck)}^2 = \abs{\psi_0}^2 \sum_{l=0}^{\infty} \sum_{l'=0}^{\infty} (2l+1)(2l'+1) \, i^{l-l'} \, \calR_{l}(kr) \, \calR_{l'}^*(kr) \, P_{l}(\cos\theta) \, P_{l'}(\cos\theta)\,.
\eea
Using the Clebsch--Gordan decomposition of spherical harmonics, the product of two Legendre polynomials can be expanded as a linear combination of Legendre polynomials:
\bea
\label{eq:Two_Pl_CG_appx}
P_{l}(\cos\theta)\,P_{l'}(\cos\theta)
  = \sum_{L=\abs{l-l'}}^{l+l'}
    \underbrace{\langle l\,0\,l'\,0 \,|\, L\,0\rangle^{2}}_{\mathcal{C}_{l\,l'\,L}}\,
    P_{L}(\cos\theta)\,.
\eea
Here the coefficients $\mathcal{C}_{l l' L}$ are determined by the Clebsch--Gordan
coefficients and are nonzero only when the selection rules
\bea
\label{eq:CG_selection_rule_appx}
\abs{l-l'} \leq L \leq l+l', \quad \quad l+l'+L \,\, \text{even}.
\eea
are satisfied. From Ref.~\cite{edmonds1996angular}, these coefficients are given explicitly by
\begin{equation}
\label{eq:linear_coefficient_appx}
\displaystyle
\begin{aligned} \\
\mathcal{C}_{l l' L} = \underbrace{\frac{1+(-1)^J}{2}}_{\text{Zero for odd $J$}}  \times (2L+1) \, \frac{(J-2l)! \, (J-2l')! \, (J-2L)!}{(J+1)!}  \left[\frac{\left(\frac{J}{2}\right)!}{\left(\frac{J}{2} - l\right)! \left(\frac{J}{2} - l'\right)! \left(\frac{J}{2} - L\right)!}\right]^2
\end{aligned}\,,
\end{equation}
where $ J = l + l' + L$. Another useful equivalent expression of $\mathcal{C}_{l l' L}$ to simplify the numerical calculation and analytical derivation is
\bea
\label{eq:linear_coefficient_appx_simple}
\mathcal{C}_{l,\,l+L-2p,\,L} = K_{L,p} \times \frac{1}{2(l-p)+1} \times  \displaystyle\prod_{i=1}^{L} \frac{2(l-p+i)}{2(l-p+i)+1}.
\eea
Note that \Eq{linear_coefficient_appx_simple} also makes explicit which coefficients $\mathcal{C}_{l l' L}$ in \Eq{linear_coefficient_appx} are nonzero. All remaining
coefficients vanish by the Clebsch--Gordan selection rules. Here, $K_{L,p}$ is defined as
\bea
K_{L,p} \equiv \frac{2L+1}{2^L} \, \mathscr{E}_p \, \mathscr{E}_{L-p}, \quad  \quad \quad \text{where} \,\quad \mathscr{E}_p \equiv \frac{(2p)!}{2^p \,(p!)^2}=\frac{\binom{2p}{p}}{2^p}. 
\eea
\begin{figure}[t!]
\centering
\includegraphics[width=0.490\linewidth]{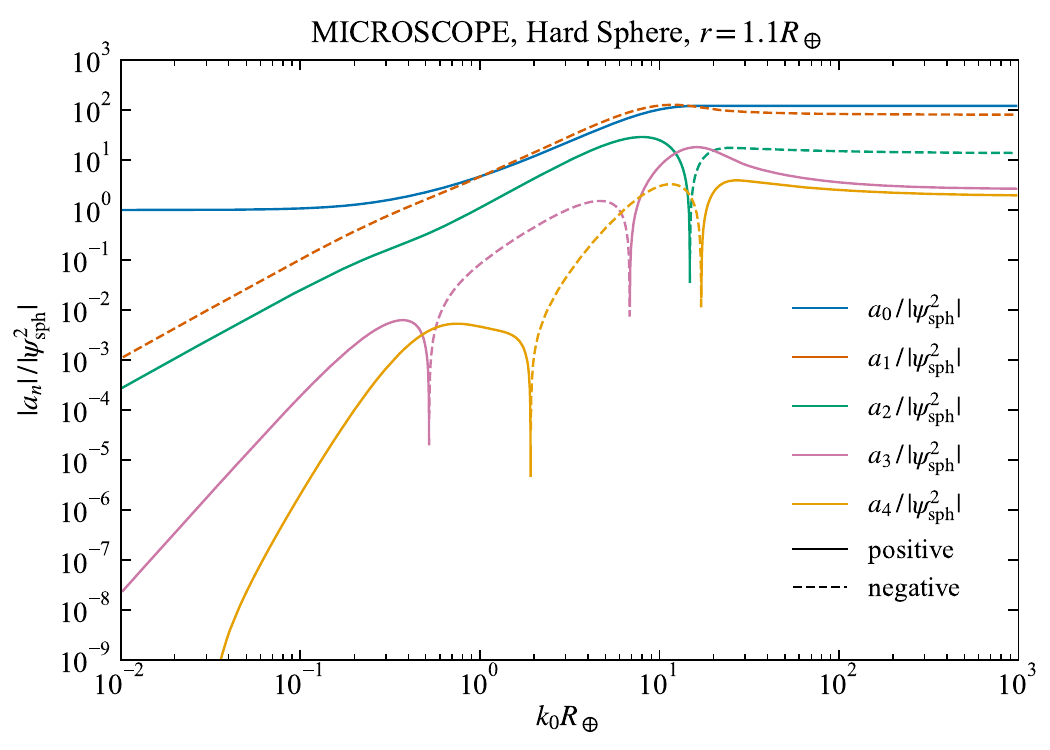}
\includegraphics[width=0.490\linewidth]{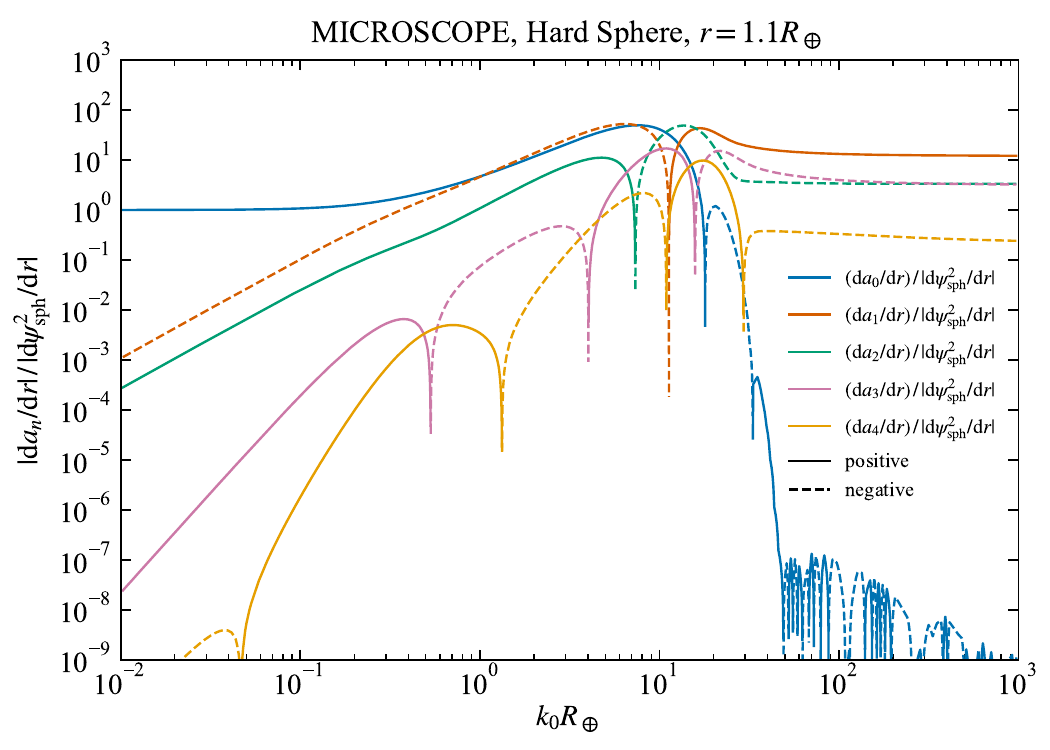}
\caption{The $a_L$-series and derivatives, same as \Fig{hard_sphere_a_series} and \Fig{hard_sphere_a_series_derivatives} but in log-scales after taking absolute values, where the solid (dashed) lines are for positive (negative) values. We use the $k_\esc \rightarrow \infty$ limit as the difference between using a finite $k_\esc$ versus an infinite cutoff does not affect our results. Note that $d a_0/dr$ is strongly suppressed when $k_0 R_\oplus \gg 10$, while other $da_L/dr$ with $L \geq 1$ tends to a constant. This behavior can be understood from the optical limit, where $a_0$ approaches a constant and therefore $d a_0/dr$ vanishes, as shown in \Appx{geometric_optics}.}
\label{fig:hard_sphere_a_series_log10_appx}
\end{figure}

Using the Legendre-product decomposition in
\Eq{Two_Pl_CG_appx}, \Eq{phi_squared_appx} becomes
\bea
\label{eq:phi_squared_2_appx}
\abs{\psi(\vecr;\veck)}^2 
= \abs{\psi_0}^2 \sum_{l=0}^{\infty} \sum_{l'=0}^{\infty} (2l+1)(2l'+1) \, i^{l-l'} \, \calR_{l}(kr) \, \calR_{l'}^*(kr) \, \sum_{L=\abs{l-l'}}^{l+l'} \mathcal{C}_{l \, l' \, L} P_L(\cos\theta)\,.
\eea
We can see that $\abs{\psi(\vecr;\veck)}^2$ can be written in terms of linear expansion of the Legendre polynomials,
\bea
\label{eq:mono_k_in_c_appx}
\abs{\psi(\vecr;\veck)}^2 = \abs{\psi_0}^2 \sum_{L=0}^\infty c_L(r;\,k,\mMearth) P_L(\cos\theta)\,,
\eea
where the coefficient $c_L$ is given by
\bea
\label{eq:cL_start_appx}
c_L(r;\,k,\mMearth) \equiv \sum_{l=0}^\infty \sum_{l'=0}^\infty (2l+1)(2l'+1) \, i^{l-l'} \, \calR_{l}(kr) \, \calR_{l'}^*(kr) \, \mathcal{C}_{l \, l' \, L}\,.
\eea
Starting from \Eq{cL_start_appx}, we keep
\bea
0 \leq p \leq \left\lfloor \frac{L}{2} \right\rfloor
\eea
based on the Clebsch--Gordan selection rules in \Eq{CG_selection_rule_appx}, and then symmetrize the summand under the interchange $l\leftrightarrow l'$ 
through the two assignments
\bea
\left\{
\begin{aligned}
& l  \rightarrow l \\
& l' \rightarrow l + L - 2p
\end{aligned}
\right.
\quad
\text{and}
\quad
\left\{
\begin{aligned}
& l  \rightarrow l + L - 2p \\
& l' \rightarrow l
\end{aligned}
\right. ,
\eea
we finally get the following general representation:
\bea
\label{eq:cL_generic_L_appx}
c_L(r;k,\mMearth) 
& = \sum_{l=0}^{\infty} \sum_{p=0}^{\left\lfloor L/2 \right\rfloor} [2l+1][2(l+L-2p)+1] \, \mathcal{C}_{l,\,(l+L-2p),\,L} \, (2-\delta_{L-2p,0}) \cdot \Re\left[i^{L-2p} \, \calR_{l+L-2p} \, \calR^*_l\right],
\eea
which enables efficient numerical computation to arbitrary orders in $L$. Here, the factor $\delta_{L-2p,0}$ prevents double counting of the diagonal terms. As examples, we list the explicit expressions for $c_L$ in \Eq{cL_generic_L_appx} for $L=0,\cdots,3$:
\begin{equation}
\label{eq:cn_low_orders}
\begin{aligned}
c_0(r;\,k,\mMearth) &= \sum_{l=0}^{\infty} (2l+1)\,\abs{\calR_l}^2\,, \\[6pt]
c_1(r;\,k,\mMearth) &= -6\sum_{l=0}^{\infty} (l+1)\,\Im\!\left(\calR_{l+1}\calR_l^{*}\right)\,, \\[6pt]
c_2(r;\,k,\mMearth) &= \sum_{l=0}^{\infty} \frac{5\,(2l+1)\,l\,(l+1)}{(2l+3)(2l-1)}\,\abs{\calR_l}^2
 - \sum_{l=0}^{\infty} \frac{15\,(l+2)(l+1)}{2l+3}\,\Re\!\left[\calR_{l+2}\calR_l^{*}\right]\,, \\[6pt]
c_3(r;k,\mMearth) &= - \sum_{l=0}^{+\infty} \frac{21 \, (l+2) \, (l+1)\, l }{(2l+5)\,(2l-1)} \Im\left[\calR_{l+1} \calR^*_l\right] + \sum_{l=0}^{+\infty} \frac{35 \, (l+3)\,(l+2)\,(l+1)}{(2l+5)\,(2l+3)} \Im\left[\calR_{l+3} \calR^*_l \right]\,, ...\\[6pt].
\end{aligned}
\end{equation}

We note that the results above focus on simplifying the squared amplitude of a monochromatic wavefunction with a single incident momentum mode $\veck$. To obtain the ensemble-averaged scalar profile $\langle |\psi|^2\rangle$, and hence the background-induced potential $V_\bg$, we must integrate over the DM phase-space distribution given in \Eq{phase_space}. As we will see below, this averaging procedure, in particular the integration over the azimuthal angle $\phi_\veck$, provides the second simplification step toward the final computational method. To prepare for the phase-space averaging, we recall from \Eq{phase_space} that the DM distribution in the Solar System is described by the SHM, a truncated Maxwell–Boltzmann distribution, which is
\bea
f_\phi(\veck) = \frac{n_\phi }{\mathcal{N}(k_\esc)} \cdot \left(\frac{2\pi}{\sigma_k^2}\right)^{3/2} \exp\left[ - \frac{(\veck-\veck_0)^2}{2\sigma_k^2} \right] \Theta(k_\esc - \abs{\veck-\veck_0})\,.
\eea
For convenience, we follow \Fig{frame} and define the spherical coordinate system with $\veck_0$ as the $z-$axis. Then we have
\bea
\veckhat = (\sin\theta_\veck \cos\phi_\veck,\, \sin\theta_\veck \sin\phi_\veck,\, \cos\theta_\veck)\,, \quad \quad 
\vecrhat = (\sin\theta_\vecr\,, 0 \,, \cos\theta_\vecr)\,.
\eea
The angle between $\hat{\veck}$ and $\hat{\vecr}$ is then given by
\bea
\cos\theta = \hat{\veck} \cdot \hat{\vecr} = \sin\theta_\veck \cos\phi_\veck \sin\theta_\vecr + \cos\theta_\veck \cos\theta_\vecr\,.
\eea
To find the expectation value of $\abs{\psi}^2$, we integrate \Eq{mono_k_in_c_appx} over $f_\phi(\veck)$,
\bea
\label{eq:phase_space_int_appx}
\langle \abs{\psi}^2 \rangle 
= & \abs{\psi_0}^2 \sum_{L=0}^\infty \, \frac{1}{n_\phi} \, \int \dbar^3 \veck \, f_\phi(\veck) \, c_L(r;\,k,\mMearth) \, P_L(\cos\theta) \\
= & \abs{\psi_0}^2 \frac{1}{\mathcal{N}(k_\esc)} \left(\frac{1}{2\pi \sigmak^2}\right)^{3/2} \sum_{L=0}^\infty \int_0^{+\infty} \dd k \, k^2 \, \, \exp\left[-\frac{k^2+k_0^2}{2\sigmak^2}\right] c_L(r;\,k,\mMearth) \\
& \times \int_{-1}^{+1} \dd \cos\theta_\veck \, \exp\left[\frac{k k_0 \cos\theta_\veck}{\sigmak^2}\right] \Theta(k_\esc - \abs{\veck-\veck_0}) \, \times \, \underbrace{\int_0^{2\pi} \dd\phi_\veck \, P_L(\cos\theta)}_{\text{$\hat{\veck}$-$\hat{\vecr}$ separation}}\,,
\eea

\begin{figure}[t!]
\centering
\includegraphics[width=0.95\linewidth]{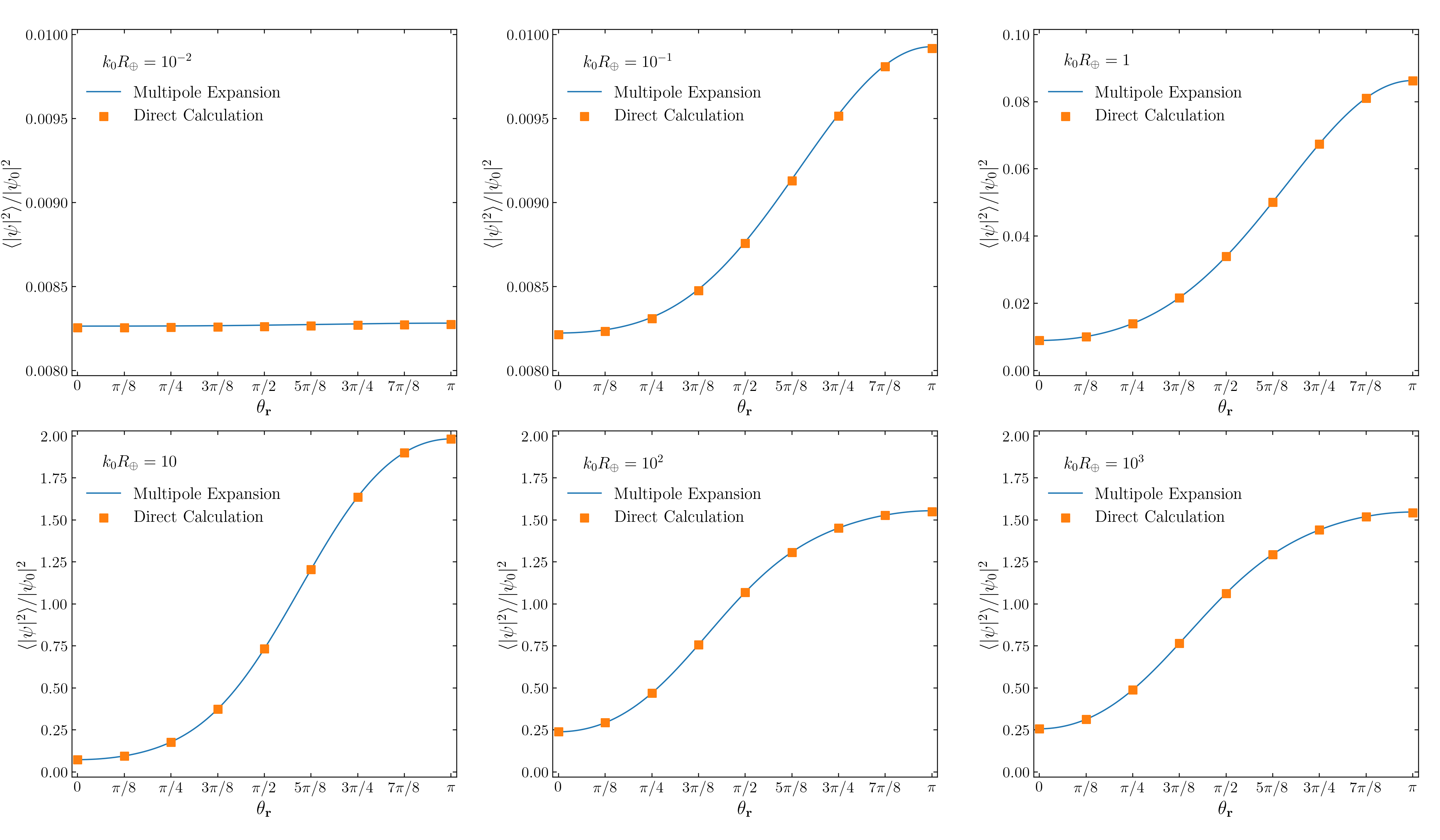}
\caption{The comparison of the $\left<\abs{\psi}^2\right>$ calculation between multipole expansion (blue curve) and direct calculation (orange squares).}
\label{fig:direct_vs_ensemble_multipole_comparison}
\end{figure}
Another important step comes from the simplification of the integration $\int_0^{2 \pi} d\phi_\veck P_L(\cos\theta)$. Using the Legendre addition theorem, we have
\bea
P_L(\underbrace{\cos\theta}_{\hat{\veck}\cdot\hat{\vecr}}) = \frac{4\pi}{2L+1} \sum_{M=-L}^{L} Y_{LM}(\hat{\vecr}) \, Y^*_{LM}(\hat{\veck}),
\eea
Because
\bea
\int_0^{2\pi} \dd \phi_\veck Y^*_{LM}(\hat{\veck})  = \sqrt{\frac{2L+1}{4\pi}} P_{L}(\cos\theta_\veck) \times 2\pi \delta_{M0},
\eea
we have
\bea
\label{eq:Legendre_phik_int}
\text{$\hat{\veck}$-$\hat{\vecr}$ separation}: \quad \quad \int_0^{2\pi} \dd\phi_\veck \, P_L(\cos\theta) = 2\pi \, P_L(\cos\theta_\veck) \, P_{L}(\cos\theta_\vecr)\,.
\eea
After applying \Eq{Legendre_phik_int} to \Eq{phase_space_int_appx} and factoring out $P_L(\cos\theta_\vecr)$, we obtain
\bea\label{eq:field_profile_from_a_appx}
\langle \abs{\psi}^2 \rangle = \abs{\psi_0}^2 \sum_{L=0}^{+\infty} a_L(r;\,k_0,\mMearth) \, P_L(\cos\theta_\vecr)\,.
\eea
Here $a_L$ is the multipole coefficient used throughout this work, given by
\bea
\label{eq:an_full_appx}
a_L(r;\,k_0,\mMearth)
& = \frac{1}{\mathcal{N}(k_\esc)} \left(\frac{1}{2\pi \sigmak^2}\right)^{3/2} \int_0^{+\infty} \dd k \, k^2 \, \, \exp\left[-\frac{k^2+k_0^2}{2\sigmak^2}\right] c_L(r;\,k,\mMearth)\\
& \quad \times 2\pi \int_{-1}^{+1} \dd \cos\theta_\veck \, \exp\left[\frac{k k_0 \cos\theta_\veck}{\sigmak^2}\right] P_L(\cos\theta_\veck) \,\Theta(k_\esc-\abs{\veck-\veck_0})\,.
\eea
In the limit where $k_\esc \gg \sigma_k$, using
\bea
\int_{0}^{\pi} \dd\theta_\veck \, \sin\theta_\veck \exp\left[\frac{k k_0 \cos\theta_\veck}{\sigmak^2}\right] P_L(\cos\theta_\veck) = 2 \, i_L\left(\frac{k k_0}{\sigmak^2}\right),
\eea
where $i_L(x) = \sqrt{\frac{\pi}{2\,x}} I_{L+\frac{1}{2}}(x)$ is the spherical modified Bessel function of the first kind,
we have
\bea
\label{eq:an_nocut_appx}
k_\esc \gg k_0: \quad \,\, a_L(r;k_0,\mMearth) = \left(\frac{1}{2\pi \sigmak^2}\right)^{3/2} 4\pi \int_0^{\infty} \dd k \, k^2 \exp\left[-\frac{k^2+k_0^2}{2 \sigmak^2}\right] c_L(r;k,\mMearth) \, i_L\left(\frac{k k_0}{\sigmak^2}\right),
\eea
which is consistent with \Eq{aL_from_cL_int_simple} in the main text. We now summarize our numerical procedure. First, we use \Eq{cL_generic_L_appx} to generate the analytical representation of $c_L$ up to the required multipole order.~\footnote{For the MICROSCOPE analysis in this work, $L_{\max}=5$ is sufficient. A larger $L_{\max}$ is needed to compute $V_\bg$ at large $r/R_\oplus$, although we have verified that $L_{\max}=8$ is sufficient for most realistic cases.} Second, for each $c_L$, we perform the partial-wave summation over $l$. Here $\calR_l$ contains both the incident plane-wave contribution $\psiinc$ and the scattered-wave contribution $\psisc$. The convergence condition for the scattered wave is $l_{\max}\sim kR_\oplus$, whereas that for the incident plane wave is $l_{\max}\sim kr$. This requires careful numerical treatment: the summation must be taken sufficiently far to capture the incident plane-wave contribution, while avoiding numerical instabilities at excessively large $l$. Third, after obtaining $c_L$ for a fixed $\veck$ through the $l$-summation, we perform the phase-space averaging using either \Eq{an_full_appx} or \Eq{an_nocut_appx}, depending on whether the velocity cutoff is included. Finally, once $a_L$ is obtained, the ensemble-averaged scalar profile $\langle |\psi|^2\rangle$ and the corresponding background-induced potential $V_\bg$ can be constructed. To compute the background-induced force $\vecF_\bg$, one also needs the radial derivatives $\dd a_L/\dd r$, as shown in \Eq{multipole_force_dimless}. Although these derivatives can equivalently be obtained by taking numerical finite differences of $V_\bg$ or $a_L$, the most convenient approach is to first take the analytical radial derivative of $c_L$ in \Eq{cL_generic_L_appx}, and then apply the same numerical procedure used for computing $a_L$.

As a validity check, we verified that, in the parameter regions of interest, the resulting $a_L$ values are negligibly affected by whether the velocity cutoff is included. The reason is that the contribution to the integration from the phase space of $\abs{\veck-\veck_0}>k_\esc$ is suppressed by the Maxwell-Boltzmann distribution ($k_\esc / \sigma_k = 3.23$). As an example, we show the numerical result of $a_L$ and $\dd a_L/\dd r$ in the hard sphere limit, with $r = 1.1 R_\oplus$, in \Fig{hard_sphere_a_series_derivatives}. We also show the results in log-scale in \Fig{hard_sphere_a_series_log10_appx}.

As a cross-check, we perform an additional check to compare the field profile $\langle \abs{\psi}^2 \rangle$ from the $a_L$ calculation and from the direct calculation:
\bea\label{eq:field_profile_brutal_force_appx}
\langle \abs{\psi}^2 \rangle 
=  \frac{1}{n_\phi} \int \dbar^3 \veck \, f_\phi(\veck) \, \abs{\psiout(\vecr;\veck)}^2 \,,
\eea
where the monochromatic wave function comes from the general expression of \Eq{psiout_appx}. To directly calculate the field profile outside, we solve the phase shift $\delta_l$ using \Eq{tan_delta_l_solid_sphere_appx}, sum over the partial waves in \Eq{psiout_appx} and integrate over the phase space according to \Eq{field_profile_brutal_force_appx} directly. We compare the field profile from the multipole expansion \Eq{field_profile_from_a_appx} and from direct calculation, with different $k_0$, at different angle of location $\theta_\vecr$ in \Fig{direct_vs_ensemble_multipole_comparison}. We can see that the results obtained from the two formalisms agree well with each other.

\section{Optical Limit}\label{appx:geometric_optics}

In this section, we show how the calculation of the scalar profile and the corresponding background-induced potential can be performed in the optical limit at high DM masses. 
When the DM wavelength is much smaller than all relevant length scales, in particular the altitude, $h$, and the radius of the Earth, $R_\oplus$, the propagation of the DM field can be treated in the formalism of geometric-optics. 
Before proceeding, we emphasize that this geometric-optics approximation is appropriate when $k_0 R_\oplus, k_0h\gg 1$ and for $\mMearth > k_0$ where the Earth can be treated as a hard sphere.

\begin{figure}[t!]
   \centering
\includegraphics[width=0.47\textwidth]{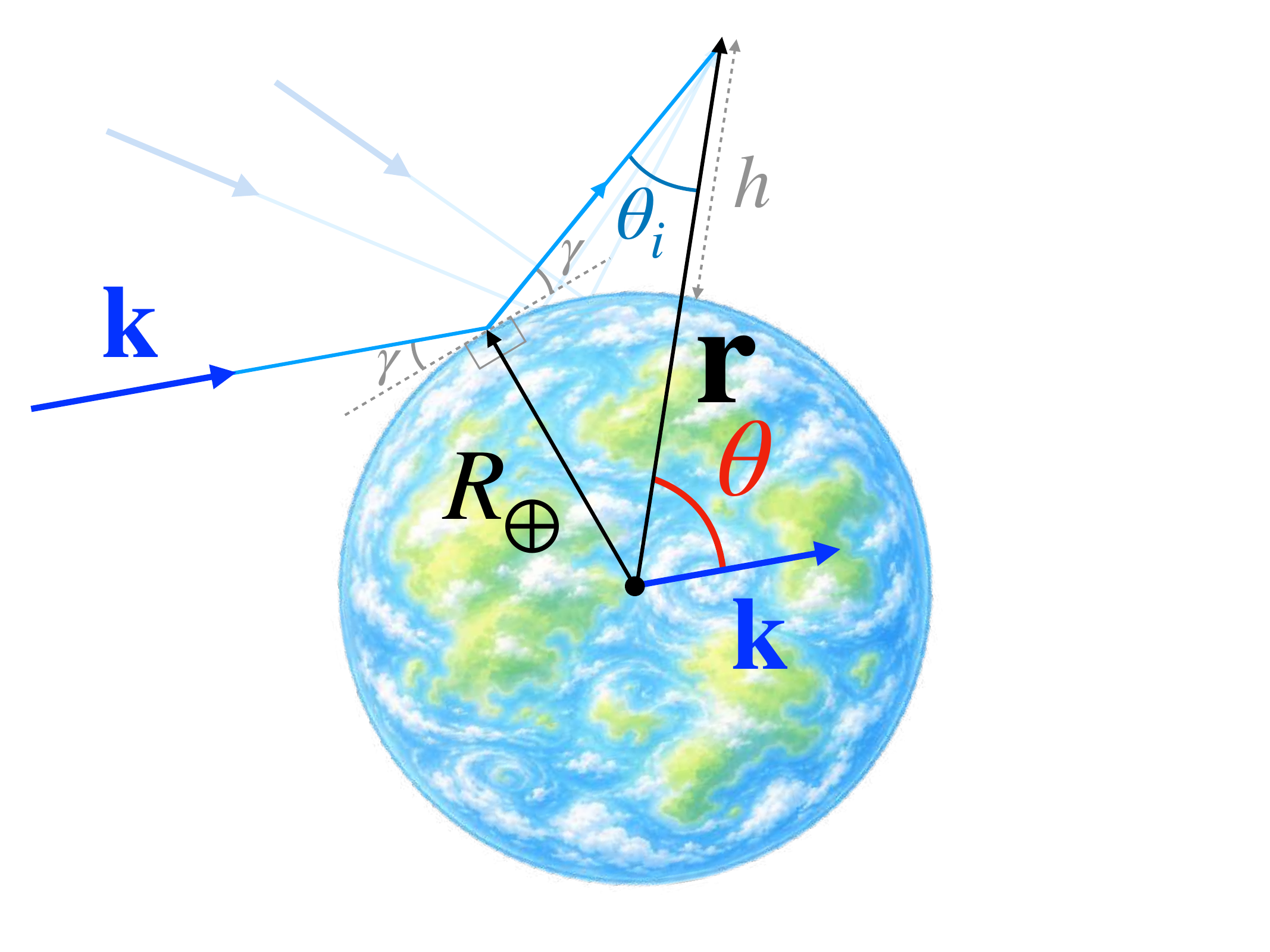}
\includegraphics[width=0.47\textwidth]{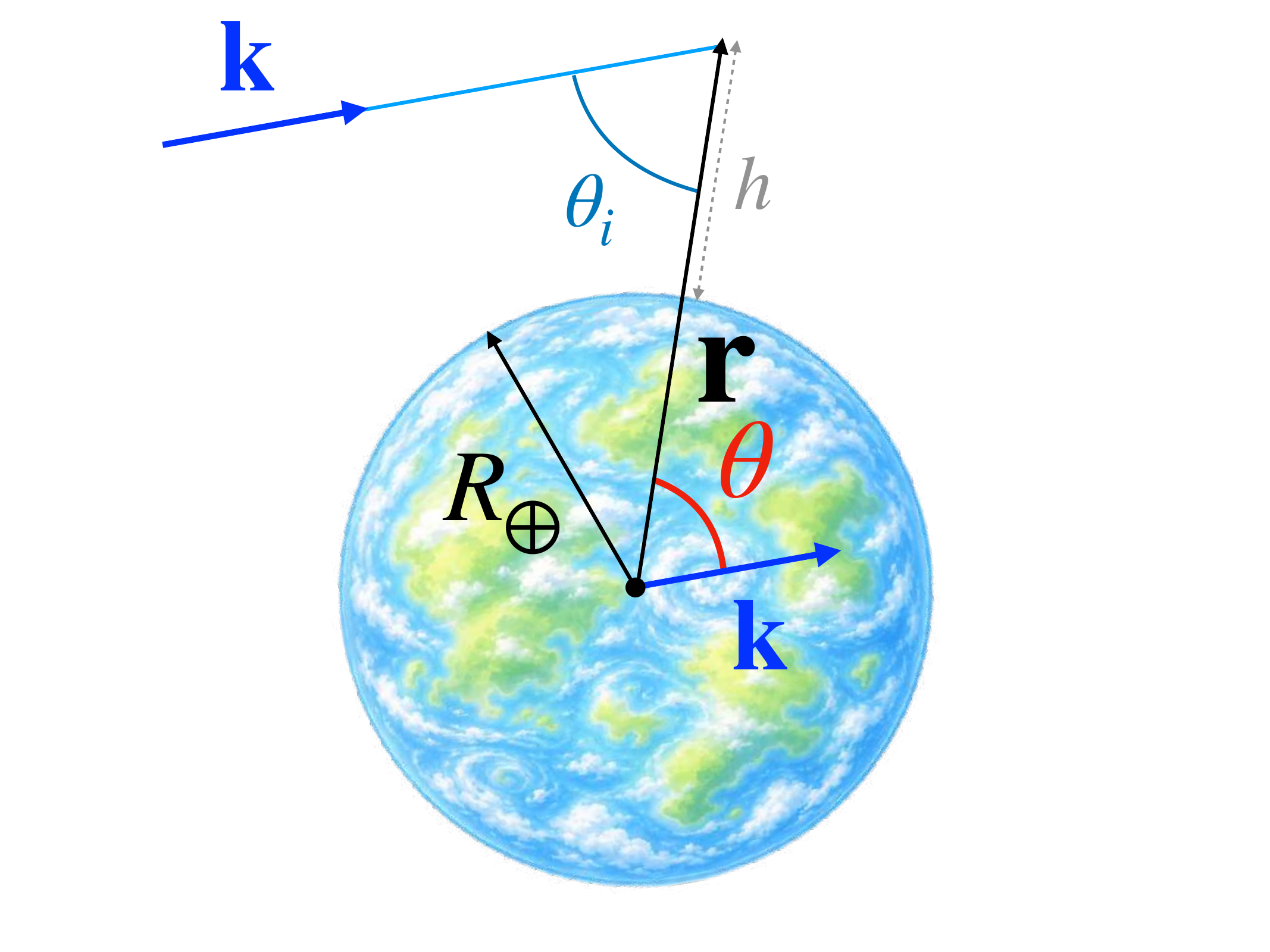}
  \caption{Ray geometry for the optical-limit calculation. {\bf Left}.  Contribution from the incident ray that reaches the observation point after specular reflection by the Earth. Applying the law of sines to the triangle formed by the Earth's center, the observation point, and the reflection point gives $\sin\theta_i/R_\oplus=\sin(\pi/2+\gamma)/r=\cos\gamma/r$. Thus $\gamma=\arccos[(r/R_\oplus)\sin\theta_i]$. Specular reflection then gives $\theta=\theta_i+2\gamma$. {\bf Right}. Contribution from the directly incident ray, which reaches the observation point without reflection. $\theta = \theta_i$.}
  \label{fig:rays}
\end{figure}

Since $\langle |\psi|^2\rangle$ is proportional to the local ULDM density, the normalized ensemble-averaged squared field amplitude is equal to the normalized local number density,
\begin{equation}
\frac{\langle |\psi|^2 \rangle (\vecr)}{|\psi_0|^2}
=
\frac{n_\phi(\vecr)}{n_{\phi}}\,,
\label{eq:ray_phi_density_relation}
\end{equation}
where the $\vecr$ dependence encodes the spatial inhomogeneity induced by the matter effect. In the optical regime where $\phi$ is treated as the collection of particles, this number density can be obtained by integrating the spectral radiance divided by the particle velocity, $\mathscr{L}/v$, over the directions of the incident rays,
\begin{equation}
  \frac{n_\phi(\vecr)}{n_\phi}
  =
  \frac{1}{n_\phi}
  \int \dd\Omega_i \int_0^\infty k^2\,\dd k\,
  \frac{\mathscr{L}(\vecr,\Omega_i,k)}{v}\, .
  \label{eq:density_from_radiance}
\end{equation}
Here $v=k/m_\phi$ is the particle velocity, and $\dd\Omega_i=\dd\phi_i\,\sin\theta_i\,\dd\theta_i$ is the solid-angle element associated with the direction $\Omega_i$, defined opposite to the ray-propagation direction and therefore labeling the incident ray reaching the point $\vecr$. 
Since the spectral radiance is conserved along each ray, including through specular reflection and direct incidence, we have
\bea
\mathscr{L}(\vecr,\Omega_i,k)
=
\mathscr{L}_\infty(\Omega_\veck,k),
\eea
where $\mathscr{L}_\infty(\Omega_\veck,k)$ denotes the asymptotic radiance at infinity. We therefore rewrite Eq.~\eqref{eq:density_from_radiance} as an integral over the direction of the incident momentum $\hat{\veck}$, which is also the asymptotic sky direction $\Omega_\veck \equiv \hat{\veck}$,
\begin{align}
  \frac{n_\phi(\vecr)}{n_\phi}
  &=
  \int \dd\Omega_\veck\,
  \left|\frac{\dd\Omega_i}{\dd\Omega_\veck}\right|
  \frac{1}{n_\phi}
  \int_0^\infty k^2\,\dd k\,
  \frac{\mathscr{L}_\infty(\Omega_\veck,k)}{k/m_\phi}
  \nonumber \\
  &=
  \int \dd\Omega_\veck\,4\pi\,T(r,\theta)\,I(\Omega_\veck)\, .
  \label{eq:density_kernel_convolution}
\end{align}
Here $\theta$ is the angle between the position vector direction $\hat{\vecr}$ and the incident momentum direction $\hat{\veck}$, as shown in \Fig{rays}. $T(r,\theta)$ is defined as
\begin{equation}
  T(r,\theta)
  \equiv
  \frac{1}{4\pi}
  \left|\frac{\dd\Omega_i}{\dd\Omega_\veck}\right|
  \label{eq:angular_kernel_definition},
\end{equation}
which is the angular redistribution kernel. With this normalization, $T=1/(4\pi)$ when the Earth does not redirect the rays. 
In the optical regime where the scalar is treated as classical particle, the spectral radiance from infinitely far divided by the particle velocity $k/m_\phi$ is precisely the DM phase-space distribution,
\bea
 \frac{\mathscr{L}_\infty(\Omega_\veck,k)}{k/m_\phi}  = f_\phi(\veck),
\eea
where $k/m_\phi$ is the particle velocity and $f_\phi(\veck)$ is the ULDM phase space distribution in \Eq{phase_space}. The angular intensity profile is therefore given by
\begin{align}
  I(\Omega_\veck)
  &=
  \frac{1}{n_\phi}
  \int_0^\infty k^2\,\dd k\,
  \frac{\mathscr{L}_\infty(\Omega_\veck,k)}{k/m_\phi} \nonumber\\
  &=
  \frac{1}{n_\phi}
  \int_0^\infty k^2\,\dd k\, f_\phi(\veck)\,,
  \label{eq:angular_intensity_profile}
\end{align} 
which captures the angular distribution of the incident DM. As shown in the second equal sign in \Eq{angular_intensity_profile}, for fixed geometric factors $\sigma_k/k_0$ and $k_0/k_\esc$, $I(\Omega_\veck)$ is independent of the DM mass $m_\phi$.

Because the kernel $T(r,\theta)$ is azimuthally symmetric around $\hat{\vecr}$, it can be expanded in terms of Legendre polynomials. We define the multipole transfer coefficients by
\begin{equation}
  T_L(r)
  =
  \frac{2L+1}{2}
  \int_0^\pi \sin\theta\,\dd\theta\,
  T(r,\theta) P_L(\cos\theta)\, .
  \label{eq:multipole_transfer_definition}
\end{equation}
Equivalently, $T(r,\theta)=\sum_{L=0}^\infty T_L(r)P_L(\cos\theta)$. Note that the coefficients $T_L(r)$, which encode how the geometry of the reflected rays modifies each angular multipole, depend only on the geometry and not on the DM distribution $f_\phi(\veck)$. 

Using the addition theorem of Legendre polynomials, we can then write the ensemble-averaged Eq.~\eqref{eq:density_kernel_convolution} over the DM distribution as
\begin{equation}
  \frac{n_\phi(\vecr)}{n_\phi}
  =
  \sum_{L=0}^{\infty}\sum_{M=-L}^{L}
  \frac{(4\pi)^2}{2L+1}
  I_L^M T_L(r) Y_L^M(\Omega_\vecr),
  \label{eq:ray_general}
\end{equation}
where the index $L$ is chosen to match the convention used in the ensemble-averaged multipole expansion in Appendix~\ref{appx:partial_wave_appx}. $\Omega_\vecr \equiv \hat{\vecr}$ denotes the direction of $\vecr$, and the multipole moments of the incident angular intensity are
\begin{equation}
  I_L^M
  =
  \int \dd\Omega_\veck\,I(\Omega_\veck)Y_L^{M*}(\Omega_\veck)\, .
  \label{eq:intensity_multipoles}
\end{equation}

\begin{table}
\centering
\setlength{\tabcolsep}{12pt}
\renewcommand{\arraystretch}{1.7}
\begin{tabular}{|c|c|c|c|c|c|c|}
\hline
Coefficient & $I^0_0$ & $I^0_1$ & $I^0_2$ & $I^0_3$ & $I^0_4$ & $I^0_5$ \\
\hline
    Value & $1/\sqrt{4\pi}$ & $0.310$ & $0.196$ & $0.0952$ & $0.0386$ & $0.0137$ \\
    \hline
\end{tabular}
\caption{The first six $M=0$ multipole moments of the incident angular intensity for the DM distribution described by the SHM.}
\label{tab:dm_multipoles}
\end{table}

We now compute the angular redistribution kernel using the geometry shown in \Fig{rays}. 
Consider a ray that reaches the observation point at radius $r$ with incidence polar angle $\theta_i$, measured from the inward radial direction. The corresponding asymptotic sky angle is
\begin{equation}
  \theta(\theta_i,r)
  =
  \begin{cases}
    \theta_i + 2\arccos\!\left(\dfrac{r}{R_\oplus}\sin\theta_i\right),
    & \theta_i < \pi/2 \ \text{and}\ \sin\theta_i < R_\oplus/r, \\[6pt]
    \theta_i,
    & \text{otherwise} .
  \end{cases}
  \label{eq:theta_of_theta_i}
\end{equation}
The first line applies to rays that intersect the Earth and undergo specular reflection before reaching the observation point, as illustrated in the left panel of \Fig{rays}. 
The second line applies to unreflected rays that reach the observation point directly, as illustrated in the right panel of \Fig{rays}.

% %%
\begin{figure}[t!]
\centering
\includegraphics[height=6cm]{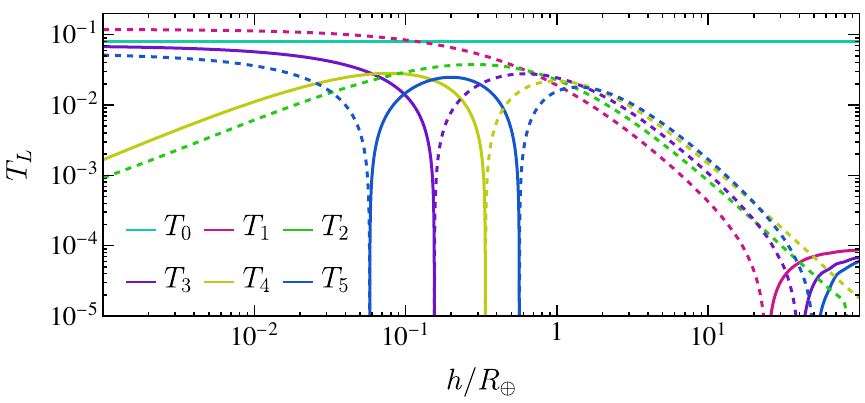}
\caption{Multipole transfer coefficients $T_L(r)$ as a function of $h/R_\oplus$ for $L=0,\ldots,5$. The solid curves correspond to cases where $T_L>0$, while the dashed curves indicate cases where $T_L<0$.}
\label{fig:multipole_transfer}
\end{figure}

\begin{figure}[t!]
\centering
\includegraphics[height=6cm]{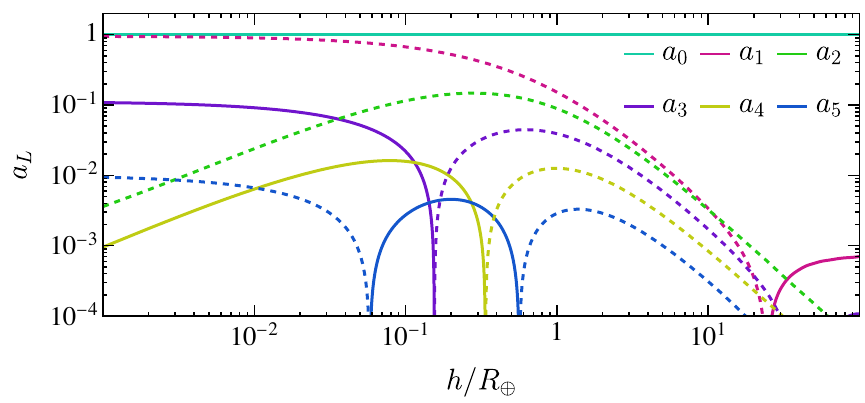}
\caption{The $a_L$-series as a function of $h/R_\oplus$ for $L=0,\ldots,5$ in the optical limit and the SHM. The solid curves correspond to cases where $a_L>0$, while the dashed curves indicate cases where $a_L<0$. We have numerically verified the consistency between the optical-limit calculation of $a_L$ and the multipole-expansion method described in \Sec{multipole}, in the regime $k_0 R_\oplus \gg 1$ and $k_0 h \gg 1$. }
\label{fig:ray_multipole}
\end{figure}

Rather than writing a closed-form expression for the kernel, we compute it as the probability density for a uniformly distributed local ray direction to map to a given point on the asymptotic sky. 
Let $p_r(\theta)$ be the one-dimensional probability density obtained by applying Eq.~\eqref{eq:theta_of_theta_i} while sampling $\cos\theta_i$ uniformly on $[-1,1]$. Equivalently,
\begin{equation}
  p_r(\theta)
  =
  \frac{1}{2}\sum_{\theta_i:\,\theta(\theta_i,r)=\theta}
  \left|\frac{\dd\cos\theta_i}{\dd\theta}\right| .
  \label{eq:pr_definition}
\end{equation}
Azimuthal symmetry then gives
\begin{equation}
  T(r,\theta)
  =
  \frac{p_r(\theta)}{2\pi\sin\theta} .
  \label{eq:kernel_from_probability_density}
\end{equation}
This form is convenient for numerical evaluation of the transfer coefficients in Eq.~\eqref{eq:multipole_transfer_definition}.

For this work, we take the SHM for the DM distribution. Figure~\ref{fig:multipole_transfer} shows the first few $T_L$ as a function of $h/R_\oplus$ while
Table~\ref{tab:dm_multipoles} lists the first six $M=0$ multipole coefficients $I^0_L$ for the SHM.
For the SHM, the incident angular distribution is axially symmetric. If the polar axis is chosen along the DM halo-wind $\veck_0$ as shown in \Fig{frame}, then only the coefficients with $M=0$ are nonzero, and Eq.~\eqref{eq:ray_general} reduces to
\begin{equation}
  \frac{n_\phi(r,\theta_\vecr)}{n_\phi}
  =
  \sum_{L=0}^{\infty}
  4\pi \, \sqrt{\frac{4\pi}{2L+1}}
  \, I_L^0 \, T_L(r) \, P_L(\cos\theta_\vecr),
  \label{eq:ray_axisymmetric}
\end{equation}
where $\theta_\vecr$ is the angle between $\vecr$ and the halo-wind axis $\veck_0$. Comparing Eq.~\eqref{eq:ray_axisymmetric} with Eq.~\eqref{eq:averaged_multipole_main}, the optical limit gives
\begin{equation}
  a_L(r)
  =
  4\pi \, \sqrt{\frac{4\pi}{2L+1}} \, 
  I_L^0 \, T_L(r)\, .
  \label{eq:ray_al}
\end{equation}

Figure~\ref{fig:ray_multipole} shows first few $a_L$ as a function of $h/R_\oplus$ in the optical limit. Recalling that $r=h+R_\oplus$, this figure shows that, in the optical limit, $a_0(r)\to 1$ so the monopole contribution to the background-induced force vanishes in contrast with the spherical limit, where the monopole provides the only contribution. This explains the strong suppression of $\dd a_0/\dd r$ in the large-$k_0 R_\oplus$ limit, in contrast to the higher-multipole derivatives $\dd a_L/\dd r$ with $L \geq 1$, which approach nonzero constants, as shown in \Fig{hard_sphere_a_series_derivatives} and \Fig{hard_sphere_a_series_log10_appx}.

As a consistency check, we numerically verified that the optical-limit calculation of $a_L$ given by \Eq{ray_al} agrees with the ensemble-averaged multipole-expansion method described in \Sec{multipole} in the regime $k_0 R_\oplus \gg 1$ and $k_0 h \gg 1$ over the full range of $h/R_\oplus$ considered. For the MICROSCOPE altitude, $h=0.1\,R_\oplus$, once $k_0 h \gg 1$, or equivalently $k_0 R_\oplus \gg 10$, the optical-limit values of $a_L$ agree with the large-$k_0 R_\oplus$ behavior shown in
\Fig{hard_sphere_a_series} and in the left panel of \Fig{hard_sphere_a_series_log10_appx}. 
The finite-difference derivatives $\dd a_L/\dd(r/R_\oplus)$ likewise agree with \Fig{hard_sphere_a_series_derivatives} and the right panel of \Fig{hard_sphere_a_series_log10_appx}.

\section{Frequency Band Structure}\label{appx:frequency_band_full_appx}

In \Subsec{force_proj_band_structure}, we present the frequency-band structure of the background-induced force projected onto the $\vecX$, $\vecY$, and $\vecZ$ axes. Here, we provide a detailed derivation of these results and give the corresponding frequency-band structure to arbitrary order.

As shown in \Eq{FXYZ_projection}, the projection of the background-induced force onto the $\vecX$ axis contains two contributions: the radial component, $(\vecX \cdot \widetilde{\vecF}_\bg)_{\vecr}$, and the polar component, $(\vecX \cdot \widetilde{\vecF}_\bg)_{\thetarhat}$. By contrast, the projection onto the $\vecY$ axis contains only the polar component, $(\vecY \cdot \widetilde{\vecF}_\bg)_{\thetarhat}$. The corresponding projection onto the $\vecZ$ axis can be obtained from the $\vecX$-projection result through the substitution $\Theta_\EP \rightarrow \Theta_\EP+\pi/2$.

We begin with the series representation of the Legendre polynomial given in \Eq{PL_cos_theta_r_expansion}, which we repeat here:
\begin{equation}
\label{eq:PL_cos_theta_r_expansion_appx}
P_L(\cos \theta_\vecr) = \frac{1}{2^L}\sum_{k=0}^{\left \lfloor{L/2}\right \rfloor}(-1)^k \binom{L}{k}\binom{2L - 2k}{L}\, (\cos \theta_\vecr)^{L - 2k}\,,
\end{equation}
where $\cos \theta_\vecr = - \cos\alpha \sin \Theta_\orb$ according to \Eq{cos_theta_r_in_orb}. To help the following calculation, we also use the following trigonometric identities:
\begin{equation}
\begin{aligned}
\label{eq:trigonometric_appx}
& (\cos E + i \sin E) \, (\sin O)^\mathscr{N} = \frac{1}{2^\mathscr{N}} \sum_{j=0}^\mathscr{N} \binom{\mathscr{N}}{j} (-1)^{j} e^{i\left[E+(\mathscr{N}-2j)\, O - \mathscr{N} \frac{\pi}{2}\right]}\\
& (\cos E + i \sin E) \, \cos O\, (\sin O)^\mathscr{N} = \frac{1}{2^{\mathscr{N}+1}} \sum_{j=0}^{\mathscr{N}+1} \frac{\mathscr{N}+1-2j}{\mathscr{N}+1} \binom{\mathscr{N}+1}{j} (-1)^j e^{i\left[ E + (\mathscr{N}-2j+1)\,O - \mathscr{N} \frac{\pi}{2} \right]}\\
\end{aligned}.
\end{equation}
These identities express products of trigonometric functions as linear superpositions of harmonic components, making the frequency structure explicit. The first identity follows from expressing $\sin O$ in exponential form and applying the binomial theorem. The second follows from applying $\partial_O$ to the first identity at order $\mathscr{N}+1$. In the following discussion, we take $E=\Theta_\EP$ and $O=\Theta_\orb$, respectively.

Substituting \Eq{PL_cos_theta_r_expansion_appx} into \Eq{FXYZ_projection}, and using the trigonometric identities in \Eq{trigonometric_appx} to decompose products of the form $\cos \Theta_\EP \, (\sin \Theta_\orb)^\mathscr{N}$ and $\sin \Theta_\EP \, \cos\Theta_\orb (\sin \Theta_\orb)^\mathscr{N}$ into sums of cosine harmonics exhibiting the frequency-band structure, we obtain
\begin{equation}
\begin{aligned}
\label{eq:XFbg_A_appx}
(\vecX \cdot \widetilde{\vecF}_\bg)_{\vecr} 
& = \sum_{L=0}^\infty \sum_{k=0}^{\lfloor \frac{L}{2} \rfloor} \sum_{j=0}^{L-2k} \frac{\dd a_L}{\dd r} \frac{1}{2^{2L-2k}} (-1)^{L-k+j+1} \binom{L}{k} \binom{2L-2k}{L} \binom{L-2k}{j}\\
& \quad \quad \quad \quad \quad \quad \times \cos^{L-2k} \alpha \,\, \cos\left[\Theta_\EP+(L-2k-2j)\Theta_\orb - (L-2k) \frac{\pi}{2} \right]\\
(\vecX \cdot \widetilde{\vecF}_\bg)_{\thetarhat} & = \sum_{L=0}^{\infty} \sum_{k=0}^{\lfloor \frac{L}{2}\rfloor} \sum_{j=0}^{L-2k} \frac{a_L}{r} \frac{1}{2^{2L-2k}} (-1)^{L-k+j} (L-2k-2j) \binom{L}{k} \binom{2L-2k}{L} \binom{L-2k}{j} \\
& \quad \quad \quad \quad \quad \quad \times \cos^{L-2k} \alpha \,\, \cos\left[ \Theta_\EP + (L-2k-2j) \Theta_\orb - (L-2k) \frac{\pi}{2} \right]
\end{aligned},
\end{equation}
for the radial and polar force projection along the $\vecX$ axis, respectively, and
\begin{equation}
\label{eq:YFbg_A_appx}
\begin{aligned}
(\vecY \cdot \widetilde{\vecF}_\bg)_{\thetarhat} 
& = \sin\alpha \,\, \sum_{L=1}^\infty \sum_{k=0}^{\lfloor \frac{L-1}{2} \rfloor} \sum_{j=0}^{L-1-2k} \frac{a_L}{r} \frac{1}{2^{2L-2k-1}} (-1)^{L-k+j} (L-2k) \binom{L}{k} \binom{2L-2k}{L} \binom{L-2k-1}{j} \\
& \quad \quad \quad \quad \quad \quad \times \cos^{L-2k-1} \alpha \,\, \cos\left[ (L-1-2k-2j) \Theta_\orb - (L-1-2k) \frac{\pi}{2} \right],
\end{aligned}
\end{equation}
for the force projection along the $\vecY$ axis. Up to this point, the background-induced force projections in \Eq{XFbg_A_appx} and \Eq{YFbg_A_appx} are still organized by the outermost multipole index $L$. While this representation is convenient for illustrating the convergence of the ensemble-averaged multipole expansion, contributions from different multipoles are mixed within the same frequency band, making the frequency-band structure less transparent.

To make the frequency-band structure in \Eq{XFbg_A_appx} and \Eq{YFbg_A_appx} more explicit, we reshuffle the sums by changing the summation indices and placing the index that labels the frequency bands in the outermost summation. For both $(\vecX \cdot \widetilde{\vecF}_\bg)_{\vecr}$ and $(\vecX \cdot \widetilde{\vecF}_\bg)_{\thetarhat}$, we introduce the change of summation variables $(L,k,j)\leftrightarrow(n,L,m)$ through the following bijective mapping:
\begin{equation}
\label{eq:X_reshuffle_index_appx}
\text{$\vecX$-Projection}: \quad \quad \left\{\begin{aligned}
& n = L-2k-2j \\
& m = L-2k \\
& L = L
\end{aligned}
\right. 
\quad \quad \text{and} \quad \quad
\left\{\begin{aligned}
& (L-m) \, \text{mod} \, 2 = 0 \\
& (m-n) \, \text{mod} \, 2 = 0 \\
& L \geq m \geq \abs{n} \\
& L \geq 0, \quad m \geq 0
\end{aligned}
\right. .
\end{equation}
Under this change of variables, the original summation in \Eq{XFbg_A_appx} can be reshuffled as
\bea
\label{eq:X_reshuffle_sum_appx}
\text{$\vecX$-Projection}: \quad \quad \sum_{L=0}^\infty \sum_{k=0}^{\lfloor \frac{L}{2} \rfloor} \sum_{j=0}^{L-2k} = \sum_{n=-\infty}^\infty \sum_{L=\abs{n}}^\infty \sum_{m=\abs{n}}^L \, \delta_{(L-m)\,\text{mod}\,2,\, 0} \,\, \delta_{(m-n)\, \text{mod}\,2,\, 0},
\eea
where $n$ directly labels the orbital main band and sidebands in both the radial and polar contributions to the background-induced force projected along the $\vecX$ axis. Here, the Kronecker deltas in \Eq{X_reshuffle_sum_appx} enforce that only terms for which $L$, $m$, and $n$ have the same parity contribute to the sum.

For $(\vecY \cdot \widetilde{\vecF}_\bg)_{\thetarhat}$, we similarly introduce the change of summation variables $(L,k,j)\leftrightarrow(n,L,m)$ through the following bijective mapping:
\begin{equation}
\label{eq:Y_reshuffle_index_appx}
\text{$\vecY$-Projection}: \quad \quad
\left\{\begin{aligned}
& n = (L-1)-2k-2j \\
& m = (L-1)-2k \\
& L = L
\end{aligned}
\right. 
\quad \quad \text{and} \quad \quad
\left\{\begin{aligned}
& (L-m) \, \text{mod} \, 2 = 1 \\
& (m-n) \, \text{mod} \, 2 = 0 \\
& L - 1 \geq m \geq \abs{n} \\
& L \geq 1,  \quad m \geq 0
\end{aligned}
\right..
\end{equation}
Accordingly, the original summation in \Eq{YFbg_A_appx} can be reorganized as
\bea
\label{eq:Y_reshuffle_sum_appx}
\text{$\vecY$-Projection}: \quad \quad \sum_{L=1}^\infty \sum_{k=0}^{\lfloor \frac{L-1}{2} \rfloor} \sum_{j=0}^{L-1-2k} = \sum_{n=-\infty}^{\infty} \sum_{L=\abs{n}+1}^\infty \sum_{m=\abs{n}}^{L-1} \delta_{(L-m)\,\text{mod}\,2,\, 1} \,\, \delta_{(m-n)\,\text{mod}\,2,\, 0},
\eea
with the outermost index $n$ labeling the orbital harmonics of the force projected along the $\vecY$ axis. Here, the Kronecker deltas in \Eq{Y_reshuffle_sum_appx} enforce that $m$ and $n$ have the same parity, whereas $L$ and $m$ have opposite parity, so that only terms satisfying both conditions contribute to the sum.

After reshuffling \Eq{XFbg_A_appx} using \Eq{X_reshuffle_index_appx} and \Eq{X_reshuffle_sum_appx}, and similarly reshuffling \Eq{YFbg_A_appx} using \Eq{Y_reshuffle_index_appx} and \Eq{Y_reshuffle_sum_appx}, we finally obtain
\begin{equation}
\begin{aligned}
& (\vecX \cdot \widetilde{\vecF}_\bg)_{\vecr} = \sum_{n=-\infty}^{\infty} R_n^X \cos\left(\Theta_{\rm EP}+n\,\Theta_{\rm orb}+\frac{n\pi}{2}\right)\\
& (\vecX \cdot \widetilde{\vecF}_\bg)_{\thetarhat} = \sum_{n=-\infty}^{\infty} T_n^X \cos\left(\Theta_{\rm EP}+n\,\Theta_{\rm orb}+\frac{n\pi}{2}\right)\\
& (\vecY \cdot \widetilde{\vecF}_\bg)_{\thetarhat} = \sum_{n=0}^\infty T_n^Y \cos\left(n\,\Theta_{\rm orb}+\frac{n\pi}{2}\right).
\end{aligned}
\end{equation}
Here, $n$ labels the frequency band. The force projected along the $\vecX$ axis is expressed in a two-sided form, since its spectrum contains sidebands at $\omega_\EP \pm n\,\omega_\orb$ around the main band at $\omega_\EP$, which is generally much larger than $\omega_\orb$~(See \Eq{spin_mode} for example). By contrast, the force projected along the $\vecY$ axis is written in a one-sided form, since its main band is the DC component and the remaining contributions occur at nonnegative integer multiples of $\omega_\orb$. The coefficients $R_n^X$ and $T_n^X$ characterize, respectively, the radial and polar components of the background-induced force projected along the $\vecX$ axis, while $T_n^Y$ characterizes the polar component projected along the $\vecY$ axis. 

Here, we provide analytical expressions for $R_n^X$, $T_n^X$, and $T_n^Y$, which make the underlying symmetries and selection rules manifest and facilitate their numerical evaluation. To express these coefficients compactly, we use the multinomial coefficient
\bea
\label{eq:multinomial_appx}
\binom{n}{n_1 \, n_2 \, n_3\,n_4} \equiv \frac{n!}{n_1!\,n_2!\,n_3!\,n_4!},
\eea
where $n=n_1+n_2+n_3+n_4$. We define the multinomial coefficient in \Eq{multinomial_appx} to vanish whenever any of the entries is not a nonnegative integer. The coefficients associated with the $\vecX$ projection are then given by
\begin{equation}
\label{eq:RnX_TnX_appx}
\begin{aligned}
& (\vecX \cdot \widetilde{\vecF}_\bg)_{\vecr}: \quad \quad R_n^X = - \sum_{L=\abs{n}}^\infty \, \frac{\dd a_L}{\dd r} \, \sum_{m=\abs{n}}^L \cos^m \alpha \, \delta_{(L-m)\,\text{mod}\,2,\, 0} \,\, \delta_{(m-n)\, \text{mod}\,2, \,0} \\
& \quad \quad \quad \quad \quad \quad \quad \quad \quad \quad \quad \quad \quad \times (-1)^{\frac{L-m}{2}} \, \frac{1}{2^{L+m}} \binom{L+m}{\frac{L-m}{2} \, \frac{L+m}{2} \, \frac{m+n}{2}\,\frac{m-n}{2} }\\
& (\vecX \cdot \widetilde{\vecF}_\bg)_{\thetarhat}: \quad \quad T_n^X = n \sum_{L=\abs{n}}^\infty \frac{a_L}{r} \sum_{m=\abs{n}}^L \cos^m \alpha \, \delta_{(L-m)\,\text{mod}\,2, \,0} \,\, \delta_{(m-n)\,\text{mod}\,2, \,0} \\
& \quad \quad \quad \quad \quad \quad \quad \quad \quad \quad \quad \quad \quad \times (-1)^{\frac{L-m}{2}} \, \frac{1}{2^{L+m}} \binom{L+m}{\frac{L-m}{2}\,\frac{L+m}{2}\,\frac{m+n}{2}\,\frac{m-n}{2}}
\end{aligned}.
\end{equation}
Several important properties follow immediately from \Eq{RnX_TnX_appx}. First, $R_n^X$ is an even function of $n$, whereas $T_n^X$ is an odd function:
\bea
R_{-n}^X = R_n^X, \quad \quad \quad T_{-n}^X = - T_n^X. 
\eea
Consequently, only $R_0^X$ contributes to the main $\omega_\EP$-band, while $T_0^X=0$. Second, for sidebands with $\abs{n}\geq1$, the $\dd a_L/\dd r$ contributions to $R_n^X$ and the $a_L/r$ contributions to $T_n^X$ arise only from multipoles satisfying $L\geq\abs{n}\geq1$. Therefore, in the long-wavelength limit $k_0R_\oplus\ll1$, where the higher multipoles are suppressed, the sidebands vanish, as shown in \Fig{X_band}. This indicates that the sidebands arise directly from the departure of the scalar distribution from spherical symmetry. Third, both $R_n^X$ and $T_n^X$ contain factors of $\cos^m\alpha$, with $m\geq\abs{n}$ as shown in \Eq{RnX_TnX_appx}. Sidebands with larger $\abs{n}$ therefore lead to higher powers of $\cos\alpha$ and consequently exhibit stronger annual modulation as $\alpha(t)$ varies with time. Fourth, the Kronecker deltas in \Eq{RnX_TnX_appx} impose the parity selection rule that $L$ and $n$ must have the same parity. Consequently, for the force projected onto the $\vecX$ axis, even~(odd) sidebands receive contributions exclusively from even~(odd) multipoles. In particular, the main band at $\omega_\EP$, corresponding to $n=0$, receives contributions only from even multipoles. Therefore, when the monopole contribution $\dd a_0/\dd r$ becomes suppressed at $k_0R_\oplus \gg 1$, as shown in \Fig{hard_sphere_a_series_derivatives} and \Fig{hard_sphere_a_series_log10_appx}, the leading contribution arises from the quadrupole term $\dd a_2/\dd r$, rather than from the dipole term $\dd a_1/\dd r$. This selection rule underlies one of the key observations made in the main text in our reanalysis of the MICROSCOPE data in \Subsec{microscope_revisit}. Furthermore, in the regime $k_0 R_\oplus \gg 1$, the first sidebands at $\omega_\EP \pm \omega_\orb$, corresponding to $n=\pm1$, are found numerically to be dominated by the dipole-derivative term
$\dd a_1/\dd r$ appearing in \Eq{X_1st_sideband_coefficient}. Since this contribution exceeds the leading quadrupole contribution $\dd a_2/\dd r$ to the main $\omega_\EP$ band, corresponding to $n=0$, the combined contribution of the first sidebands to the sensitivity is larger than that of the main band, as shown in \Fig{X_band} and \Fig{ng_full_compare}. Consequently, when all relevant frequency bands are included, the overall sensitivity in the high-momentum regime, $k_0 R_\oplus \gg 1$, is dominated by the first sidebands. Because the force projection along the $\vecZ$ axis is obtained from that along the $\vecX$ axis by the substitution $\Theta_\EP \rightarrow \Theta_\EP+\pi/2$, the corresponding coefficients satisfy $R_n^Z=R_n^X$ and $T_n^Z=T_n^X$.

The coefficient associated with the $\vecY$ projection is given by
\begin{equation}
\label{eq:TnY_appx}
\begin{aligned}
& (\vecY \cdot \widetilde{\vecF}_\bg)_{\thetarhat}: \quad \quad T_n^Y =  (2-\delta_{n,\,0}) \,\, \sin \alpha \, \sum_{L=n+1}^\infty \frac{a_L}{r} \sum_{m=n}^{L-1} \cos^m \alpha \, \delta_{(L-m)\,\text{mod}\,2, \,1} \,\, \delta_{(m-n)\,\text{mod}\,2, \,0} \\
& \quad \quad \quad \quad \quad \quad \quad \quad \quad \quad \quad \quad \quad \times (-1)^{\frac{L-m+1}{2}} \, \frac{L+m+1}{2^{L+m}} \binom{L+m}{ \frac{L-m-1}{2} \, \frac{L+m+1}{2}\, \frac{m-n}{2}\,\frac{m+n}{2} }
\end{aligned}.
\end{equation}
Here, the factor $(2-\delta_{n,0})$ accounts for the one-sided representation by doubling the contributions from the nonzero harmonics while leaving the DC component unchanged. Several important properties follow directly from \Eq{TnY_appx}. First, the lowest multipole contributing to $T_n^Y$ is $L=n+1$. Consequently, in the long-wavelength limit $k_0R_\oplus\ll1$, all frequency bands vanish because they receive contributions only from multipoles with $L\geq1$, for which $a_L$ is suppressed in this limit as shown in \Fig{hard_sphere_a_series_log10_appx}. Equivalently, from an intuitive perspective, the frequency-band structure in the $\vecY$ projection arises only from departures from spherical symmetry. Second, in addition to the factors of $\cos^m\alpha$, $T_n^Y$ contains an overall factor of $\sin\alpha$. The frequency bands in the $\vecY$ projection therefore exhibit a particularly strong dependence on the annually varying angle $\alpha(t)$ and can undergo more pronounced annual modulation than those in the $\vecX$ projection. In particular, because $\alpha(t)$ crosses zero during the MICROSCOPE mission, all $T_n^Y$ coefficients vanish at the crossing and subsequently reverse sign, giving rise to a distinctive annual modulation. Third, the Kronecker deltas in \Eq{TnY_appx} impose the parity selection rule that $L$ and $n$ must have opposite parity. Consequently, for the force projected onto the $\vecY$ axis, even~(odd) sidebands receive contributions exclusively from odd~(even) multipoles.

Equations \Eq{RnX_TnX_appx} and \Eq{TnY_appx} provide the generic representations, valid to arbitrary frequency-band order, of the coefficients associated with the main band and sidebands in the $\vecX$ and $\vecY$ projections. In \Subsec{force_proj_band_structure} in the main text, we give the first concrete representation of $R_n^X$, $T_n^X$, as well as $T_n^Y$ to leading orders. 

\section{Matter Effect from Atmosphere}\label{appx:atm_matter_effect}

\begin{figure}[t]
\centering
\includegraphics[width=0.5\linewidth]{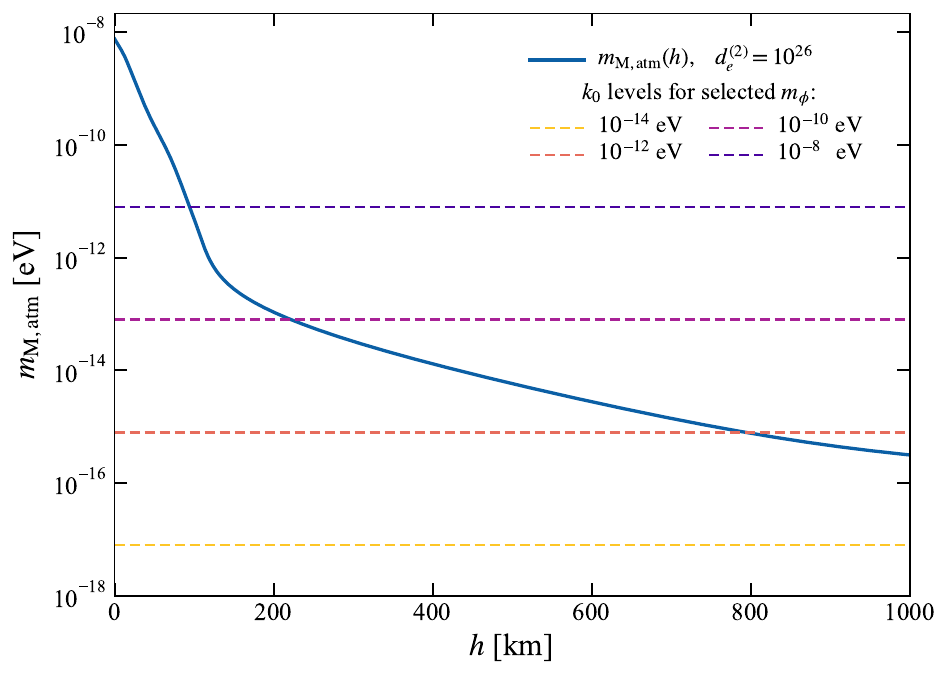}
\caption{Matter-induced scalar mass $\mMatm$ as a function of altitude $h$ for the benchmark photon coupling $d_e^{(2)}=10^{26}$.}
\label{fig:atm_mM_profile}
\end{figure}

In this section, we examine the impact of atmospheric matter effects on space-based EP tests. We find that these effects become qualitatively important for the MICROSCOPE constraint only in the ultra-strong-coupling regime, beyond the strong-coupling ceiling imposed by satellite screening. Atmospheric matter effects can therefore be safely neglected throughout the parameter space relevant to our analysis.

We evaluate the transparency of the atmosphere at different altitudes. We begin by computing the matter-induced scalar mass $\mMatm$ as a function of altitude $h$, using the atmospheric density profile from NRLMSIS 2.1~\cite{Emmert2022NRLMSIS21}, implemented through \texttt{pymsis}~\cite{Lucas2022pymsis}. The matter-induced mass is given by
\bea
\mMatm^2(h) = \frac{4 \pi}{\Mpl^2} \alpha^{(2)}_\atm \,\, \rho_\atm(h)\,,
\eea
where $\rho_\atm$ is the averaged matter density at height $h$ obtained from NRLMSIS. For illustration, we show the altitude dependence of $\mMatm$ in
\Fig{atm_mM_profile}.
\begin{table}[t!]
\centering
\newcommand{\atmTabWidth}{0.82\columnwidth}
\setlength{\tabcolsep}{4.5pt}
\renewcommand{\arraystretch}{1.7}
% ----------------------------
\begin{tabular}{|c|c|c|c|c|c|}
\hline
\makecell{Atmosphere\,$[\mathrm{km}]$}
& $0\mbox{--}100$
& $100\mbox{--}200$
& $200\mbox{--}300$
& $300\mbox{--}600$
& $600\mbox{--}1000$
\\
\hline
\makecell{$\tau_{\rm atm}$}
& $6\times10^{2}$
& $4\times10^{-1}$
& $3\times10^{-2}$
& $2\times10^{-2}$
& $10^{-3}$
\\
\hline
\end{tabular}%
\caption{Atmospheric optical depth $\tau_{\rm atm}$ for the benchmark photon coupling $d_e^{(2)}=10^{26}$. In this case, only the atmospheric layer at altitudes of $0\mbox{--}100\,\mathrm{km}$ is optically thick, and it has a negligible impact on the resulting constraint. The optical depth for other couplings can be obtained by rescaling according to \Eq{optical_depth_appx}. We find that atmospheric matter effects become non-negligible only for sufficiently large couplings, beyond the strong-coupling ceiling imposed by satellite screening.}
\label{tab:atm_tau_examples_appx}
\end{table}

In the strong-coupling limit, $k_0 \ll \mMatm$, the suppression of the scalar amplitude across an atmospheric shell extending from $h_1$ to $h_2$ can be estimated by counting the number of penetration depths, $\mMatm^{-1}$, contained within the atmospheric layer. The resulting suppression factor is $\exp(-\tau_\text{atm})$, where the optical depth is defined as
\bea
\label{eq:optical_depth_appx}
\tau_\text{atm}(h_1, h_2) = \int_{h_1}^{h_2} \dd h \, \mMatm(h) 
= \sqrt{\frac{4 \pi \alpha^{(2)}_\atm}{\Mpl^2}} \int_{h_1}^{h_2} \dd h \, \sqrt{\rho_\atm(h)}\,.
\eea
The optical depth scales as $\sqrt{\alpha^{(2)}_\atm}$ and depends on the altitude interval considered. To account for the full atmospheric profile, we divide the atmosphere into several altitude intervals and evaluate the optical depth of each layer, as summarized in Table~\ref{tab:atm_tau_examples_appx}. To recast these results in the $m_\phi$--$d^{(2)}_i$ plane through $\alpha_\atm^{(2)} \sim (Q_\atm)_i \, d^{(2)}_i$, we adopt dilaton charges for air, taking volume fractions of $78\%$ for $N_2$ and $22\%$ for $O_2$. The resulting charges are
\begin{equation}
\label{eq:atm_dilaton_charge_appx}
\begin{array}{@{}lll@{}}
\quad \quad \quad (Q_\atm)_g = 1, 
& \quad \quad \quad (Q_\atm)_e \simeq 1.3 \times 10^{-3},
& \quad \quad \quad (Q_\atm)_{m_e} \simeq 2.7 \times 10^{-4}, \\[6pt]
\quad \quad \quad (Q_\atm)_{\hat{m}} \simeq 9.4 \times 10^{-2},
& \quad \quad \quad (Q_\atm)_{\delta m} \simeq 4.2 \times 10^{-7},
& \quad \quad \quad (Q_\atm)_{H} \simeq 0.3.
\end{array}
\end{equation}
In the following discussion, we take the scalar--photon interaction, parameterized by $d_e^{(2)}$, as a representative example and set all other couplings to zero. The corresponding results for a different coupling $d_i^{(2)}$ can be obtained by keeping $\alpha_\atm^{(2)}$ fixed and applying the rescaling $d_i^{(2)} \sim d^{(2)}_e \times (Q_\atm)_e /(Q_\atm)_i$.

We first consider the lower, weak-coupling boundary of the excluded region in \Fig{classification}. As a conservative benchmark, we take $d_e^{(2)}=10^{26}$, which is larger than the coupling along most of this boundary. At this benchmark, we find that the optical depth from the Earth's surface to an altitude of $100\,\km$ reaches $\mathcal{O}(10^2)$, while the optical depths at higher altitudes rapidly decrease, becoming $\mathcal{O}(10^{-2})$ or smaller above $200\,\km$ and negligible for the atmospheric column above MICROSCOPE. This implies that, for this fiducial value of $d_e^{(2)}$, the scalar-field profile around the Earth is only slightly modified by the atmosphere, leading to an $\mathcal{O}(10\%)$ correction to the final result. Since the coupling along most of the lower boundary in \Fig{classification} is considerably weaker than this benchmark value, the atmospheric matter effect is negligible over most of this boundary. We next examine the coupling strength at which atmospheric screening may become relevant, taking $d_e^{(2)}=10^{30}$ as a representative benchmark. At this coupling, the atmospheric layer below the 
MICROSCOPE altitude only begins to develop a non-negligible optical depth, marking the onset of atmospheric screening rather than a regime of substantial attenuation. Its impact on the MICROSCOPE constraint therefore remains limited. A significant modification would arise only for $d_e^{(2)}\gg 10^{30}$, when even the atmosphere above the MICROSCOPE orbit becomes optically thick. However, this regime already lies above the strong-coupling ceiling of the MICROSCOPE exclusion region imposed by satellite screening. Our optical-depth criterion provides an approximate
estimate of the relevance of atmospheric screening,
rather than a complete calculation of the scalar-field
profile and the resulting MICROSCOPE constraints.
A complete treatment of atmospheric screening,
including the nonspherically symmetric structure and
temporal variations of the atmosphere, would require
a dedicated numerical analysis~\footnote{Atmospheric screening can also be estimated by propagating scalar partial waves through a spherically averaged
atmospheric potential, as done in
Ref.~\cite{Brzeminski:2026pgz}. Although the two approaches are technically different,
they provide a qualitatively similar picture of atmospheric
screening, indicating that its effects are negligible
over most of the parameter space of interest.}.

\FloatBarrier
\bibliography{References}

\end{document}